\documentclass[sigconf, nonacm]{acmart}

\newcommand\vldbdoi{XX.XX/XXX.XX}
\newcommand\vldbpages{XXX-XXX}
\newcommand\vldbvolume{14}
\newcommand\vldbissue{1}
\newcommand\vldbyear{2022}
\newcommand\vldbauthors{\authors}
\newcommand\vldbtitle{\shorttitle} 
\newcommand\vldbavailabilityurl{URL_TO_YOUR_ARTIFACTS}
\newcommand\vldbpagestyle{plain} 

\usepackage{tikz}
\usepackage{listings}
\usepackage{multirow}
\usepackage{rotating}
\usepackage{pifont}
\usepackage{colortbl}
\usepackage{tabulary}

\usepackage{todonotes}
\usepackage{xspace}
\newcommand{\name}{SparqLog\xspace}

\newif\ifFullVersion

\FullVersiontrue

\newcommand{\mset}[1]{ \{ \hskip-2pt  \{ #1 \} \hskip-2pt\} }
\newcommand{\epag}[2]{\llbracket #1 \rrbracket_{#2}} 
\newcommand{\lojoin}{\sqsupset \hskip -1.7pt \Join}
\newcommand\rsb{\rule{0pt}{2.6ex}} 
\newcommand\rsa{\rule[-1.2ex]{0pt}{0pt}} 
\newcommand\rs{\rsb \rsa} 
\newcommand{\names}{\mathit{names}}

\usepackage{svg}

\usepackage{mathtools} 
\usepackage[inline]{enumitem}
\usepackage{subcaption}

\newcommand{\cmark}{\ding{51}}
\newcommand{\xmark}{\ding{55}}
\newcommand*\rot{\rotatebox{90}}

\newcommand{\code}[1]{\texttt{\small#1}} 

\newcommand{\var}{\operatorname{var}}
\newcommand{\graph}{\operatorname{gr}}
\newcommand{\dom}{\operatorname{dom}}

\newcommand{\false}{\operatorname{\emph{false}}}
\newcommand{\true}{\operatorname{\emph{true}}}

\newcommand{\bound}{\operatorname{bound}}
\newcommand{\SELECT}{\operatorname{SELECT}}
\newcommand{\CONSTRUCT}{\operatorname{CONSTRUCT}}
\newcommand{\DESCRIBE}{\operatorname{DESCRIBE}}
\newcommand{\ASK}{\operatorname{ASK}}
\newcommand{\DISTINCT}{\operatorname{DISTINCT}}
\newcommand{\FROM}{\operatorname{FROM}}
\newcommand{\FROMN}{\operatorname{FROM~NAMED}}
\newcommand{\WHERE}{\operatorname{WHERE}}
\newcommand{\AAND}{\operatorname{{.}}}
\newcommand{\JOIN}{\operatorname{{.}}}
\newcommand{\UNION}{\operatorname{UNION}}
\newcommand{\OPT}{\operatorname{OPT}}
\newcommand{\OPTIONAL}{\operatorname{OPT}}
\newcommand{\FILTER}{\operatorname{FILTER}}
\newcommand{\NEX}{\operatorname{FILTER~NOT~EXISTS}}
\newcommand{\MINUS}{\operatorname{MINUS}}

\newcommand{\GRAPH}{\operatorname{GRAPH}}
\newcommand{\LIMIT}{\operatorname{LIMIT}}
\newcommand{\OFFSET}{\operatorname{OFFSET}}
\newcommand{\COUNT}{\operatorname{COUNT}}
\newcommand{\GBY}{\operatorname{GROUP~BY}}
\newcommand{\OBY}{\operatorname{ORDER~BY}}
\newcommand{\isI}{\operatorname{isIRI}}
\newcommand{\isL}{\operatorname{isLiteral}}
\newcommand{\isB}{\operatorname{isBlank}}
\newcommand{\regex}{\operatorname{regex}}

\newcommand{\inv}{\string^}

\newcommand{\revision}[1]{#1}

\usepackage{fancyvrb}

\begin{document}
\title{SparqLog: A System for Efficient Evaluation of SPARQL 1.1 Queries via Datalog 
[Experiment, Analysis \& Benchmark]}

\author{Renzo Angles}
\affiliation{\institution{Universidad de Talca, Chile}
}

\author{Georg Gottlob}
\affiliation{\institution{University of Oxford, UK}
}

\author{Aleksandar Pavlovi\'{c}}
\affiliation{\institution{TU Wien, Austria}
}

\author{Reinhard Pichler}
\affiliation{\institution{TU Wien, Austria}
}

\author{Emanuel Sallinger}
\affiliation{\institution{TU Wien, Austria}
}

\begin{abstract}
Over the past decade, Knowledge Graphs have received enormous interest both from industry and
from academia. Research in this area has been driven, above all, by 
the Database (DB) community and the Semantic Web  (SW) community. 
However, there still remains a certain divide between
approaches coming from these two communities.
For instance, while languages such as SQL or Datalog are widely used in the DB area, a different set of languages such as SPARQL and OWL is used in the SW area. Interoperability between such technologies
is still a challenge. 
The goal of this work is to present a uniform and consistent framework 
meeting important requirements
from both, the SW and DB field.
\end{abstract}

\maketitle

\pagestyle{\vldbpagestyle}
\begingroup\small\noindent\raggedright\textbf{PVLDB Reference Format:}\\
\vldbauthors. \vldbtitle. PVLDB, \vldbvolume(\vldbissue): \vldbpages, \vldbyear.\\
\href{https://doi.org/\vldbdoi}{doi:\vldbdoi}
\endgroup
\begingroup
\renewcommand\thefootnote{}\footnote{\noindent
This work is licensed under the Creative Commons BY-NC-ND 4.0 International License. Visit \url{https://creativecommons.org/licenses/by-nc-nd/4.0/} to view a copy of this license. For any use beyond those covered by this license, obtain permission by emailing \href{mailto:info@vldb.org}{info@vldb.org}. Copyright is held by the owner/author(s). Publication rights licensed to the VLDB Endowment. \\
\raggedright Proceedings of the VLDB Endowment, Vol. \vldbvolume, No. \vldbissue\ ISSN 2150-8097. \\
\href{https://doi.org/\vldbdoi}{doi:\vldbdoi} \\
}\addtocounter{footnote}{-1}\endgroup

\ifdefempty{\vldbavailabilityurl}{}{
\vspace{.3cm}
\begingroup\small\noindent\raggedright\textbf{PVLDB Artifact Availability:}\\
The source code, data, and/or other artifacts have been made available at \url{\vldbavailabilityurl}.
\endgroup
}

\section{Introduction}
\label{sect:Introduction}

Since Google launched its Knowledge Graph (KG) roughly a decade ago, we have seen intensive work on 
this topic both in industry and in academia. However, there are two research communities working mostly isolated from each other on the development of KG management systems, namely the \emph{Database} and the \emph{Semantic Web} community. Both of them come with their specific key requirements and they have introduced their own approaches. 

Of major importance to the {\em Semantic Web} (SW) community is the compliance with the 
relevant W3C standards:

\smallskip
\noindent
\textbf{[RQ1] SPARQL Feature Coverage}. The query language SPARQL is one of the major Semantic Web standards. Therefore, we require the support of the 
most commonly used SPARQL features. 
\\[1.1ex]
\textbf{[RQ2] Bag Semantics}. SPARQL employs per default \emph{bag semantics} (also referred to as \emph{multiset semantics}) 
unless specified otherwise in a query.
We therefore require the support of this.

\smallskip
\noindent
\textbf{[RQ3] Ontological Reasoning}. OWL 2 QL to support ontological reasoning is a major Semantic Web standard. Technically, for rule-based languages, this means that 
existential quantification (i.e., ``object invention'') in the rule heads is required.

\medskip

The {\em  Database} (DB) community puts particular emphasis on the expressive power and efficient evaluation of query languages. This leads us to the following additional requirement:

\smallskip
\noindent
\textbf{[RQ4] Full Recursion}. Full recursion is vital to 
provide the expressive power needed to support complex querying in business applications and sciences (see e.g., \cite{DBLP:conf/fgit/PrzymusBBS10})
and it is the main feature of the relational query language Datalog \cite{DBLP:conf/pods/Vianu21}. 
Starting with SQL-99, recursion has also been integrated into the 
SQL standard and most relational database management systems have meanwhile incorporated recursion capabilities to increase their expressive power.

\medskip

Finally, for an approach to be {\em accepted and used in practice}, 
we formulate the following requirement for both communities: 

\smallskip
\noindent
\textbf{[RQ5] Implemented System}. 
Both communities require an implemented system. 
This makes it possible to verify if the theoretic results are applicable in practice and to evaluate the usefulness of the approach under real-world settings.

\smallskip
The above listed requirements explain why 
there exists a certain gap between the SW and DB communities. 
There have been several attempts to close this gap. However, as will be detailed in 
Section~\ref{sect:RelatedApproaches}, 
no approach has managed to fulfil the requirements of both sides so far. 
Indeed, while existing solutions individually satisfy some of the requirements
listed above, all of them fail to satisfy central other requirements. 
The goal of this  work is to 
develop one uniform and consistent framework that 
satisfies the requirements of both communities.
\noindent
More specifically, our contributions are as follows: 

\smallskip
\noindent
\textbf{Theoretical Translation}.
We provide a uniform and complete framework to 
integrate SPARQL support into a KG language 
that meets all of the above listed requirements RQ1--RQ5. 
We have thus extended, simplified and -- in some cases -- corrected previous approaches of translating SPARQL queries (under both set and bag semantics) to various Datalog dialects \cite{Polleres2007FromST,AnswerSetTrans,Angles2016TheMS}.
For instance, to the best of our knowledge, all previous translations have missed or did not consider correctly certain aspects of the SPARQL standard of the 
zero-or-one and zero-or-more property paths. 
        
\smallskip
\noindent
\textbf{Translation Engine.}    
We have developed the translation engine \name on top of the Vadalog system
that covers most of the considered SPARQL 1.1 functionality.
We thus had to fill several gaps between the abstract theory and the practical development of the translation engine. 
For instance, to support bag semantics, we have designed specific Skolem functions to generate a universal duplicate preservation process. On the other hand, the use of the Vadalog system as the 
basis of our engine made significant simplifications possible (such as letting Vadalog take care of complex filter constraints) and we also get ontological reasoning ``for free''.
\name therefore supports both query answering and ontological reasoning
in a single uniform and consistent system.

\smallskip
\noindent
\textbf{Experimental Evaluation}.
We carry out an extensive empirical evaluation on multiple benchmarks with 
two main goals in mind: to  verify the compliance of 
\name with the SPARQL standard 
as well as to compare the performance of our system with 
comparable ones. It turns out that, while \name covers a great part of 
the selected SPARQL 1.1 functionality in the correct way, 
some other systems (specifically Virtuoso) employ a non-standard behaviour on queries containing property paths. As far as query-execution times are concerned, 
the performance of \name is, in general, comparable to other systems such as the SPARQL system Fuseki or the querying and reasoning system Stardog and it significantly outperforms 
these systems on complex queries containing recursive property paths and/or involving ontologies. 

\smallskip

\noindent
{\em Structure of the paper.}
After 
a review of existing approaches in Section \ref{sect:RelatedApproaches}, 
and the  preliminaries in Section \ref{sect:Preliminaries}, 
we present our main results: 
the general principles of our 
\name system
in Section \ref{sec:integration}, 
a more detailed look into the
translation engine in 
Section~\ref{sect:TranslationEngine},
and an experimental evaluation 
in Section \ref{sect:ExperimentalEvaluation}. 
We conclude with  Section \ref{sect:Conclusion}.
\ifFullVersion
Further details on our theoretical translation, implementation, and 
experimental results are provided in the appendix.
\fi

\section{Related Approaches}
\label{sect:RelatedApproaches} 

We review several approaches -- both from the Semantic Web and the Database community. 
This discussion of related approaches is divided into 
theoretical and practical aspects of our work. 

\subsection{Theoretical Approaches}
\label{sec:SotATranslation}

Several theoretical research efforts have 
aimed at bridging the gap between the DB and SW 
communities. 

\smallskip
\noindent
\textbf{Translations of SPARQL to Answer Set Programming.} 
In a series of papers, Polleres et al.~presented translations of SPARQL and SPARQL 1.1 
to various extensions of Datalog. The first translation from SPARQL to Datalog \cite{Polleres2007FromST} 
converted SPARQL queries into Datalog programs by employing negation as failure. 
This translation was later extended by the addition of new features of SPARQL 1.1 and by considering its bag semantics in \cite{AnswerSetTrans}. Thereby, Polleres and Wallner created a nearly complete translation of SPARQL 1.1 queries to Datalog with disjunction (DLV) programs. However, the translation had two major drawbacks: On the one hand, the chosen target language DLV does not support ontological reasoning as it does not contain existential quantification, thereby missing a key requirement (RQ3) of the Semantic Web community. On the other hand, the requirement of an implemented system (RQ5) is only partially fulfilled, 
since the prototype implementation \textit{DLVhex-SPARQL Plugin} 
\cite{DLVHexSPARQLPlugin} of the SPARQL to Datalog translation of \cite{Polleres2007FromST}
has not been extended to cover also SPARQL 1.1 and bag semantics.
\\[1.05ex]
\textbf{Alternative Translations of SPARQL to Datalog.} 
An alternative approach of relating SPARQL to non-recursive Datalog with stratified negation (or, equivalently, to Relational Algebra) 
was presented by Angles and Gutierrez in 
\cite{DBLP:conf/semweb/AnglesG08}. The peculiarities of negation 
in SPARQL were treated in a separate paper \cite{DBLP:conf/amw/AnglesG16}. The authors later extended this line of research to an 
exploration of the bag semantics of SPARQL 
and a characterization of the structure of its algebra and logic
in \cite{Angles2016TheMS}. 
They translated a few SPARQL features into a Datalog dialect with bag semantics (multiset non-recursive Datalog with safe negation). 
This work considered only a small set of SPARQL functionality on a very abstract level and used again a target language that does not support ontological reasoning, failing to meet 
important requirements (RQ1, RQ3) of the SW community. Most importantly, no implementation
exists of the translations provided by Angles and Gutierrez, thus failing to fulfil RQ5.
\\[1.05ex]
\textbf{Supporting Ontological Reasoning via Existential Rules.} 
In \cite{DBLP:conf/icdt/CaliGL09}, Datalog$^{\pm}$ was presented as a family of languages that are particularly well suited for capturing 
ontological reasoning. The ``$+$'' in Datalog$^{\pm}$ refers to the 
crucial extension compared with Datalog by {\em existential rules},
that is, allowing existentially quantified variables in the rule heads.
However, without restrictions,
basic reasoning tasks such as answering Conjunctive Queries w.r.t.\ 
an ontology given by a set of existential rules  become
undecidable \cite{DBLP:journals/jcss/JohnsonK84}. 
Hence, numerous restrictions have been proposed~\cite{DBLP:journals/tcs/FaginKMP05,DBLP:journals/jair/CaliGK13,pvldb/CaliGP10,DBLP:conf/rr/CaliGP10,BagetLMS09,BagetLMS11}
to ensure decidability of such tasks, which 
led to the ``$-$`` in Datalog$^{\pm}$.
Of all variants of Datalog$^{\pm}$, Warded Datalog$^{\pm}$ 
\cite{Arenas2014ExpressiveLF} ultimately 
turned out to constitute the best compromise between complexity and expressiveness and it  has been implemented in an industrial-strength system -- the Vadalog system~\cite{Vadalog}, 
thus fulfilling requirement RQ5. However, the requirement of supporting SPARQL (RQ1) with or without bag semantics (RQ2) 
have not been fulfilled up to now.
\\[1.05ex]
\textbf{Warded Datalog$^{\pm}$  with Bag Semantics.} 
In \cite{BagSemantic}, it was shown that Warded Datalog$^{\pm}$ using set semantics can be used to represent Datalog using bag semantics by using existential quantification to introduce
new tuple IDs. It was assumed that these results could be leveraged for future translations from SPARQL with bag semantics to Warded Datalog$^{\pm}$ with set semantics. However, the theoretical translation of SPARQL to Vadalog 
(RQ1) using these results and also implementation (RQ5) by extending Vadalog were left open in  \cite{BagSemantic}  and considered of primary importance for future work. 

\subsection{Practical Approaches}
\label{sec:DatalogSPARQLCombiningSystems}

Several systems have aimed at bridging the gap between DB and SW
technologies. The World Wide Web Consortium (W3C) lists \emph{StrixDB}, \emph{DLVhex SPARQL-engine} and \emph{RDFox} as systems that support SPARQL in combination with Datalog\footnote{\href{https://www.w3.org/wiki/SparqlImplementations}{https://www.w3.org/wiki/SparqlImplementations}}. Furthermore, we 
also have a look at ontological reasoning systems 
\emph{Vadalog}, \emph{Graal} and \emph{VLog}, 
which either understand SPARQL to some extent or, at least in principle, could be extended in order
to do so. 
\\[1.05ex]
\textbf{DLVhex-SPARQL Plugin.}
As mentioned above, 
the DLVhex-SPARQL Plugin \cite{DLVHexSPARQLPlugin} 
is a prototype implementation 
of the SPARQL to Datalog translation in \cite{Polleres2007FromST}. 
According to the repository's ReadMe file\footnote{\href{https://sourceforge.net/p/dlvhex-semweb/code/HEAD/tree/dlvhex-sparqlplugin/trunk/README}{https://sourceforge.net/p/dlvhex-semweb/code/HEAD/tree/dlvhex-sparqlplugin/trunk/README}}, it supports basic graph patterns, simple conjunctive \emph{FILTER} expressions (such as \emph{ISBOUND}, \emph{ISBLANK}, and arithmetic comparisons), the \emph{UNION}, \emph{OPTIONAL}, and \emph{JOIN} operation. Other operations, language tags, etc.~are not supported and query results do not conform to the SPARQL protocol, according to the ReadMe file. Moreover, the 
underlying logic programming language DLV provides only domain specific existential quantification (described in \cite{HexExistentials}), 
produced e.g.~by hash-functions. Hence, it only provides very limited support of existential quantification, which 
does not suffice for ontological reasoning as required by the OWL 2 QL standard
(RQ3). Also the support of bag semantics is missing (RQ2).
\\[1.05ex]
\textbf{RDFox.} 
RDFox is an RDF store developed and maintained at the University of Oxford \cite{RDFox}. It reasons over OWL 2 RL ontologies in Datalog and computes/stores materialisations of the inferred consequences for efficient query answering \cite{RDFox}. The answering process of SPARQL queries is not explained in great detail, except stating that queries are evaluated on top of these materialisations, by employing different scanning algorithms \cite{RDFox}.
\revision{However, translating SPARQL to Datalog -- one of the main goals of this paper -- is not supported\footnote{see \href{https://docs.oxfordsemantic.tech/reasoning.html}{https://docs.oxfordsemantic.tech/reasoning.html} for the documentation discussing the approach}.}
Moreover, RDFox does currently not support property paths and some other SPARQL 1.1 features\footnote{\href{https://docs.oxfordsemantic.tech/3.1/querying-rdfox.html\#query-language}{https://docs.oxfordsemantic.tech/3.1/querying-rdfox.html\#query-language}}.

\smallskip
\noindent
\textbf{StrixDB.} 
StrixDB is an RDF store developed as a simple tool for working with middle-sized RDF graphs,
supporting SPARQL 1.0 and Datalog reasoning capabilities\footnote{\href{http://opoirel.free.fr/strixDB/}{http://opoirel.free.fr/strixDB/}}.
To the best of our knowledge, there is no academic paper or technical report that explains the capabilities of the system in greater detail -- leaving us with the web page as the only source of information
on StrixDB.
The \emph{StrixStore} documentation page\footnote{\href{http://opoirel.free.fr/strixDB/DOC/StrixStore_doc.html}{http://opoirel.free.fr/strixDB/DOC/StrixStore\_doc.html}} 
lists examples of how to integrate Datalog rules into SPARQL queries, to query graphs enhanced by Datalog ontologies.
\revision{However, translating SPARQL to Datalog -- one of the main goals of this paper -- is not supported\footnote{see \href{http://opoirel.free.fr/strixDB/dbfeatures.html}{http://opoirel.free.fr/strixDB/dbfeatures.html} for the documentation discussing capabilities}.}
Moreover, 
important SPARQL 1.1 features such as aggregation and property paths are not 
supported by StrixDB.
\\[1.05ex]
\textbf{Graal.} 
Graal was developed as a toolkit for querying ontologies with existential rules \cite{Graal}. The system does not focus on a specific storage system, however specializes in algorithms that can answer queries regardless of the underlying database type \cite{Graal}. It reaches this flexibility, by translating queries from their host system language into Datalog$^{\pm}$. However, it pays the trade-off of restricting itself to answering conjunctive queries only \cite{Graal} and therefore supports merely a small subset of SPARQL features\footnote{\href{https://graphik-team.github.io/graal/}{https://graphik-team.github.io/graal/}} --- e.g. not even being able to express basic features such as \emph{UNION} or \emph{MINUS}.
Clearly, the goal of 
developing a uniform and consistent framework for both, 
the Semantic Web and Database communities, 
cannot be achieved without supporting at least the most vital features of SPARQL (RQ1).

\smallskip
\noindent
\textbf{VLog.} 
VLog is a rule engine, developed at the TU Dresden \cite{VLog}.
The system transfers incoming SPARQL queries to specified external SPARQL endpoints such as Wikidata and DBpedia and incorporates the received query results into their knowledge base \cite{VLog}. Therefore, the responsibility of query answering is handed over to RDF triple stores that provide a SPARQL query answering endpoint, thus failing to provide a uniform, integrated framework for combining 
query answering with ontological reasoning (RQ5).

\smallskip
\noindent
\textbf{The Vadalog system \cite{Vadalog}} is a KG management system
implementing the logic-based language Warded Datalog$^{\pm}$. It extends Datalog by including existential quantification necessary for ontological reasoning, while maintaining 
reasonable 
complexity. As an extension of Datalog, it supports full recursion. 
Although Warded Datalog$^{\pm}$ has the capabilities to support SPARQL 1.1 under the OWL 2 QL entailment regime \cite{Arenas2014ExpressiveLF} (considering set semantics though!),
no complete theoretical nor any practical translation from SPARQL 1.1 to Warded Datalog$^{\pm}$ exists. 
Therefore, the bag semantics (RQ2) and SPARQL feature coverage (RQ1) 
requirements are not met.

\section{Preliminaries}
\label{sect:Preliminaries}

\subsection{RDF and SPARQL}
\label{sect:sparql}
RDF \cite{91402} is a W3C standard that defines a graph data model for describing Web resources. 
The RDF data model assumes three data domains: \emph{IRIs} that identify Web resources, \emph{literals} that represent simple values, and \emph{blank nodes} that identify anonymous resources. 
An \emph{RDF triple} is a tuple $(s, p, o)$,
where $s$ is the subject, $p$ is the predicate, $o$ is the object, all the components can be IRIs, the subject and the object can alternatively be a blank node, and the object can also be a literal.
An \emph{RDF graph} is a set of RDF triples.
A \emph{named graph} is an RDF graph identified by an IRI. 
An \emph{RDF dataset} is a structure formed by a default graph and zero or more named graphs.

For example, consider that \code{<http://example.org/graph.rdf>} is an IRI that identifies an RDF graph with the following RDF triples:
\footnotesize
\begin{verbatim}
<http://ex.org/glucas> <http://ex.org/name> "George" 
<http://ex.org/glucas> <http://ex.org/lastname> "Lucas"
_:b1 <http://ex.org/name> "Steven"
\end{verbatim}
\normalsize
This graph describes information about film directors.
Each line is an RDF triple, \code{<http://ex.org/glucas>} is an IRI, \code{"George"} is a literal, and 
\code{\_:b1} is a blank node.

\smallskip
SPARQL \cite{10155,90699} is the standard query language for RDF.
The general structure of a SPARQL query is shown in Figure \ref{fig:sparql-query}, where: the \code{SELECT} clause defines the output of the query, the \code{FROM} clause defines the input of the query (i.e. an RDF dataset), and the \code{WHERE} clause defines a graph pattern.

\begin{figure}[b]
\centering
\begin{Verbatim}[numbers=left,fontsize=\footnotesize]
SELECT ?N ?L
FROM <http://example.org/graph.rdf>
WHERE { ?X <http://ex.org/name> ?N 
         . OPTIONAL { ?X <http://ex.org/lastname> ?L }}
ORDER BY ?N 
\end{Verbatim}
\vspace{-5pt}
\caption{Example of SPARQL query.}
\label{fig:sparql-query}
\end{figure}

The evaluation of a query begins with the construction of the RDF dataset to be queried, whose graphs are defined by 
one or more dataset clauses. A \emph{dataset clause}  is either an expression $\FROM$ $u$ or $\FROMN$ $u$, where $u$ is an IRI that refers to an RDF graph. The former clause merges a graph into the default graph of the dataset, and the latter adds a named graph to the dataset. 

\smallskip
The $\WHERE$ clause defines a graph pattern (GP). There are many types of GPs: triple patterns (RDF triples extended with variables), basic GPs (a set of GPs), optional GPs, alternative GPs (UNION),  GPs on named graphs (GRAPH), negation of GPs (NOT EXISTS and MINUS), GPs with constraints (FILTER), existential GPs (EXISTS), and nesting of GPs (SubQueries).
A property path is a special GP which allows to express different types of reachability queries.

The result of evaluating a graph pattern is a multiset of solution mappings. A \emph{solution mapping} is a set of variable-value assignments. E.g.,  the evaluation of the query in Figure \ref{fig:sparql-query} 
over the above RDF graph
returns two mappings $\{\mu_1,\mu2\}$ with 
$\mu_1$(\code{?N}) $=$ \code{"George"},
$\mu_1$(\code{?L}) $=$ \code{"Lucas"}  and
$\mu_1$(\code{?N}) $=$ \code{"Steven"}.

\smallskip
The graph pattern matching step returns a multiset whose solution mappings are  treated as a sequence without specific order. 
Such a sequence can be arranged by using solution modifiers: $\OBY$ allows to sort the solutions; $\DISTINCT$ eliminates duplicate solutions; $\OFFSET$ allows to skip a given number of solutions; and $\LIMIT$ restricts the number of 
output solutions.

\smallskip
Given the multiset of solution mappings, the final output is defined by a \emph{query form}: $\SELECT$ projects the variables of the solutions; $\ASK$ returns $\true$ if the multiset of solutions is non-empty and $\false$ otherwise; $\CONSTRUCT$ returns an RDF graph whose content is determined by a set of triple templates; and $\DESCRIBE$ returns an RDF graph that describes the resources found. 

\subsection{Warded Datalog\texorpdfstring{$^{\pm}$}{} and the Vadalog System}
\label{sect:warded}

In \cite{DBLP:conf/icdt/CaliGL09}, Datalog$^{\pm}$ was presented as a family of languages 
that
extend Datalog (whence the $+$) to increase its expressive power 
but also impose restrictions (whence the $-$) to ensure decidability 
of answering Conjunctive Queries (CQs). The extension most relevant for our purposes 
is allowing {\em existential rules} of the form

\begin{center}
$\exists \bar{z} P(\bar{x}',\bar{z})  \ \leftarrow \ P_1(\bar{x}_1), \ldots, P_n(\bar{x}_n)$,  
\end{center}

\noindent
with $\bar{x}' \subseteq \bigcup_i\bar{x}_i$, and $\bar{z} \cap \bigcup_i\bar{x}_i = \emptyset$.
Datalog$^{\pm}$ is thus well suited to capture ontological reasoning. Ontology-mediated query answering is 
defined by considering a given database $D$ and program $\Pi$ as logical theories. 
The answers to a CQ $Q(\vec{z})$ with free variables $\vec{z}$ over database $D$ under the ontology expressed by 
Datalog$^{\pm}$  program $\Pi$ are defined as $\{\vec{a} \mid \Pi \cup D \models Q(\vec{a})\}$, where
$\vec{a}$ is a tuple of the same arity as $\vec{z}$ with values from the domain of $D$.

Several subclasses of  Datalog$^{\pm}$ have been 
presented 
\cite{DBLP:journals/tcs/FaginKMP05,DBLP:journals/jair/CaliGK13,pvldb/CaliGP10,DBLP:conf/rr/CaliGP10,BagetLMS09,BagetLMS11,Arenas2014ExpressiveLF}
that ensure decidability of CQ answering 
(see \cite{DBLP:journals/jair/CaliGK13} for an overview).
One such subclass is 
{\em Warded} Datalog$^\pm$  
\cite{Arenas2014ExpressiveLF}, which makes 
CQ answering 
even tractable (data complexity).
For a formal definition of {\em Warded} Datalog$^\pm$, see
\cite{Arenas2014ExpressiveLF}. We give the intuition of 
{\em Warded} Datalog$^\pm$ here.
First, for all positions in rules of a program $\Pi$, distinguish if they are 
{\em affected} or not: a position is affected, if the chase may introduce a labelled null here, i.e., a position in a head atom either with an existential variable or with a 
variable that occurs only in affected positions in the body. 
Then, for variables occurring in 
a rule $\rho$ of $\Pi$, we identify the {\em dangerous} ones: a variable is dangerous in $\rho$, 
if it may propagate a null in the chase, i.e., 
it appears in the head and all its occurrences in the body of $\rho$ are at 
affected positions. 
A Datalog$^\pm$ program $\Pi$ is {\em warded} if all rules $\rho \in \Pi$ 
satisfy: either $\rho$ contains no dangerous variable 
or all dangerous variables of $\rho$ occur in a single body atom 
$A$ (= the ``ward'') such that the variables shared by $A$ and the remaining body
occur in at least one non-affected position (i.e., they cannot propagate nulls). 

Apart from the favourable computational properties, 
another important aspect of Warded Datalog$^\pm$ is that a
full-fledged engine (even with further extensions) exists: the 
Vadalog system~\cite{Vadalog}.  
It combines full support of Warded Datalog$^\pm$ plus a number of extensions needed for practical use, including (decidable) arithmetics, aggregation, and other features. It has been deployed in numerous industrial scenarios, including the finance sector as well as the supply chain and logistics sector.

\section{The \name System}
\label{sec:integration}
This section introduces \name, a system that allows to evaluate SPARQL 1.1 queries on top of the Vadalog system.
To the best of our knowledge, \name is the first system that provides a complete translation engine from SPARQL 1.1 with bag semantics to Datalog.
In order to obtain a functional and efficient system, we combined the knowledge provided by the theoretical work with database implementation techniques.  

\name implements three translation methods: 
(i) a \emph{data translation method} $T_D$ which generates Datalog$^{\pm}$ rules from an RDF Dataset;
(ii) a \emph{query translation} method $T_Q$ which generates Datalog$^{\pm}$ rules from a SPARQL query; and 
(iii) a \emph{solution translation method} $T_S$ which generates a SPARQL solution from a Datalog$^{\pm}$ solution.
Hence, given an RDF dataset $D$ and a SPARQL query $Q$, \name generates a Datalog\texorpdfstring{$^{\pm}$}{} program $\Pi$ as the union of the rules returned by $T_D$ and $T_Q$, then evaluates the program $\Pi$, and uses $T_S$ to transform the resulting Datalog\texorpdfstring{$^{\pm}$}{} solution into a SPARQL solution.

\subsection{Example of Graph Pattern Translation}
\label{sect:gp-translation}
In order to give a general idea of the translation, we will sketch the translation of the RDF graph and the SPARQL query presented in Section \ref{sect:sparql}. 
To facilitate the notation, we will abbreviate the IRIs by using their prefix-based representation. 
For example, the IRI \code{http://ex.org/name} will be represented as \code{ex:name}, where \code{ex} is a prefix bound to the namespace \code{http://ex.org/}. Additionally, we will use \code{graph.rdf} instead of \code{http://example.org/graph.rdf}.

\subsubsection{Data translation}
Consider the RDF graph $G$ presented in Section \ref{sect:sparql}.
First, the data translation method $T_D$ generates a special fact for every RDF term (i.e., IRI, literal, and blank node) in $G$: 

\small
\begin{verbatim}
iri("ex:glucas"). iri("ex:name"). iri("ex:lastname"). 
literal("George"). literal("Lucas"). literal("Steven").
bnode("b1").
\end{verbatim}
\normalsize

\smallskip
\noindent
These facts are complemented by the following rules, which represent the domain of RDF terms:

\small
\begin{verbatim}
term(X) :- iri(X). 
term(X) :- literal(X). 
term(X) :- bnode(X).
\end{verbatim}
\normalsize

\smallskip
\noindent
For each RDF triple \code{(s,p,o)} in graph $G$ with IRI \code{g}, 
$T_D$ generates a fact \code{triple(s,p,o,g)}. Hence, 
in our example, $T_D$ produces:    

\smallskip

\small
\begin{verbatim}
triple("ex:glucas", "ex:name", "George", "graph.rdf").
triple("ex:glucas", "ex:lastname", "Lucas", "graph.rdf").
triple("b1", "ex:name", "Steven", "graph.rdf").
\end{verbatim}
\normalsize

\subsubsection{Query translation}
Assume that $Q$ is the SPARQL query presented in Figure \ref{fig:sparql-query}. 
The application of the query translation method $T_Q$ over $Q$ returns the Datalog$^{\pm}$ rules shown in Figure \ref{fig:program1}.
The general principles of the translation will be discussed  
in Section~\ref{sect:basic-rules}.
In the interest of readability, we slightly simplify the presentation, e.g., by omitting language tags and type definitions and using simple (intuitive) variable names (rather than more complex ones as would be generated by \name to rule out name clashes).

\begin{figure}
\small
\centering
\begin{Verbatim}[numbers=left,fontsize=\small]
// SELECT ?N ?L
ans(ID, L, N, D) :- ans1(ID1, L, N, X, D), 
  ID = ["f", L, N, X, ID1].
// P3 = { P1 . OPTIONAL { P2 } } 
ans1(ID1, V2_L, N, X, D) :- ans2(ID2, N, X, D), 
   ans3(ID3, V2_L, V2_X, D), comp(X, V2_X, X), 
   ID1 = ["f1a", X, N, V2_X, V2_L, ID2, ID3].
ans1(ID1, L, N, X, D) :- ans2(ID2, N, X, D), 
   not ans_opt1(N, X, D), null(L), 
   ID1 = ["f1b", L, N, X, ID2].
ans_opt1(N, X, D) :- ans2(ID2, N, X, D), 
   ans3(ID3, V2_L, V2_X, D), comp(X, V2_X, X).
// P2 = ?X ex:name ?N
ans2(ID2, N, X, D) :- 
   triple(X, "ex:name",  N, D), 
   D = "default",
   ID2 = ["f2", X, "ex:name",  N, D].
// P3 = ?X ex:lastname ?L 
ans3(ID3, L, X, D) :- 
   triple(X, "ex:lastname",  L, D), 
   D = "default",
   ID3 = ["f3", X, "ex:lastname",  L, D].
@post("ans", "orderby(2)").
@output("ans").
\end{Verbatim}
\vspace{-5pt}
\caption{Datalog$^{\pm}$ rules for SPARQL query $Q$ 
in Figure \ref{fig:sparql-query}.}
\label{fig:program1}
\end{figure}

The query translation method $T_Q$ produces rules for each language construct of SPARQL 1.1 plus rules defining several auxiliary predicates. In addition, also system instructions (e.g., to indicate the answer predicate or ordering requirements) are generated. 
The translation begins with the \code{WHERE} clause, then continues with the \code{SELECT} clause, and finalizes with the \code{ORDER BY} clause. 

The most complex part of $T_Q$ is the translation of the graph pattern defined in the \code{WHERE} clause. 
In our example, the graph pattern defined by the \code{WHERE} clause is of the form 
$P_1 = P_2$ \code{OPTIONAL} $P_3$
with triple patterns $P_2 = $ \code{?X ex:name ?N} and 
$P_3 = $ \code{?X ex:lastname ?L}.
The instruction \code{@output} (line 24) is used to define the literal of the goal rule \code{ans}. It realises the projection defined by the \code{SELECT} clause.  The instruction \code{@post("ans","orderby(2)")} (line 23) realises 
the \code{ORDER BY} clause; it indicates a sort operation over the elements in the second position of the goal rule \code{ans(ID,L,N,D)}, i.e. sorting by \code{N} (note that \code{ID} is at position 0). The \code{ans} predicate is defined 
 (lines 2--3)
by projecting out the \code{X} variable from the \code{ans1} relation, which 
contains the result of evaluating pattern $P_1$. 
The tuple IDs are generated as Skolem terms 
(line 3 for \code{ans}; likewise lines 7, 10, 17, 22).
In this example, we assume that the pattern $P_1$ and its subpatterns $P_2$ and $P_3$ 
are evaluated over the default graph. This is explicitly defined for the basic graph
patterns (lines 15, 20) and propagated by the last argument \code{D} 
of the answer predicates.

The OPTIONAL pattern $P_1$ gives rise to 3 rules defining
the 
predicate \code{ans1}: a rule (lines 11--12) to define the predicate \code{ans\_opt1}, which 
computes those mappings for pattern $P_2$ that can be extended to mappings of $P_3$; 
a rule (lines 5--7) to compute those tuples of \code{ans1} that are obtained by extending mappings of $P_2$ to mappings of $P_3$; and finally a rule (lines 8--10) to compute those tuples of \code{ans1} that are obtained from mappings of $P_2$ that have no extension to mappings of $P_3$. In the latter case, the additional variables of $P_3$ (here: only variable \code{L}) are set to \code{null} (line 9).
The two basic graph patterns $P_2$ and $P_3$ are translated to rules for the 
predicates \code{ans2} (lines 14--17) and \code{ans3} (lines 19--22) in the obvious way.

\subsubsection{Solution translation}
The evaluation of the program $\Pi$ produced by the 
data translation and query translation methods 
yields a set of ground atoms 
for the goal predicate $p$. 
In our example, we thus get two ground atoms: 
\code{ans(id1, "George","Lucas"},
\code{"graph.rdf")} and
\code{ans(id2, "Steven","null","graph.rdf")}. Note that 
the ground atoms are guaranteed to have pairwise distinct tuple IDs.
These ground atoms can be easily translated to the {\em multiset} of solution mappings by projecting out the tuple ID. Due to the simplicity of our example, we only get a {\em set} 
$\{\mu_1,\mu2\}$ 
of 
solution mappings 
with 
$\mu_1$(\code{?N}) $=$ \code{"George"},
$\mu_1$(\code{?L}) $=$ \code{"Lucas"}  and
$\mu_1$(\code{?N}) $=$ \code{"Steven"}.

\subsection{Example of Property Path Translation}
\label{sect:pp-translation}
A property path is a feature of the SPARQL query language that allows the user to query for complex paths between nodes, instead of being limited to graph patterns with a fixed structure.
SPARQL defines different types of property path, named: PredicatePath, InversePath, SequencePath, AlternativePath, ZeroOrMorePath, OneOrMorePath, ZeroOrOnePath and NegatedPropertySet.
Next we present an example to show the translation of property paths.

Assume that \code{<http://example.org/countries.rdf>} identifies an RDF graph with the following prefixed RDF triples:

\small
\begin{verbatim}
@prefix ex: <http://ex.org/> . 
ex:spain ex:borders ex:france . 
ex:france ex:borders ex:belgium . 
ex:france ex:borders ex:germany . 
ex:belgium ex:borders ex:germany .
ex:germany ex:borders ex:austria .
\end{verbatim}
\normalsize

\noindent
Note that each triple describes two bordered countries in Europe.
Recall that \code{ex} is a prefix for the namespace \code{http://ex.org/}, meaning, e.g., that  \code{ex:spain} is the abbreviation of \code{http://ex.org/spain}. 

A natural query could be asking for the countries than can be visited by starting a trip in $Spain$. In other terms, we would like the get the nodes (countries) reachable from the node representing $Spain$.
Although the above query could be expressed by computing the union of different fixed patterns (i.e. one-country trip, two-country trip, etc.), 
the appropriate way is to use the SPARQL query shown in Figure \ref{fig:pp-query}.
The result of this query is the set 
$\{\mu_1, \mu_2,\mu_3,\mu_4\}$
of mappings with 
$\mu_1(?B)$ = \code{ex:france}, $\mu_2(?B)$ = \code{ex:germany}, $\mu_3(?B)$ = \code{ex:austria}, and $\mu_4(?B)$ = \code{ex:belgium}.

\begin{figure}[b]
\centering
\begin{Verbatim}[numbers=left,fontsize=\small]
PREFIX ex: <http://ex.org/>
SELECT ?B
FROM <http://example.org/countries.rdf>
WHERE { ?A ex:borders+ ?B . FILTER (?A = ex:spain) }
\end{Verbatim}
\vspace{-5pt}
\caption{Example of SPARQL property path query.}
\label{fig:pp-query}
\end{figure}

A property path pattern is a generalization of a triple pattern $(s,p,o)$ where the predicate $p$ is extended to be a regular expression called a property path expression.
Hence, the expression \code{?A ex:borders+ ?B} shown in Figure \ref{fig:pp-query} is a property path pattern, where 
the property path expression \code{ex:borders+} allows to return all the nodes \code{?B} reachable from node \code{?A} by following one or more matches of edges with \code{ex:borders} label. The \code{FILTER} condition restricts the solution mappings to those where variable $?A$ is bound to \code{ex:spain}, i.e. pairs of nodes where the source node is $spain$. Finally, the \code{SELECT} clause projects the result 
to variable \code{?B}, i.e., the  target nodes.

\begin{figure}
\small
\centering
\begin{Verbatim}[numbers=left,fontsize=\small]
// P1 = "{?A ex:borders+ ?B . FILTER (?A = ex:spain)}" 
ans1(ID1,A,B,D) :- ans2(ID2,A,B,D), 
                   X = "ex:spain", ID1 = [...].
// P2 = "?A ex:borders+ ?B"
ans2(ID2,X,Y,D) :- ans3(ID3,X,Y,D), ID2 = [...].
// PP3 = "ex:borders+"
ans3(ID3,X,Y,D) :- ans4(ID4,X,Y,D), ID4 = [].
ans3(ID3,X,Z,D) :- ans4(ID4,X,Y,D),
                   ans3(ID31,Y,Z,D), ID4 = []. 
// PP4 = "ex:borders"                   
ans4(ID4,X,Y,D) :- triple(X,"ex:borders",Y,D),
                   D = "default", ID4 = [...].
@output("ans1").
\end{Verbatim}
\vspace{-5pt}
\caption{Datalog$^{\pm}$ rules obtained after translating the SPARQL property path pattern shown in Figure \ref{fig:pp-query}.}
\label{fig:program2}
\end{figure}

In Figure \ref{fig:program2}, we show the Datalog$^{\pm}$ rules obtained by translating the graph pattern shown in Figure \ref{fig:pp-query}. 
The rule in line 2 corresponds to the translation of the filter graph pattern.
The rule in line 5 is the translation of the property path pattern \code{?A ex:borders+ ?B}.
The rules shown in lines 8 and 9 demonstrate the use of recursion to emulate the property path expression \code{ex:borders+}. 
The rule in line 11 is the translation of \code{ex:borders} which is called a link property path expression.
The general principles of the translation of property paths 
will be discussed in Section~\ref{sect:translation-propertypaths}.

\subsection{Coverage of SPARQL 1.1 features}
\label{sec:coverage}
In order to develop a realistic integration framework between SPARQL and Vadalog, we conduct a prioritisation of SPARQL features. We first lay our focus on basic features, 
such as \emph{terms} and \emph{graph patterns}. 
Next, we prepare a more detailed prioritisation by considering the results of 
Bonifati et al.~\cite{SPARQLFeatureUsage}, who examined the real-world adoption of SPARQL features by analysing a massive amount of real-world query-logs from different well-established Semantic Web sources. 
Additionally, we study further interesting properties of SPARQL, for instance SPARQL's approach to support partial recursion (through the addition of property paths) or interesting edge cases (such as the combination of \emph{Filter} and \emph{Optional} features) for which a ``special'' treatment is required.

The outcome of our prioritisation step is shown in Table \ref{tab:coverage}. 
For each feature, 
we present its real-world usage according to \cite{SPARQLFeatureUsage}
and its current implementation status in our 
\name system. 
The table represents the real-world usage by a percentage value 
(drawn from \cite{SPARQLFeatureUsage}) 
in the feature usage field, if  \cite{SPARQLFeatureUsage} covers the feature, 
``Unknown'' if \cite{SPARQLFeatureUsage} does not cover it, and 
``Basic Feature'' if we consider the feature as fundamental to SPARQL.
Note that some features are supported by \name with minor restrictions, such as ORDER BY for which we did not re-implement the sorting strategy defined by the SPARQL standard, but directly use the sorting strategy employed by the Vadalog system. 
Table \ref{tab:coverage} reveals that our \name engine covers 
all features that are used in more than 5\% of the queries in practice and are deemed therefore to be of highest relevance to SPARQL users.
Some of these features have a rather low usage in practice ($< 1\%$), however are still supported by our engine. These features include \emph{property paths} and $\GBY$. 
We have chosen to add \emph{property paths} to our engine, as they are not only interesting for being SPARQL's approach to support partial recursion but, according to \cite{SPARQLFeatureUsage}, there are datasets that make extensive use of them. Moreover, we have chosen to add $\GBY$ and some aggregates (e.g. $\COUNT$), as they are very important in traditional database settings, and thus are important to establish a bridge between the Semantic Web and Database communities.

\revision{
In addition to these most widely used features, we have covered all features occurring in critical benchmarks (see Section 6.1 for a detailed discussion). Specifically, as used in the FEASIBLE benchmark, we cover the following features: ORDER BY with complex arguments (such as ORDER BY with BOUND conditions), functions on strings such as UCASE, the DATATYPE function, LIMIT, and OFFSET. 
For the gMark benchmark, we cover the ``exactly n occurrences'' property path, ``n or more occurrences'' property path, and the ``between 0 and n occurrences'' property path. 
}

Among our contributions, concerning the translation of SPARQL to Datalog, are: the available translation methods have been combined into a uniform and practical framework for translating RDF datasets and SPARQL queries to Warded Datalog$\pm$ programs;
we have developed simpler translations for $\MINUS$ and $\OPTIONAL$, compared with those presented in \cite{AnswerSetTrans};
we provide translations for both bag and set semantics, thus covering queries
with and without the DISTINCT keyword;
we have enhanced current translations by adding partial support for data types and language tags;
we have developed a novel duplicate preservation model based on the abstract theories of ID generation (this was required because plain existential ID generation turned out 
to be problematic due to 
\ifFullVersion
its dependence on a very specific 
chase algorithm of the Vadalog system);
\fi
and we propose a complete method for translating property paths, including zero-or-one and zero-or-more property paths.

\begin{table}[t!]
\centering
\resizebox{\columnwidth}{!}{\begin{tabular}{|c|c|c|c|}
\hline
\multicolumn{1}{|c|}{\textbf{General Feature}} & \multicolumn{1}{c|}{\textbf{Specific Feature}}  & \multicolumn{1}{c|}{\textbf{Feature Usage}} & \multicolumn{1}{c|}{\textbf{Status}} \\ \hline
Terms & IRIs, Literals, Blank nodes  & Basic Feature & \cmark \\ \hline
Semantics & Sets, Bags & Basic Feature & \cmark \\ \hline
\multirow{5}{*}{Graph patterns}                         
 & Triple pattern & Basic Feature & \cmark \\
 & AND / JOIN & 28.25\% & \cmark \\
 & OPTIONAL & 16.21\% & \cmark \\
 & UNION & 18.63\% & \cmark \\
 & GROUP Graph Pattern
 & < 1\% &  \xmark \\ \hline
\multirow{6}{*}{Filter constraints} 
 & Equality / Inequality & \multirow{6}{*}{\shortstack{All Constraints\\ 40.15\%}} & \cmark \\
 & Arithmetic Comparison &  & \cmark \\
 & bound, isIRI, isBlank, isLiteral & & \cmark \\
 & Regex & & \cmark \\
 & AND, OR, NOT & & \cmark \\ \hline
\multirow{4}{*}{Query forms}                                    & SELECT                         & 87.97\%                                   & \cmark                           \\
                                               & ASK                                             & 4.97\%                                 & \cmark                           \\
                                               & CONSTRUCT                                       & 4.49\%                                    &                   \xmark                   \\
                                               & DESCRIBE                                        & 2.47\%                                    &                   \xmark                   \\ \hline
\multirow{4}{*}{Solution modifiers}                             & ORDER BY                       & 2.06\%                                   & \cmark                           \\
                                               & DISTINCT                                        & 21.72\%                                   & \cmark                           \\
                                               & LIMIT                                           & 17.00\%                                   & \cmark                           \\ 
                                               & OFFSET                                          & 6.15\%                                 & \cmark                               \\ \hline
\multirow{2}{*}{RDF datasets}                                   & GRAPH ?x \{ … \}               & 2.71\%                                 & \cmark                           \\ 
											   & FROM (NAMED)                                    & Unknown                                 & \xmark                                  \\ \hline
Negation                 & MINUS                               & 1.36\%                                  & \cmark                           \\
                         & FILTER NOT EXISTS                   & 1.65\%                                  & \xmark                       \\ \hline
Property paths           & LinkPath (X exp Y)                  & < 1\%                                   & \cmark                           \\
                         & InversePath (\textasciicircum{}exp) & < 1\%                                   & \cmark                           \\
                         & SequencePath (exp1 / exp2)          & < 1\%                                   & \cmark                           \\
                         & AlternativePath (exp1 | exp2)       & < 1\%                                   & \cmark                           \\
                         & ZeroOrMorePath (exp*)               & < 1\%                                   & \cmark                           \\
                         & OneOrMorePath (exp+)                & < 1\%                                   & \cmark                           \\
                         & ZeroOrOnePath (expr?)               & < 1\%                                   & \cmark                           \\
                         & NegatedPropertySet (!expr)          & < 1\%                                   & \cmark                           \\ \hline
Assignment               & BIND                                & < 1\%                                   & \xmark                                     \\
                         & VALUES                              & < 1\%                                   & \xmark                                     \\ \hline
Aggregates               & GROUP BY                            & < 1\%                                   & \cmark                           \\
                         & HAVING                              & < 1\%                                   & \xmark                                     \\ \hline
Sub-Queries               & Sub-Select Graph Pattern             & < 1\%                                   & \xmark                                     \\
                         & FILTER EXISTS                       & < 1\%                                   & \xmark                                     \\ \hline
Filter functions         & Coalesce                            & Unknown                                 & \xmark                                    \\
                         & IN / NOT IN                         & Unknown                                 & \xmark                                    \\ \hline
										
\end{tabular}}
\caption{Selected SPARQL features, including their real-world usage 
according to \cite{SPARQLFeatureUsage}
and the current status in \name.}
\label{tab:coverage}
\end{table}

There are also a few features that have a real-world usage of slightly above one percent and which are currently not supported by \name. Among these features are $\CONSTRUCT$, $\DESCRIBE$, 
and $\NEX$. We do not support features $\CONSTRUCT$ and $\DESCRIBE$, as these solution modifiers do not yield any interesting theoretical or practical challenges and they did not occur in any of the benchmarks chosen for our experimental evaluation. 
The features for query federation are out of the considered the scope, as our translation engine demands RDF datasets to be translated to the Vadalog system for query answering. Furthermore, SPARQL query federation is used in less than 1\% of SPARQL queries \cite{SPARQLFeatureUsage}.

\section{SPARQL to Datalog$^{\pm}$ Translation}
\label{sect:TranslationEngine}

In this section, we present some general principles of our translation 
from SPARQL queries into Datalog$^{\pm}$ programs. 
We thus first discuss the translation of graph patterns (Section \ref{sect:basic-rules}), and then treat the translation of a property paths separately 
(Section \ref{sect:translation-propertypaths}).
We conclude this section with a discussion of the correctness of our translation (Section \ref{sect:Correctness}).  
Full details of the translation and its correctness proof are given 
\ifFullVersion
in Appendix~\ref{sec:rules}.
\fi

\subsection{Translation of Graph Patterns}
\label{sect:basic-rules}
Let $P$ be a SPARQL graph pattern and a $D$ be an RDF dataset 
$D = \langle G, G_{named} \rangle$ where $G$ is the default graph and $G_{named}$ is the set of named graphs.
The translation of graph patterns is realised by
the translation function $\tau(P, dst, D, \textit{NodeIndex})$ where: 
$P$ is the graph pattern that should be translated next;
$dst$ (short for ``distinct'') is a Boolean value that describes whether the result should have set semantics ($dst = true$) or bag semantics ($dst = false$);
$D$ is the graph on which the pattern should be evaluated;
$\textit{NodeIndex}$ is the index of the pattern $P$ to be translated; and the output of function $\tau$ is a set of  Datalog$^{\pm}$ rules.

\begin{figure}[t]
\begin{flushleft}
{\bf Triple pattern.}
Let $P_i$ be the i-th subpattern of P and let $P_i$ be a triple pattern $(s, p, o)$. Then $\tau(P_i, \mathit{false}, D, i)$ is defined as:

\medskip

$ans_i(Id, \overline{var}(P_{i}), D)$ :- $triple(s, p, o, D)$.

\medskip

$ans_i(Id, \overline{var}(P_{i}), g)$ :-
$ans_{2i}(Id_1, \overline{var}(P_1), g), named(g).$\\
$\tau(P_1, \mathit{false}, g, 2i)$

\medskip
{\bf Join.}
Let $P_i$ be the i-th subpattern of P and let $P_i$ be of the form 
$(P_1 \JOIN P_2)$. Then $\tau(P_i, \mathit{false}, D, i)$ is defined as:

\medskip

$ans_i(Id, \overline{var}(P_{i}), D)$ :- $ans_{2i}(Id_1, v_1(\overline{var}(P_1)), D)$, \\
\hskip20pt $ans_{2i+1}(Id_2, v_2(\overline{var}(P_2)), D)$,\\
\hskip20pt $comp(v_1(x_1), v_2(x_1), x_1)$, \dots, $comp(v_1(x_n), v_2(x_n), x_n)$. \\
$\tau(P_1, \mathit{false}, D, 2i)$. \\
$\tau(P_2, \mathit{false}, D, 2i + 1)$.

\medskip

Here we are using the following notation:
\begin{itemize}[noitemsep, nolistsep]
    \item
    $\overline{var}(P_{i}) = \overline{var(P_1) \cup var(P_2)}$
    \item
    $\{x_1, \dots, x_n\} = var(P_1) \cap var(P_2)$
    \item
    $v_1, v_2: var(P_1) \cap var(P_2) \rightarrow V$, such that\\
    $Image(v_1) \cap Image(v_2) = \emptyset$
\end{itemize}

\medskip

\noindent
{\bf Filter.}
Let $P_i$ be the i-th subpattern of P and let $P_i$ be of the form 
$(P_1 \FILTER C)$.
Then  $\tau(P_i, \mathit{false}, D, i)$ is defined as:

\medskip

$ans_i(id, \overline{var}(P_{i}), D)$ :- 
 $ans_{2i}(id_1, \overline{var}(P_1), D), C$. \\
 $\tau(P_1, \mathit{false}, D, 2i)$

\medskip

{\bf Optional.}
Let $P_i$ be the i-th subpattern of $P$ and furthermore let $P_i$ be  $(P_1 \OPT P_2)$, then $\tau(P_i, \mathit{false}, D, i)$ is defined as:

\medskip

$ans_{opt-i}(\overline{var}(P_1), D)$ :- $ans_{2i}(Id_1, \overline{var}(P_1), D)$, \\
\hskip20pt $ans_{2i+1}(Id_2, v_2(\overline{var}(P_2)), D)$,\\
\hskip20pt  $comp(x_1, v_2(x_1), z_1)$, 
\dots, $comp(x_n, v_2(x_n), z_n)$.\\[1.1ex]
$ans_{i}(Id, \overline{var}(P_{i}), D)$ :- $ans_{2i}(Id_1, v_1(\overline{var}(P_1)), D)$, \\
\hskip20pt  $ans_{2i+1}(Id_2, v_2(\overline{var}(P_2)), D)$,\\
\hskip20pt  $comp(v_1(x_1), v_2(x_1), x_1)$, 
\dots, $comp(v_1(x_n), v_2(x_n), x_n)$. \\[1.1ex]
$ans_{i}(Id, \overline{var}(P_{i}), D)$ :- $ans_{2i}(Id_1, \overline{var}(P_1), D)$, \\
\hskip20pt   $\mathit{not} \ ans_{opt-i}(\overline{var}(P_1), D)$,\\
\hskip20pt   $\mathit{null}(y_1), \dots, \mathit{null}(y_m)$.\\[1.1ex]
$\tau(P_1, \mathit{false}, D, 2i)$. \\
$\tau(P_2, \mathit{false}, D, 2i + 1)$.

\medskip

\noindent
{\bf Union.}
Let $P_i$ be the i-th subpattern of P and let $P_i$ be 
of the form $(P_1 \UNION P_2)$. Then $\tau(P_i, \mathit{false}, D, i)$ is defined as:

\medskip

$ans_i(Id,\overline{var}(P_{i}), D)$ :- 
    $ans_{2i}(Id_1, \overline{var}(P_1), D)$, \\ 
\hskip20pt  
$\mathit{null}(x_1), \dots \mathit{null}(x_n)$.\\
$ans_i(Id, \overline{var}(P_{i}), D)$ :- 
    $ans_{2i+1}(Id_2, \overline{var}(P_2), D)$,\\
\hskip20pt      
$\mathit{null}(y_1), \dots \mathit{null}(y_m)$. \\
$\tau(P_1, \mathit{false}, D, 2i)$\\
$\tau(P_2, \mathit{false}, D, 2i + 1)$

\end{flushleft}
\vspace{-5pt}
\caption{Translation rules for SPARQL graph patterns.}
\label{fig:basictranslation}
\end{figure}

The  function $\tau$ for different types of graph patterns is presented in Figure \ref{fig:basictranslation}.
In the sequel, we concentrate on bag semantics (i.e., \emph{dst = false}), since this is the more complex case.
To improve readability, we apply the simplified notation used in Figure~\ref{fig:program1} now also to Figure~\ref{fig:basictranslation}. Additionally,  we omit the explicit generation of IDs via Skolem functions and simply put a fresh ID-variable in the first position of the head atoms of the rules.

\smallskip \noindent
\textbf{General strategy of the translation}.
Analogously to \cite{Polleres2007FromST,AnswerSetTrans}, 
our translation proceeds by recursively traversing the parse tree of a SPARQL 1.1 query and translating each subpattern into its respective Datalog$^{\pm}$ rules.
Subpatterns of the parse tree are indexed. The root has index 1, the left child of the $i$-th node has index $2 * i$, the right child has index $2 * i + 1$. During the translation, bindings of the $i$-th subpattern are represented by the predicate $ans_i$.
In all answer predicates $ans_i$, we have the current graph as last component. It can be changed by the GRAPH construct; for all other SPARQL constructs, it is transparently passed on from the children 
to the parent in the parse tree. 
Since the order of variables in predicates is relevant, some variable sets will need to be lexicographically ordered, which we denote by $\overline{x}$ as in \cite{AnswerSetTrans}.
We write $\overline{var}(P)$ to denote the lexicographically ordered tuple of variables of $P$.
Moreover a variable renaming function
$v_j: V \rightarrow V$ is defined.

\smallskip \noindent
\textbf{Auxiliary Predicates}.
The translation generates several auxiliary predicates. 
Above all, we need a predicate 
\code{comp} for testing if two mappings are
{\em compatible}. 
The notion of compatible mappings is fundamental for the 
evaluation of SPARQL graph patterns. 
Two mappings $\mu_1$ and $\mu_2$ are \emph{compatible}, denoted $\mu_1 \sim \mu_2$, 
if for all $?X \in \dom(\mu_1) \cap \dom(\mu_2)$ it is satisfied that $\mu_1(?X)=\mu_2(?X)$.
The auxiliary predicate $comp(X_1,X_2,X_3)$ checks if two values $X_1$ and $X_2$
are compatible. The third position $X_3$ represents the value 
that is used in the result tuple 
when joining over $X_1$ and $X_2$:

\small
\begin{verbatim}
null("null").
comp(X,X,X) :- term(X).
comp(X,Z,X) :- term(X), null(Z).
comp(Z,X,X) :- term(X), null(Z).
comp(Z,Z,Z) :- null(Z).
\end{verbatim}
\normalsize

\smallskip \noindent
\textbf{Bag semantics}.
For bag semantics, (i.e., $dst = false$) all answer predicates contain a fresh existential variable 
when they occur in the head of a rule. In this way, whenever such a rule fires, a fresh tuple ID is generated. This is particularly important for the translation of the UNION construct. In contrast to 
\cite{AnswerSetTrans}, we can thus distinguish duplicates without the need to increase the arity of the answer predicate.
We have developed a novel duplicate preservation model 
based on the abstract theories of ID generation of \cite{BagSemantic}. 
As mentioned above, plain existential ID generation turned out to be problematic due to peculiarities of the Vadalog system. 
Therefore, our ID generation process is abstracted away by using a Skolem function generator and representing nulls (that correspond to tuple IDs) as specific Skolem terms. 

\smallskip \noindent
\textbf{Filter constraints}.
Note how we treat filter conditions in FILTER constructs: building our translation engine on top of the Vadalog system allows us to literally copy (possibly complex) filter conditions into the rule body and let the Vadalog system evaluate them. For instance, 
the regex functionality uses the corresponding Vadalog function, which 
makes direct use of the Java regex library.
For evaluating filter functions isIRI, isURI, isBlank, isLiteral, isNumeric, and bound expressions, our translation engine uses the corresponding auxiliary predicates generated in our data translation method. 

\subsection{Translation of Property Paths}
\label{sect:translation-propertypaths}

Property paths are an important feature, introduced in SPARQL 1.1. 
A translation of property paths to Datalog  was presented in \cite{AnswerSetTrans} -- 
but not fully compliant with the SPARQL 1.1 standard: the main problem 
in \cite{AnswerSetTrans} was the way how \textit{zero-or-one} and 
\textit{zero-or-more} property paths were handled. In particular, the case that 
a path of zero length from $t$ to $t$ also exists for those terms $t$
which occur in the query but not in the current graph, was omitted in \cite{AnswerSetTrans}.
A {\em property path pattern} is given in the 
form $s, p, o$, where $s, o$ are the usual
subject and object and $p$ is a 
{\em property path expression}. That is, 
$p$ is either an IRI (the base case) or 
composed from one or two other property path
expressions
$p_1, p_2$ as: 
$\inv  p_1$ (inverse path expression),
$p_1 \mid p_2$ (alternative path expression),
$p_1 / p_2$ (sequence path expression),
$p_1?$ (zero-or-one path expression),
$p_1+$ (one-or-more path expression),
$p_1*$ (zero-or-more path expression), or
$!p_1$ (negated path expression).
A property path pattern $s,p,o$ is translated by 
first translating the property path expression $p$  into rules for each subexpression of $p$.
The endpoints $s$ and $o$ of the overall path are only applied to the 
top-level expression $p$. Analogously to our translation function 
$\tau(P, dst, D,  \textit{NodeIndex})$ for graph patterns, we 
now 
also introduce a translation function 
$\tau_{PP}(PP, dst, S, O, D,  \textit{NodeIndex})$ for property path expressions $PP$, 
where 
$S,O$, are the subject and object of the top-level property path expression
that have to be
kept track of during the entire evaluation as will become clear when we highlight our translation 
in Figure~\ref{fig:propertyPathTranslation}.

\begin{figure}[t]
\begin{flushleft}
{\bf Property path.}
Let $P_i$ be the i-th subpattern of P and let $P_i$ be a property path pattern
of the form $(S, P_1, O)$ where $P_1$ is a property path expression. 
Then $\tau(P_i, \mathit{false}, D, i)$ is defined as:

\medskip

$ans_i(Id, \overline{var}(P_i), D)$ :- $ans_{2i}(Id_1, S, O, D)$. \\
$\tau_{PP}(P_1, \mathit{false}, S, O, D, 2i)$.

\medskip

\noindent
{\bf Link property path}.
Let $PP_i$ be the $i$-th subexpression of a property path 
expression $PP$ and let 
$PP_i = p_1$ be a link pro\-per\-ty path expression. Then
$\tau_{PP}(PP_i, \mathit{false}, S, O, D, i)$ is defined as:

\medskip

$ans_i(Id, X, Y, D)$ :- $triple(X, p1, Y, D)$.

\medskip

\noindent
{\bf One-or-more path}.
Let $PP_i$ be the $i$-th subexpression of a property path 
$PP$ and let 
$PP_i = PP_1\hspace{-1pt}+ $
be a one-or-more property path expression. 
Then $\tau_{PP}(PP_i, \mathit{false}, S, O, D, i)$ is defined as:

\medskip

\noindent
{\bf Zero-or-one path}.
Let $PP_i$ be the $i$-th subexpression of a property path 
expression $PP$ and 
let $PP_i = PP_1? $ 
be a zero-or-one property path expression. 
Then $\tau_{PP}(PP_i, \mathit{false}, S, O, D, i)$ is defined as:

\medskip

$ans_i(Id, X, X, D)$ :- $subjectOrObject(X)$, $Id = [ ]$.\\
$ans_i(Id, X, Y, D)$ :-  $ans_{2i}(Id_1, X, Y, D)$, $Id = [ ]$.\\
$\tau_{PP}(PP_1, \mathit{false}, S, O, D, 2i)$

\end{flushleft}
\caption{Translation rules for SPARQL property paths.}
\label{fig:propertyPathTranslation}
\end{figure}

Again we restrict ourselves to the more interesting case of 
bag semantics.
The translation of a property path pattern $S, P_1, O$ for some property path expression $P_1$ consists of two parts: the translation of $P_1$ by the translation function $\tau_{PP}$ 
and the translation $\tau$ of $S, P_1, O$ -- now applying the endpoints $S$ and $O$ 
to the 
top-level property path expression $P_1$.  
The base case of 
$\tau_{PP}$ is a link property path 
$PP_i = p_1$ (i.e., simply an IRI), 
which returns all pairs $(X,Y)$ that occur as 
subject and object in a triple with predicate $p_1$. 
Equally simple translations apply to 
inverse paths (which swap start point 
and end point), alternative  paths (which are treated 
similarly to UNION 
in Figure~\ref{fig:basictranslation}), and 
sequence paths (which combine 
two paths by identifying the end point of the first path with the 
start point of the second path).

For zero-or-one paths (and likewise for zero-or-more paths), 
we need to collect all terms that occur as subjects or objects in the 
current graph by an auxiliary predicate \code{subjectOrObject}:

\smallskip

\noindent
\hskip10pt $subjectOrObject(X)$ :- $triple(X, P, Y, D)$.\\
\mbox{}\hskip10pt $subjectOrObject(Y)$ :- $triple(X, P, Y, D)$.

\smallskip

This is needed to produce paths of length zero (i.e., from $X$ to $X$) for all 
these terms occurring in the current graph. Moreover, if exactly one of $S$ and $O$
is not a variable or if both are the same non-variable, then also for these nodes we have to 
produce paths of zero length. 
It is because of this special treatment of 
zero-length paths that subject $S$ and object $O$ from the top-level 
property path expression have to be propagated through all recursive calls 
of the translation function $\tau_{PP}$.
In addition to the zero-length paths, of course, also paths of length one have to be 
produced by recursively applying the translation $\tau_{PP}$ to $PP_1$ if 
$PP_i$ is of the 
form $PP_i = PP_1? $. 
Finally, one-or-more paths are realised in the 
usual style of transitive closure programs in Datalog.

It should be noted that, according to the SPARQL semantics of property paths\footnote{\href{https://www.w3.org/TR/SPARQL11-query/\#defn_PropertyPathExpr}{https://www.w3.org/TR/SPARQL11-query/\#defn\_PropertyPathExpr}}, 
zero-or-one, zero-or-more, and one-or-more property paths always have set semantics.
This is why the Datalog$^{\pm}$ rules for these three path expressions 
contain a body literal $Id = []$. By forcing the tuple ID to the same value whenever one of these rules
fires, multiply derived tuples are indistinguishable for our system and will, therefore, 
never give rise to duplicates.

\subsection{Correctness of our Translation}
\label{sect:Correctness}

To ensure the correctness of our translation, 
we have applied a two-way strategy -- consisting of an extensive empirical evaluation and  a formal analysis. For the empirical evaluation, we have run  \name as well as Fuseki and Virtuoso on several benchmarks, which provide a good coverage of SPARQL~1.1. 
The results are summarized in 
Section~\ref{sec:Compliance}. In a nutshell, \name and Fuseki turn out to 
fully comply with the SPARQL~1.1 standard, while Virtuoso shows deviations from the standard on quite some queries.

For the formal analysis, we juxtapose our translation with the formal semantics of the various language constructs of 
SPARQL~1.1. 
Below we briefly outline our proof strategy: 
Following 
\cite{DBLP:conf/semweb/AnglesG08,DBLP:books/sp/virgilio09/ArenasGP09,AnswerSetTrans}
for SPARQL graph patterns and 
\cite{DBLP:conf/semweb/KostylevR0V15,AnswerSetTrans}
for property path expressions, we first of all 
provide a formal definition of the semantics of the various SPARQL~1.1 
features.\footnote{We note that the semantics definitions in all of these sources either only cover a rather small subset of SPARQL~1.1 or contain erroneous definitions. The most complete exposition is given in 
\cite{AnswerSetTrans} with some inaccuracies in the treatment of Optional Filter patterns and of zero-length property paths.}Given a SPARQL graph pattern $P$ and a graph $D$, we write 
$\epag{P}{D}$ to denote the result of evaluating $P$ over $D$. The semantics
$\epag{PP}{D,s,o}$ 
of property path expressions $PP$ is defined in a similar way, but now also taking the top level start and end points $s,o$ of the property path 
into account. 

Both 
$\epag{P}{D}$ and 
$\epag{PP}{D,s,o}$ 
are defined inductively on the structure of the expression 
$P$ or $PP$, respectively, with triple patterns
$P = (s,p,o)$ and link property paths $PP = p_1$ as base cases. 
For instance, for a join pattern $P_i = (P_1 \JOIN P_2)$ and 
optional pattern $P_j = (P_1 \OPT P_2)$, 
the semantics is defined as follows: 

\begin{tabbing}  
$\epag{P_i}{D}$ \= $=$ \=
$\mset{\mu_1 \cup \mu_2 \mid \mu_1 \in \epag{P_1}{D} \text{ and }
\mu_2 \in \epag{P_2}{D} \text{ and } \mu_1 \sim \mu_2}$ \\
$\epag{P_j}{D}$ \> $=$ \> 
$\mset{\mu \mid 
\mu \in \epag{P_1 \JOIN P_2}{D} \text{ and } 
\mu \models C } \cup \mbox{}$ \\
\>\>
$\{ \hskip-2pt \{
\mu_1 \mid \mu_1 \in \epag{P_1}{D} \text{ and for all } 
\mu_2  \in \epag{P_2}{D}$:  $\mu_1 \not\sim \mu_2 
\} \hskip-2pt \}$
\end{tabbing}

Now consider the Datalog$^{\pm}$ rules generated by our translation. 
For a join pattern,  the two body atoms 
$ans_{2i}(Id_1, v_1(\overline{var}(P_1)), D)$
and 
$ans_{2i+1}(Id_2, v_2(\overline{var}(P_2)), D)$
yield, by induction, the sets of mappings 
$\epag{P_1}{D}$ and
$\epag{P_2}{D}$.
The variable renamings
$v_1$ and $v_2$ make sure that there is no interference
between the evaluation of 
$\epag{P_1}{D}$ (by the first body atom) and the
evaluation of $\epag{P_2}{D}$ (by the second body atom). 
The $comp$-atoms in the body of the rule make sure 
that $\mu_1$ and $\mu_2$ are compatible on all common variables. Moreover, they bind the common variables 
$\{x_1, \dots, x_n\}$ to the correct value according 
to the definition of the $comp$-predicate.

The result of evaluating an optional pattern consists of 
two kinds of mappings: (1) the mappings $\mu$ in $\epag{(P_1 \AAND P_2)}{D}$  and
(2) the mappings $\mu_1$ in $\epag{P_1}{D}$ which are not 
compatible with any mapping
$\mu_2$ in $\epag{P_2}{D}$.
Analogously to join patterns, the second rule generated by our 
translation produces the mappings of type (1). 
The first rule generated by our translation computes those 
mappings in $\epag{P_1}{D}$ which are compatible with 
some mapping in $\epag{P_2}{D}$.
Hence, the third rule produces the mappings of type (2). Here the negated second body literal 
removes all those mappings from $\epag{P_1}{D}$ which are 
compatible with some mapping in $\epag{P_2}{D}$.

Full details of the semantics definitions $\epag{P}{D}$ and $\epag{PP}{D,s,o}$
and of the juxtaposition with the rules generated by our translation 
are provided 
\ifFullVersion
in Appendix \ref{app:correctness}.
\fi

\section{Experimental Evaluation}
\label{sect:ExperimentalEvaluation}

In this section, we report on the experimental evaluation of the \name system. 
We want to give a general understanding of the behaviour of \name in the following three areas: (1) we first analyse various benchmarks available in the area to identify \textbf{coverage} of SPARQL features and which benchmarks to use subsequently in our evaluation, (2) we analyse the \textbf{compliance} of our system with the SPARQL standard using the identified benchmarks, and set this in context with the two state-of-the-art systems Virtuoso and Fuseki, 
and, finally, (3) we evaluate the \textbf{performance} of query execution of \name
and compare it with state-of-the-art systems for SPARQL query answering and reasoning 
over ontologies, respectively. We thus put particular emphasis on property paths and their combination with ontological reasoning.
Further details on our experimental evaluation -- in particular, how we 
set up the analysis of different benchmarks and of the standard-compliance
of various systems -- are provided 
\ifFullVersion
in Appendix~\ref{appExp}.
\fi

\vspace*{-.5em}
\subsection{Benchmark Analysis}
\label{sec:BenchmarkAnalysis}

In this subsection, we analyse current state-of-the-art 
benchmarks for SPARQL engines. 
Table~\ref{tab:BenchmarkFeatureUagse} is based on the analysis of \cite{HowRepIsSPARQLBench} and represents the result of our exploration of the SPARQL feature coverage of the considered benchmarks. Furthermore, it was adjusted and extended with additional features by us. Particularly heavily used SPARQL features are marked in \fcolorbox{blue!30}{blue!30}{blue}, while missing SPARQL features are marked in \fcolorbox{orange!60}{orange!60}{orange}. The abbreviations of the columns represent the following SPARQL 
features: DIST[INCT], FILT[ER], REG[EX], OPT[IONAL], UN[ION], GRA[PH], 
P[roperty Path] Seq[uential], P[roperty Path] Alt[ernative], GRO[UP BY].
Note that, in Table~\ref{tab:BenchmarkFeatureUagse},
we do not display explicitly basic features, such as \emph{Join}, \emph{Basic Graph pattern}, etc., since these
are of course covered by every benchmark considered here. Morerover, we have not included the SPARQL features \emph{MINUS} and the \emph{inverted}, \emph{zero-or-one}, \emph{zero-or-more}, \emph{one-or-more}, and \emph{negated property path} in 
Table~\ref{tab:BenchmarkFeatureUagse}, as none of the selected benchmarks covers any of these SPARQL features. 
\ifFullVersion
More details on the setup of our benchmark analysis can be found in Appendix \ref{app:BenchAnalysis}.
\fi

\begin{table}[h!]
\centering
\resizebox{\columnwidth}{!}{\begin{tabular}{cl|c|c|c|c|c|c|c|c|c}
& Benchmark & DIST & FILT & REG & OPT & UN & GRA & PSeq & PAlt & GRO \\
\hline
\multirow{7}{*}{\rot{\textbf{Synthetic}}}
& Bowlogna  & 5.9 & 41.2 & 11.8 & \cellcolor{orange!60} 0.0 & \cellcolor{orange!60} 0.0 & \cellcolor{orange!60} 0.0 & \cellcolor{orange!60} 0.0 & \cellcolor{orange!60} 0.0 & 76.5 \\ 
& TrainBench  & \cellcolor{orange!60} 0.0 & 41.7 & \cellcolor{orange!60} 0.0 & \cellcolor{orange!60} 0.0 & \cellcolor{orange!60} 0.0 & \cellcolor{orange!60} 0.0 & \cellcolor{orange!60} 0.0 & \cellcolor{orange!60} 0.0 & \cellcolor{orange!60} 0.0 \\ 
& BSBM  & 25.0 & 37.5 & \cellcolor{orange!60} 0.0 & 54.2 & 8.3 & \cellcolor{orange!60} 0.0 & \cellcolor{orange!60} 0.0 & \cellcolor{orange!60} 0.0 & \cellcolor{orange!60} 0.0 \\ 
& SP2Bench  & 35.3 & 58.8 & \cellcolor{orange!60} 0.0 & 17.6 & 17.6 & \cellcolor{orange!60} 0.0 & \cellcolor{orange!60} 0.0 & \cellcolor{orange!60} 0.0 & \cellcolor{orange!60} 0.0 \\ 
& WatDiv  & \cellcolor{orange!60} 0.0 & \cellcolor{orange!60} 0.0 & \cellcolor{orange!60} 0.0 & \cellcolor{orange!60} 0.0 & \cellcolor{orange!60} 0.0 & \cellcolor{orange!60} 0.0 & \cellcolor{orange!60} 0.0 & \cellcolor{orange!60} 0.0 & \cellcolor{orange!60} 0.0 \\
& SNB-BI  & \cellcolor{orange!60} 0.0 & 66.7 & \cellcolor{orange!60} 0.0 & 45.8 & 20.8 & \cellcolor{orange!60} 0.0 & 16.7 & \cellcolor{orange!60} 0.0 & \cellcolor{blue!30} 100.0 \\ 
& SNB-INT & \cellcolor{orange!60} 0.0 & 47.4 & \cellcolor{orange!60} 0.0 & 31.6 & 15.8 & \cellcolor{orange!60} 0.0 & 5.3 & 10.5 & 42.1 \\ 
\hline
\multirow{5}{*}{\rot{\textbf{Real}}} 
& \mbox{FEASIBLE (D)}  & 56.0 & 58.0 & 14.0 & 28.0 & 40.0 & \cellcolor{orange!60} 0.0 & \cellcolor{orange!60} 0.0 & \cellcolor{orange!60} 0.0 & \cellcolor{orange!60} 0.0 \\ 
& \mbox{FEASIBLE (S)}  & 56.0 & 27.0 & 9.0 & 32.0 & 34.0 & 10.0 & \cellcolor{orange!60} 0.0 & \cellcolor{orange!60} 0.0 & 25.0 \\ 
& Fishmark  & \cellcolor{orange!60} 0.0 & \cellcolor{orange!60} 0.0 & \cellcolor{orange!60} 0.0 & 9.1 & \cellcolor{orange!60} 0.0 & \cellcolor{orange!60} 0.0 & \cellcolor{orange!60} 0.0 & \cellcolor{orange!60} 0.0 & \cellcolor{orange!60} 0.0 \\ 
& DBPSB  & \cellcolor{blue!30} 100.0 & 44.0 & 4.0 & 32.0 & 36.0 & \cellcolor{orange!60} 0.0 & \cellcolor{orange!60} 0.0 & \cellcolor{orange!60} 0.0 & \cellcolor{orange!60} 0.0 \\ 
& BioBench & 39.3 & 32.1 & 14.3 & 10.7 & 17.9 & \cellcolor{orange!60} 0.0 & \cellcolor{orange!60} 0.0 & \cellcolor{orange!60} 0.0 & 10.7 \\  
\end{tabular}
}
    \caption{Feature Coverage of SPARQL Benchmarks \cite{HowRepIsSPARQLBench}}
    \label{tab:BenchmarkFeatureUagse}
\end{table}

\noindent
Table \ref{tab:BenchmarkFeatureUagse} reveals that no 
 benchmark covers all SPARQL features. Even more, SNB-BI and SNB-INT are the only benchmarks that contain property paths. Yet, they cover merely the \emph{sequential} (PSeq) and \emph{alternative property path} (PAlt), which in principle correspond to the \emph{JOIN} and \emph{UNION} operator. This means that  no existing benchmark covers recursive property paths (though we will talk about the benchmark generator gMark \cite{gMark} later), which are one of the most significant extensions provided by SPARQL 1.1. 
 Our analysis of SPARQL benchmarks leads us to the following conclusions
 for testing the compliance with the SPARQL standard and for planning the performance tests with \name and state-of-the-art systems.
 
\paragraph{Evaluating compliance with the SPARQL standard} Based on the results of Table \ref{tab:BenchmarkFeatureUagse}, we have chosen the following three benchmarks to evaluate the compliance of our \name system with the SPARQL standard: (1) We have identified \emph{FEASIBLE (S)}~\cite{Feasible} as the real-world benchmark of choice, as it produces the most diverse test cases \cite{HowRepIsSPARQLBench} and covers the highest amount of features; (2) \emph{SP2Bench}~\cite{sp2bench} is identified as the synthetic benchmark of choice, since it produces synthetic datasets with 
the most realistic characteristics \cite{HowRepIsSPARQLBench}; (3) finally, since no benchmark that employs real-world settings provides satisfactory coverage of property paths, 
we have additionally chosen \emph{BeSEPPI}~\cite{BeSEPPI} -- a simplistic, yet very extensive benchmark specifically designed for testing the correct and complete processing of property paths. We report on the results of testing the compliance of our \name system as well as Fuseki and Virtuoso in Section \ref{sec:Compliance}.

\paragraph{Performance benchmarking.} 
For the empirical evaluation of query execution times 
reported in Section \ref{sec:QueryExec},
we have identified SP2Bench as the 
most suitable benchmark, as it contains hand-crafted queries that were specifically designed to target
query optimization. 
Since none of the existing benchmarks for SPARQL performance measurements 
contains recursive property paths, we have included instances generated by the benchmark generator gMark \cite{gMark}, and report extensive results of this important aspect. 
In order to include in our tests also the performance measurements for the combination of property paths with 
ontologies, we have further extended the SP2Bench with an ontology containing subPropertyOf and subClassOf statements.

\subsection{SPARQL Compliance}
\label{sec:Compliance}

\noindent
As discussed in the previous section, we have identified three benchmarks 
(FEASIBLE(S), SP2Bench, BeSEPPI) 
for the evaluation of the standard compliance of our \name system and 
two state-of-the-art SPARQL engines.
More details on the compliance evaluation as well as some challenges encountered 
by this evaluation (such as the comparison of results 
in the presence of null nodes) are  discussed 
\ifFullVersion
in Appendix~\ref{appExp}. 
\fi
Below, we summarize the results: 

\revision{The FEASIBLE(S) benchmark contains 77 queries that we used for
testing the standard-confor\-mant behaviour.} 
It turned out that both \name and Fuseki fully comply to the standard \revision{on each of the 77 queries}, whereas Virtuoso does not. 
More specifically, for 14 queries, Virtuoso returned an erroneous result by either 
wrongly outputting duplicates (e.g., ignoring DISTINCTs) or omitting duplicates
(e.g., by handling UNIONs incorrectly). Moreover, in 18 cases, 
Virtuoso was unable to evaluate the query and produced an error. 

The SP2Bench benchmark contains 17 queries, specifically designed to test the scalability of SPARQL engines. All 3 considered systems produce the correct result for all 17 queries.

The BeSEPPI benchmark contains 236 queries, specifically designed to evaluate the correct and complete support of property path features. Table \ref{tab:BeSEPPIAnalysisM} shows the detailed results of the experimental evaluation of the 3 considered systems on this benchmark. We distinguish 4 types of erroneous behaviour: correct but incomplete results (i.e., the mappings returned are correct but there are further correct mappings missing), 
complete but incorrect (i.e., no correct mapping is missing but the answer falsely contains additional mappings), incomplete and incorrect, or failing to evaluate the query and returning an error instead. 
The entries in the table indicate the number of cases for each of the error types. 
We see that Fuseki and \name produce the correct result in all 236 cases. 
Virtuoso only handles the queries with inverse, sequence and negated path expressions 100\% correctly. For queries containing alternative, zero-or-one, one-or-more, or 
zero-or-more path expressions, Virtuoso is not guaranteed to produce the 
correct result. The precise number of queries handled erroneously is 
shown in the cells marked \fcolorbox{red!40}{red!40}{red}.

\begin{table}[h!]
 \centering
\resizebox{\columnwidth}{!}{\begin{tabular}{@{\extracolsep{0pt}}|c||cccc||cccc||cccc||c|@{}}
\hline
Stores & \multicolumn{4}{c||}{Virtuoso} & \multicolumn{4}{c||}{Jena Fuseki} & \multicolumn{4}{c||}{Our Solution} & \multirow{2}{*}{\rot{Total \#Queries \hspace{1.2em}}}\\
\cline{1-13} 
Expressions & \rot{Incomp. \& Correct} & \rot{Complete \& Incor.} & \rot{Incomp. \& Incor.} & \rot{Error} & \rot{Incomp. \& Correct} & \rot{Complete \& Incor.} & \rot{Incomp. \& Incor.} & \rot{Error} & \rot{Incomp. \& Correct} & \rot{Complete \& Incor.} & \rot{Incomp. \& Incor.} & \rot{Error} & \\ \hline \hline
Inverse & 0 & 0 & 0 & 0 & 0 & 0 & 0 & 0 & 0 & 0 & 0 & 0 & 20 \\ \hline
Sequence & 0 & 0 & 0 & 0 & 0 & 0 & 0 & 0 & 0 & 0 & 0 & 0 & 24 \\ \hline
Alternative & \cellcolor{red!40} 3 & 0 & 0 & 0 & 0 & 0 & 0 & 0 & 0 & 0 & 0 & 0 & 23 \\ \hline
Zero or One & 0 & 0 & 0 & \cellcolor{red!40} 3 & 0 & 0 & 0 & 0 & 0 & 0 & 0 & 0 & 24 \\ \hline
One or More & \cellcolor{red!40} 10 & 0 & 0 & \cellcolor{red!40} 8 & 0 & 0 & 0 & 0 & 0 & 0 & 0 & 0 & 34 \\ \hline
Zero or More & 0 & 0 & 0 & \cellcolor{red!40} 7 & 0 & 0 & 0 & 0 & 0 & 0 & 0 & 0 & 38 \\ \hline
Negated & 0 & 0 & 0 & 0 & 0 & 0 & 0 & 0 & 0 & 0 & 0 & 0 & 73 \\ \hline  \hline
Total & \cellcolor{red!40} 13 & 0 & 0 & \cellcolor{red!40} 18 & 0 & 0 & 0 & 0 & 0 & 0 & 0 & 0 & 236 \\ \hline
\end{tabular}
}
\caption{Compliance Test Results with BeSEPPI}
\label{tab:BeSEPPIAnalysisM}
\end{table}

To conclude, while \name and Fuseki handle all considered queries from the 3 chosen benchmarks correctly, Virtuoso produces a significant number of errors. 

\subsection{Performance Measurements}
\label{sec:QueryExec}

\paragraph{Experimental Setup}

\revision{Our benchmarks were executed on a system running openSUSE Leap 15.2 with dual Intel(R) Xeon(R) Silver 4314 16 core CPUs, clocked at 3.4 GHz, with 512GB RAM of which 256GB reserved for the system under test, and 256GB for the operating system.  
For each system under test, we set a time-out of 900s.} We start each benchmark by repeating the same warm-up queries $5$ times and by $5$ times loading and deleting the graph instance. Furthermore, we did $5$ repetitions of each query (each time deleting and reloading the dataset).
For our experiments we use Apache Jena Fuseki 3.15.0, Virtuoso Open Source Edition 7.2.5, and Stardog 7.7.1. Vadalog loads and queries the database simultaneously. 
Hence, to perform a fair comparison 
with competing systems, we compare their total loading and querying time to the total time that \name needs to answer the query. Since, loading includes index building and many more activities, we delete and reload the database each time, when we run a query (independent of warm-up or benchmark queries).

\paragraph{Performance on general SPARQL queries}

\begin{figure*}[h!]
  \centering
  \includegraphics[scale=0.8]{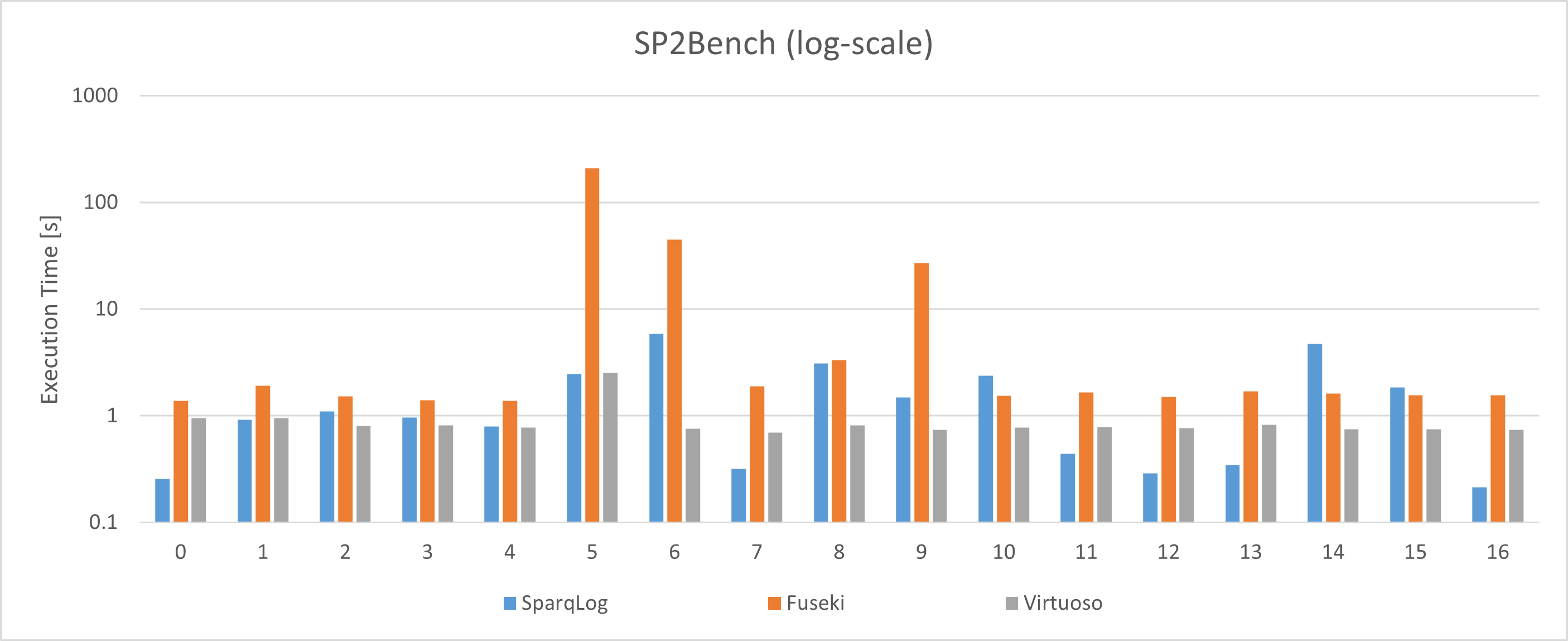}
  \caption{SP2Bench Benchmark (Log Scale)}
  \label{fig:SP2Bench}
\end{figure*}

SP2Bench is a benchmark that particularly targets query optimization and computation-intensive queries. 
We have visualized the result in Figure \ref{fig:SP2Bench} and found that SparqLog reaches highly competitive performance with Virtuoso and significantly outperforms Fuseki on most queries.

 \revision{\textit{gMark.} Since current SPARQL benchmarks provide only rudimentary coverage of property path expressions,  we have evaluated \name, Fuseki, and Virtuoso using the gMark benchmark generator \cite{gMark}, a domain- and language-independent graph instance and query workload generator which specifically focuses on path queries, i.e., queries over property paths. We have evaluated \name's, Fuseki's, and Virtuoso's path query performance on the \emph{test}\footnote{\url{https://github.com/gbagan/gMark/tree/master/demo/test}} and \emph{social}\footnote{\url{https://github.com/gbagan/gMark/tree/master/demo/social}} demo scenarios. Each of these two demo scenarios provides 50 SPARQL queries and a graph instance.
} 
\ifFullVersion
Appendix~\ref{sec:addBench} provides further details on the benchmarks that we used for evaluating a system's query execution time.
Full details on the results can be found in Section~\ref{sec:addBench}
\fi
In the following, we compare the results of the three systems on gMark:

\revision{\textbf{Virtuoso} could not (correctly) answer $48$ of the in total $100$ queries of the gMark Social and Test benchmark. Thus, it could not correctly answer almost half of the queries provided by both gMark benchmarks, which empirically reveals its dramatic limitations in answering complex property path queries. In $20$ of these $48$ cases, Virtuoso returned an incomplete result. While in solely $3$ incomplete result cases Virtuoso missed solely one tuple in the returned result multi-set, in the remaining $17$ incomplete result cases; Virtuoso produces either the result tuple $\textit{null}$ or an empty result multi-set instead of the correct non-null/non-empty result multi-set. In the other $28$ cases Virtuoso failed either due to a time-, mem-out or due to not supporting a property path with two variables. This exemplifies severe problems with handling property path queries.}

\revision{\textbf{Fuseki} suffered on $37$ of the in total $100$ queries of the gMark Social and Test benchmark a time-out (i.e., took longer than $900s$ for answering the queries). Thus, it timed-out on more than a third of gMark queries, which empirically reveals its significant limitations in answering complex property path queries.}

\begin{figure*}[h!]
  \centering
  \includegraphics[scale=0.8]{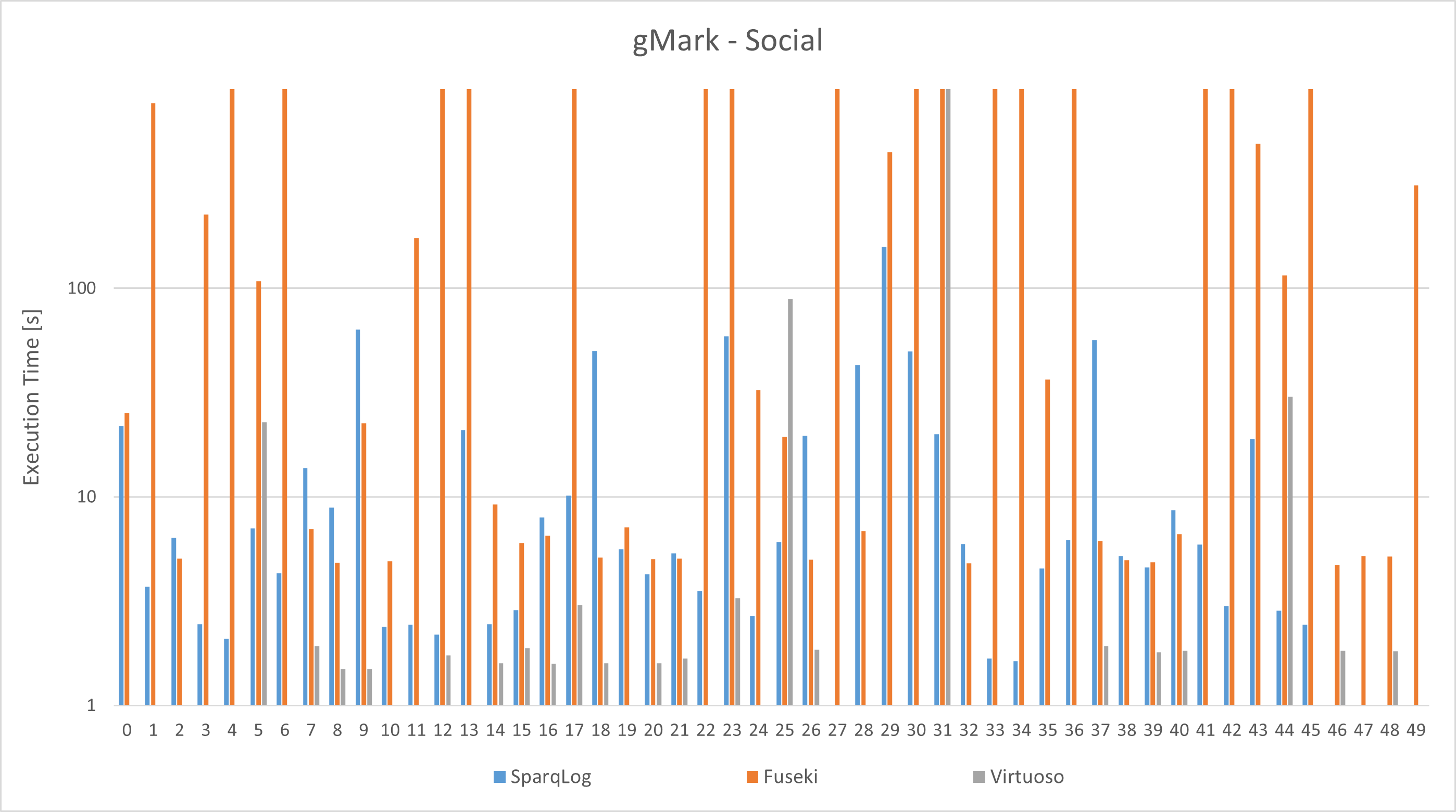}
  \caption{gMark Social Benchmark (Log Scale)}
  \label{fig:gMarkSocial}
\end{figure*}

\revision{\textbf{\name} managed to answer $98$ of gMark's (in total $100$) queries within less than $200s$ and timed out on solely $2$ queries.
The results on the gMark Social benchmark are shown in 
Figures~\ref{fig:gMarkSocial};
\ifFullVersion
the results on the gMark Test benchmark are given in 
Figure~\ref{fig:gMarkTest} in the appendix. 
\fi
These results reveal the strong ability of our system in answering queries that contain complex property paths. Furthermore, each time when both Fuseki and \name returned a result, the results were equal, even further empirically confirming the correctness of our system (i.e., that our system follows the SPARQL standard).}

\revision{In conclusion, these three benchmarks show that SparqLog (1) is highly competitive with Virtuoso on regular queries with respect to query execution time, 
(2) follows the SPARQL standard much more accurately than Virtuoso and supports more property path queries than Virtuoso,
and (3) dramatically outperforms Fuseki on query execution, while keeping its ability to follow the SPARQL standard accurately.}

\paragraph{Ontological reasoning}
One of the main advantages of our \name system is that it provides 
a uniform and consistent framework for reasoning and querying Knowledge Graphs. 
We therefore wanted to measure the performance of query answering in the 
presence of an ontology. Since Fuseki and Virtuoso do not provide such support, we compare \name with Stardog, which is a commonly accepted state-of-the-art system for reasoning and querying within the Semantic Web. 
Furthermore, we have created a benchmark based on SP2Bench's dataset that contains property path queries and ontological concepts such as subPropertyOf and subClassOf and provide this benchmark in the supplementary material.

\ifFullVersion
Figure \ref{fig:OntologyBenchM} in Appendix~\ref{app:ontologicalReasoning} shows the outcome of these experiments. 
\fi
In summary, we note that \name is faster than Stardog on most queries. Particularly interesting are queries 4 and 5, which 
contain recursive property path queries with two variables. Our engine needs on query 4 only about a fifth of the execution time of Stardog and it can even answer query 5, on which Stardog times outs (using a timeout of $900s$). On the other queries, Stardog and \name perform similarly.

To conclude, our new \name system does not only follow the SPARQL standard, but it also shows good performance. 
Even though \name is a full-fledged, general-purpose Knowledge Graph management system and neither a specialized SPARQL engine nor a specialized ontological reasoner, it is highly competitive to state-of-the-art SPARQL engines and reasoners and even outperforms them on answering property path queries and particularly hard cases.

\section{Conclusion}
\label{sect:Conclusion}

In this work we have taken a step towards bringing SPARQL-based systems and Datalog$^\pm$-based systems closer together. In particular, we have provided 
(i) a uniform and fairly complete theoretical translation of SPARQL into Warded Datalog$^\pm$, (ii) a practical translation engine that covers most of the SPARQL 1.1 functionality, and (iii) an extensive experimental evaluation.

We note that the contribution of the engine SparqLog we provided in this paper can be seen in two ways: (1) as a stand-alone translation engine for SPARQL into Warded Datalog$^\pm$, and (2) as a full Knowledge Graph engine by using our translation engine together with the Vadalog system.

However, our work does not stop here. 
As next steps, we envisage of course 100\% or close to 100\% SPARQL coverage. Possibly more (scientifically) interestingly, we plan to expand on the finding that query plan optimization provides a huge effect on performance, and investigate SPARQL-specific query plan optimization in a unified SPARQL-Datalog$^\pm$ system. 
Finally, we note that work on a unified benchmark covering all or close to 
all of the SPARQL 1.1 features would be desirable. As observed in 
Section \ref{sec:BenchmarkAnalysis}, no such benchmark currently exists.

\ifFullVersion
\begin{acks}
This work has been funded by the Vienna Science and 
Technology Fund (WWTF) [10.47379/VRG18013, 10.47379/NXT22018, 
\linebreak 
10.47379/ICT2201]; 
Renzo Angles was supported by ANID FONDECYT Chile through grant 1221727.
Georg Gottlob is a Royal Society Research Professor and acknowledges support by the Royal Society in this role through the  “RAISON DATA” project  (Reference No. RP\textbackslash{}R1\textbackslash{}201074). 
\end{acks}
\fi

\clearpage
\clearpage

\bibliographystyle{ACM-Reference-Format}
\bibliography{main}

\ifFullVersion

\clearpage

\noindent
{\huge\bf Appendix}

\appendix

\section{Translating SPARQL 1.1 to Warded Datalog$^\pm$}
\label{sec:rules}

In this section, we provide a detailed description of our translation 
from SPARQL 1.1 to Warded Datalog$^\pm$.
Note that many of the rules thus generated are simple Datalog rules, i.e., they do not have existentially quantified variables in the head. In such cases, we shall interchangeably 
refer to these rules as ```Datalog rules'' or 
``Datalog$^\pm$ rules''. Of course, if existentially quantified variables are indeed used in the head, we shall always speak of 
``Datalog$^\pm$ rules''.

We start with the translation of RDF graphs to Datalog rules in Section~\ref{app:translate-graphs}. 
We then detail our translation of graph patterns 
and the specific translation rules for property path
expressions in Sections \ref{app:translate-patterns} and
\ref{app:translate-paths}. Finally, in Section~\ref{app:translate-forms}, 
we consider query forms.

\subsection{Translation of RDF Graphs}
\label{app:translate-graphs}
Assume that $I$, $L$ and $B$ are disjoint infinite sets corresponding to IRIs, literals and blank nodes. 
An RDF term is an element in the set $T = I \cup L \cup B$.
An \emph{RDF triple} is a tuple $(s,p,o) \in T \times I \times T$ where $s$ is called the \emph{subject}, $p$ is the \emph{predicate} and $o$ is the \emph{object}.
An \emph{RDF graph} $G$ is a set of RDF triples. 
An \emph{RDF dataset} $D$ is a collection of graphs including a default graph $G_0$ and zero or more named graphs, such that a named graph is a pair $(u,G)$ where $u$ is an IRI which identifies the RDF graph $G$.

Let $G$ be a given RDF graph, the translation of the graph to Datalog facts is defined as follows:
\begin{enumerate}
    \item 
    For each IRI, constant and blank node in $G$ the corresponding facts $iri(X)$, $literal(X)$ and $bnode(X)$ are generated.
    \item
    For each named graph $g$ a tuple $named(g)$ and for each triple $(s,p,o)$ of graph $g$ a fact $triple(s,p,o,g)$ where $g$ is either ''default'' for the default graph or the IRI of a named graph is created.
    \item
     A term is either an IRI, a literal or a blank node:
    The set of terms is represented by the predicate $term$. 
\end{enumerate}
    \begin{definition}[Terms]
The predicate $terms$ is defined as follows:    
    \begin{align*}
    term(X) :-& \hspace{4pt} iri(X).\\
    term(X) :-& \hspace{4pt} literal(X).\\
    term(X) :-& \hspace{4pt} bnode(X).
    \end{align*}
    \end{definition}

\subsection{Translation of SPARQL Graph Patterns}
\label{app:translate-patterns}
Assume the existence of an infinite set $V$ of variables disjoint from $T$. We will use $\var(\alpha)$ to denote the set of variables occurring in any structure $\alpha$.
A \emph{graph pattern} is defined recursively as follows:
a tuple from $(T \cup V) \times (I \cup V) \times (T \cup V)$ is a triple pattern;
if $P_1$ and $P_2$ are graph patterns, $C$ is a filter constraint, and $g \in I$ then
$\{ P_1 \JOIN P_2 \}$,
$\{ P_1 \UNION P_2 \}$,
$\{ P_1 \OPT P_2 \}$, 
$\{ P_1 \MINUS P_2 \}$,
$\{ P_1 \FILTER C \}$, and;
$\{ \GRAPH~g~P_1 \}$ are graph patterns.
A \emph{filter constraint} is defined recursively as follows:
(i) If $?X,?Y \in V$, $c \in I \cup L$ and $r$ is a regular expression then $\true$, $\false$, $?X = c$, $?X = ?Y$, $\bound(?X)$, $\isI(?X)$, $\isB(?X)$, $\isL(?X)$ and $\regex(?X,r)$ are \emph{atomic filter constraints};
(ii) If $C_1$ and $C_2$ are filter constraints then 
$(!C_1)$, $(C_1~\&\&~C_2)$ and $(C_1~||~C_2)$ 
are \emph{Boolean filter constraints}.

A \emph{subpattern} $P'$ of a graph pattern $P$ is defined to be any substring of P that is also a graph pattern. Furthermore $P'$ is defined to be an immediate subpattern of $P$ if it is a subpattern of $P$ and if there is no other subpattern of $P$, different from $P$, that contains $P'$.
A \emph{parse tree} is specified as a tree $<V, E>$ with the set of nodes $V$ being the subpatterns of a graph pattern $P$ and the set of edges $E$ containing an edge $(P_1, P_2)$ if $P_2$ is an immediate subpattern of $P_1$.

The evaluation of a graph pattern results in a multiset of solution mappings.
A \emph{solution mapping} is a partial function $\mu : V \to T$, i.e. an assignment of variables to RDF terms.
The domain of $\mu$, denoted $\dom(\mu)$, is the subset of $V$ where $\mu$ is defined. 
The \emph{empty mapping}, denoted $\mu_0$, is the mapping satisfying that $\dom(\mu_0)= \emptyset$. 
A \emph{multiset of solution mappings} $\Omega$ is an unsorted list of solution mappings where duplicates are allowed. 
The domain of $\Omega$ is the set of variables occurring in the solution mappings of $\Omega$. 

The notion of compatible mappings is fundamental to evaluating SPARQL graphs patterns. 
Two mappings $\mu_1$ and $\mu_2$, are \emph{compatible}, denoted $\mu_1 \sim \mu_2$, when for all $?X \in \dom(\mu_1) \cap \dom(\mu_2)$ it satisfies that $\mu_1(?X)=\mu_2(?X)$.
Note that two mappings with disjoint domains are always compatible, and the empty mapping $\mu_0$ is compatible with any other mapping.

\begin{definition}[Compatibility]
The notion of compatible mappings is simulated in Datalog by using the following rules: 
\begin{align*}
\mathit{null}("\mathit{null}"). \\[1.1ex]
comp(X,X,X) :-& \hspace{4pt} term(X).\\
comp(X,\mathit{Z},X) :-& \hspace{4pt} term(X), \mathit{null}(\mathit{Z}).\\
comp(\mathit{Z},X,X) :-& \hspace{4pt} term(X), \mathit{null}(\mathit{Z}).\\
comp(\mathit{Z},\mathit{Z},\mathit{Z}) :-& \hspace{4pt} \mathit{null}(\mathit{Z}).
\end{align*}
\end{definition}

The predicate $comp(X_1,X_2,X_3)$ describes the compatibility of the values $X_1$ and $X_2$. The third position $X_3$ represents the value, that would be used in the result tuple of a join operation.

A given SPARQL graph pattern is translated into 
Warded Datalog$^\pm$
by recursively walking through the parse tree of the query and translating each subpattern into its respective Datalog$^{\pm}$ rules (as outlined in \cite{AnswerSetTrans}).
Subpatterns of the parse tree are indexed. The root has index 1, the left child of the $i$-th node has index $2 * i$, the right child has index $2 * i + 1$. During the translation, bindings of the $i$-th subpattern are represented by the predicate $ans_i$. 
Since the order of variables in predicates is relevant, some variable sets will need to be lexicographically ordered, which is denoted by $\overline{x}$ following the notation of \cite{AnswerSetTrans}.
$\overline{var}(P)$ shall represent the lexicographically ordered tuple of variables of $P$.
Moreover a renaming function $v_j: V \rightarrow V$ is defined.

For a given graph pattern $P$ and a given dataset 
$D = \langle G, G_{named} \rangle$ (where $G$ is the default graph and $G_{named}$ is the set of named graphs) the 
core translation function $\tau(P, dst, D,  \textit{NodeIndex})$ is defined as follows:
\begin{enumerate}
    \item $P$ is the graph pattern that should be translated next.
    \item $dst$ is a Boolean value that describes whether the results should have set semantics ($dst = True$) or bag semantics ($dst = False$).
    \item $D$ is the graph on which the query should be evaluated.
    \item $\textit{NodeIndex}$ is the index of the pattern $P$ that should be translated.
\end{enumerate}
In Definitions~\ref{app:def:triple} to 
\ref{def:PropertyPath} below, we present the translation function 
$\tau$ 
for the 
various language constructs of graph patterns in SPARQL 1.1.

\begin{definition}[Triple]
\label{app:def:triple}
Let $P_i$ be the i-th subpattern of P and furthermore let $P_i$ be a triple pattern $(s, p, o)$, then $\tau(P_i, \mathit{true}, D, i)$ is defined as:
\begin{align*}
ans_i(\overline{var}(P_{i}), D) :- \hspace{4pt} triple(s, p, o, D).
\end{align*}
And $\tau(P_i, \mathit{false}, D, i)$ is defined as:
\begin{align*}
ans_i(Id, \overline{var}(P_{i}), D) :- \hspace{4pt} triple(s, p, o, D).
\end{align*}
\end{definition}

\begin{definition}[Graph]
\label{app:def:graph}
Let $P_i$ be the i-th subpattern of P and furthermore let $P_i$ be $(\GRAPH$ $g$ $P_1)$, then $\tau(P_i, \mathit{true}, D, i)$ is defined as:
\begin{align*}
ans_i(\overline{var}(P_{i}), D) :-\hspace{4pt} & ans_{2i}(\overline{var}(P_1), g),\\ &named(g).\\
\tau(P_1, \mathit{true}, g, 2i).
\end{align*}
And $\tau(P_i, \mathit{false}, D, i)$ is defined as:
\begin{align*}
ans_i(Id, \overline{var}(P_{i}), D) :-\hspace{4pt}&
ans_{2i}(Id_1, \overline{var}(P_1), g),\\ &named(g).\\
\tau(P_1, \mathit{false}, g, 2i)
\end{align*}
\end{definition}

\begin{definition}[Join]
\label{app:def:join}
Let $P_i$ be the i-th subpattern of P and furthermore let $P_i$ be $(P_1 \JOIN P_2)$, then $\tau(P_i, \mathit{true}, D, i)$ is defined as:
\begin{align*}
ans_i(\overline{x}, D) :- \hspace{4pt} &
ans_{2i}(v_1(\overline{var}(P_1)), D), \\
&ans_{2i+1}(v_2(\overline{var}(P_2)), D),\\
&comp(v_1(x_1), v_2(x_1), x_1), \\
&\dots , 
comp(v_1(x_n), v_2(x_n), x_n).\\\\
\tau(P_1, \mathit{true}, D, 2i)\\
\tau(P_2, \mathit{true}, D, 2i + 1)
\end{align*}
\\
And $\tau(P_i, \mathit{false}, D, i)$ is defined as:
\begin{align*}
ans_i(Id, \overline{x}, D) :- \hspace{4pt} &
ans_{2i}(Id_1, v_1(\overline{var}(P_1)), D), \\
&ans_{2i+1}(Id_2, v_2(\overline{var}(P_2)), D),\\
&comp(v_1(x_1), v_2(x_1), x_1), \\
&\dots ,
comp(v_1(x_n), v_2(x_n), x_n).\\\\
\tau(P_1, \mathit{false}, D, 2i)\\
\tau(P_2, \mathit{false}, D, 2i + 1)
\end{align*}
with
\begin{itemize}[noitemsep, nolistsep]
    \item
    $\overline{x} = \overline{var(P_1) \cup var(P_2)}$
    \item
    $\{x_1, \dots, x_n\} = var(P_1) \cap var(P_2)$
    \item
    $v_1, v_2: var(P_1) \cap var(P_2) \rightarrow V$, such that\\
    $Image(v_1) \cap Image(v_2) = \{\}$
\end{itemize}
\end{definition}

\begin{definition}[Union]
\label{app:def:union}
Let $P_i$ be the i-th subpattern of P and furthermore let $P_i$ be $(P_1 \UNION P_2)$, then $\tau(P_i, \mathit{true}, D, i)$ is defined as:
\begin{align*}
ans_i(\overline{var}(P_{i}), D) :- \hspace{4pt}& ans_{2i}(\overline{var}(P_1), D), \\
&\mathit{null}(x_1), \dots \mathit{null}(x_n).\\
ans_i(\overline{var}(P_{i}), D) :- \hspace{4pt}& ans_{2i+1}(\overline{var}(P_2), D),\\
&\mathit{null}(y_1), \dots \mathit{null}(y_m).\\\\
\tau(P_1, \mathit{true}, D, 2i)\\
\tau(P_2, \mathit{true}, D, 2i + 1)
\end{align*}\\
And $\tau(P_i, \mathit{false}, D, i)$ is defined as:
\begin{align*}
ans_i(Id, \overline{var}(P_{i}), D) :- \hspace{4pt}& ans_{2i}(Id_1, \overline{var}(P_1), D), \\
&\mathit{null}(x_1), \dots \mathit{null}(x_n).\\
ans_i(Id, \overline{var}(P_{i}), D) :- \hspace{4pt}& ans_{2i+1}(Id_2, \overline{var}(P_2), D),\\
&\mathit{null}(y_1), \dots \mathit{null}(y_m).\\\\
\tau(P_1, \mathit{false}, D, 2i)\\
\tau(P_2, \mathit{false}, D, 2i + 1)
\end{align*}
with 
\begin{itemize}[noitemsep, nolistsep]
    \item
    $\{x_1, \dots, x_n\} = var(P_2) \setminus var(P_1)$
    \item
    $\{y_1, \dots, y_m\} = var(P_1) \setminus var(P_2)$
\end{itemize}

\begin{definition}[Optional]
\label{app:def:optional}
Let $P_i$ be the i-th subpattern of P and furthermore let $P_i$ be $(P_1 \OPT P_2)$, then $\tau(P_i, \mathit{true}, D, i)$ is defined as:
\begin{align*}
ans_{opt-i}(\overline{var}(P_1), D) :- \hspace{4pt}& ans_{2i}(\overline{var}(P_1), D), \\
&ans_{2i+1}(v_2(\overline{var}(P_2)), D),\\
&comp(x_1, v_2(x_1), z_1), \\
&\dots , comp(x_n, v_2(x_n), z_n).\\\\
ans_{i}(\overline{var}(P_{i}), D) :- \hspace{4pt}& ans_{2i}(v_1(\overline{var}(P_1)), D), \\
&ans_{2i+1}(v_2(\overline{var}(P_2)), D),\\
&comp(v_1(x_1), v_2(x_1), x_1), \\
&\dots , comp(v_1(x_n), v_2(x_n), x_n).\\\\
ans_{i}(\overline{var}(P_{i}), D) :- \hspace{4pt}& ans_{2i}(\overline{var}(P_1), D), \\
&not \hspace{4pt} ans_{opt-i}(\overline{var}(P_1), D),\\
&\mathit{null}(y_1), \dots, \mathit{null}(y_m).\\\\
\tau(P_1, \mathit{true}, D, 2i)\\
\tau(P_2, \mathit{true}, D, 2i + 1)
\end{align*}\\
And $\tau(P_i, \mathit{false}, D, i)$ is defined as:
\begin{align*}
ans_{opt-i}(\overline{var}(P_1), D) :- \hspace{4pt}& 
ans_{2i}(Id_1, \overline{var}(P_1), D), \\
&ans_{2i+1}(Id_2, v_2(\overline{var}(P_2)), D),\\
&comp(x_1, v_2(x_1), z_1), \\
&\dots , comp(x_n, v_2(x_n), z_n).\\\\
ans_{i}(Id, \overline{var}(P_{i}), D) :- \hspace{4pt}& ans_{2i}(Id_1, v_1(\overline{var}(P_1)), D), \\
&ans_{2i+1}(Id_2, v_2(\overline{var}(P_2)), D),\\
&comp(v_1(x_1), v_2(x_1), x_1), \\
&\dots , comp(v_1(x_n), v_2(x_n), x_n).\\\\
ans_{i}(Id, \overline{var}(P_{i}), D) :- \hspace{4pt}& ans_{2i}(Id_1, \overline{var}(P_1), D), \\
&not \hspace{4pt} ans_{opt-i}(\overline{var}(P_1), D),\\
&\mathit{null}(y_1), \dots, \mathit{null}(y_m).\\\\
\tau(P_1, \mathit{false}, D, 2i)\\
\tau(P_2, \mathit{false}, D, 2i + 1)
\end{align*}
with 
\begin{itemize}[noitemsep, nolistsep]
    \item 
    $\overline{var}(P_i) = \overline{var(P_1) \cup var(P_2)}$
    \item
    $\{x_1, \dots, x_n\} = var(P_1) \cap var(P_2)$
    \item
    $\{y_1, \dots, y_m\} = var(P_2) \setminus var(P_1)$
    \item
    $v_1, v_2: var(P_1) \cap var(P_2) \rightarrow V$, such that\\
    $Image(v_1) \cap Image(v_2) = \{\}$
\end{itemize}
\end{definition}

\end{definition}

\begin{definition}[Filter]
\label{app:def:filter}
Let $P_i$ be the i-th subpattern of P and furthermore let $P_i$ be $(P_1 \FILTER C)$, then $\tau(P_i, \mathit{true}, D, i)$ is defined as:
\begin{align*}
ans_i(\overline{var}(P_{i}), D) :- \hspace{4pt}& ans_{2i}(\overline{var}(P_1), D), C. \\
\tau(P_1, \mathit{true}, D, 2i)
\end{align*}

And $\tau(P_i, \mathit{false}, D, i)$ is defined as:
\begin{align*}
ans_i(Id, \overline{var}(P_{i}), D) :- \hspace{4pt}& ans_{2i}(Id_1, \overline{var}(P_1), D), C. \\
\tau(P_1, \mathit{false}, D, 2i)
\end{align*}

\end{definition}

\begin{definition}[Optional Filter]
\label{app:def:optionalFilter}
Let $P_i$ be the i-th subpattern of P and furthermore let $P_i$ be $(P_1 \OPT (P_2 \FILTER C))$, then $\tau(P_i, \mathit{true}, D, i)$ is defined as:

\begin{align*}
ans_{opt-i}(\overline{var}(P_1), D) :- \hspace{4pt}& ans_{2i}(\overline{var}(P_1), D), \\
&ans_{2i+1}(v_2(\overline{var}(P_2)), D),\\
&comp(x_1, v_2(x_1), z_1), \\
&\dots , comp(x_n, v_2(x_n), z_n), C.\\\\
ans_{i}(\overline{var}(P_{i}), D) :- \hspace{4pt}& ans_{2i}(v_1(\overline{var}(P_1)), D), \\
&ans_{2i+1}(v_2(\overline{var}(P_2)), D),\\
&comp(v_1(x_1), v_2(x_1), x_1), \\
&\dots , comp(v_1(x_n), v_2(x_n), x_n), C.\\\\
ans_{i}(\overline{var}(P_{i}), D) :- \hspace{4pt}& ans_{2i}(\overline{var}(P_1), D), \\
&not \hspace{4pt} ans_{opt-i}(\overline{var}(P_1), D),\\
&\mathit{null}(y_1), \dots, \mathit{null}(y_m).\\\\
\tau(P_1, \mathit{true}, D, 2i)\\
\tau(P_2, \mathit{true}, D, 2i + 1)
\end{align*}

And $\tau(P_i, \mathit{false}, D, i)$ is defined as:

\begin{align*}
ans_{opt-i}(\overline{var}(P_1), D) :- \hspace{4pt}& ans_{2i}(Id_1, \overline{var}(P_1), D), \\
&ans_{2i+1}(Id_2, v_2(\overline{var}(P_2)), D),\\
&comp(x_1, v_2(x_1), z_1), \\
&\dots , comp(x_n, v_2(x_n), z_n), C.\\\\
ans_{i}(Id, \overline{var}(P_{i}), D) :- \hspace{4pt}& ans_{2i}(Id_1, v_1(\overline{var}(P_1)), D), \\
&ans_{2i+1}(Id_2, v_2(\overline{var}(P_2)), D),\\
&comp(v_1(x_1), v_2(x_1), x_1), \\
&\dots , comp(v_1(x_n), v_2(x_n), x_n), C.\\\\
ans_{i}(Id, \overline{var}(P_{i}), D) :- \hspace{4pt}& ans_{2i}(Id_1, \overline{var}(P_1), D), \\
&not \hspace{4pt} ans_{opt-i}(\overline{var}(P_1), D),\\
&\mathit{null}(y_1), \dots, \mathit{null}(y_m).\\\\
\tau(P_1, \mathit{false}, D, 2i)\\
\tau(P_2, \mathit{false}, D, 2i + 1)
\end{align*}
with 
\begin{itemize}[noitemsep, nolistsep]
    \item 
    $\overline{var}(P_i) = \overline{var(P_1) \cup var(P_2)}$
    \item
    $\{x_1, \dots, x_n\} = var(P_1) \cap var(P_2)$
    \item
    $\{y_1, \dots, y_m\} = var(P_2) \setminus var(P_1)$
    \item
    $v_1, v_2: var(P_1) \cap var(P_2) \rightarrow V$, such that\\
    $Image(v_1) \cap Image(v_2) = \{\}$
\end{itemize}
\end{definition}

\newpage

\begin{definition}[Minus]
\label{app:def:minus}
Let $P_i$ be the i-th subpattern of P and furthermore let $P_i$ be $(P_1 \MINUS P_2)$, then $\tau(P_i, \mathit{true}, D, i)$ is defined as:
\begin{align*}
ans_{join-i}(\overline{x}, D) :- \hspace{4pt} &
ans_{2i}(\overline{var}(P_1), D), \\
&ans_{2i+1}(v_2(\overline{var}(P_2)), D),\\
&comp(x_1, v_2(x_1), z_1), \\
&\dots , 
comp(x_n, v_2(x_n), z_n).\\
ans_{equal-i}(\overline{var}(P_1), D) :- \hspace{4pt} & ans_{join-i}(\overline{x}, D), \\
&x_1 = v_2(x_1),\hspace{3pt} not \hspace{4pt} null(x_1).\\
\dots\\
ans_{equal-i}(\overline{var}(P_1), D) :- \hspace{4pt} & ans_{join-i}(\overline{x}, D), \\
&x_n = v_2(x_n),\hspace{3pt} not \hspace{4pt} null(x_n).\\
ans_i(\overline{var}(P_1), D) :- \hspace{4pt} &
ans_{2i}(\overline{var}(P_1), D), \\
&not \hspace{4pt} ans_{equal-i}(\overline{var}(P_1), D).
\\\\
\tau(P_1, \mathit{true}, D, 2i)\\
\tau(P_2, \mathit{true}, D, 2i + 1)
\end{align*}
\\
And $\tau(P_i, \mathit{false}, D, i)$ is defined as:
\begin{align*}
ans_{join-i}(\overline{x}, D) :- \hspace{4pt} &
ans_{2i}(Id_1, \overline{var}(P_1), D), \\
&ans_{2i+1}(Id_2, v_2(\overline{var}(P_2)), D),\\
&comp(x_1, v_2(x_1), z_1), \\
&\dots , 
comp(x_n, v_2(x_n), z_n).\\
ans_{equal-i}(\overline{var}(P_1), D) :- \hspace{4pt} & ans_{join-i}(\overline{x}, D), \\
&x_1 = v_2(x_1),\hspace{3pt} not \hspace{4pt} null(x_1).\\
\dots\\
ans_{equal-i}(\overline{var}(P_1), D) :- \hspace{4pt} & ans_{join-i}(\overline{x}, D), \\
&x_n = v_2(x_n),\hspace{3pt} not \hspace{4pt} null(x_n).\\
ans_i(Id, \overline{var}(P_1), D) :- \hspace{4pt} &
ans_{2i}(Id_1, \overline{var}(P_1), D), \\
&not \hspace{4pt} ans_{equal-i}(\overline{var}(P_1), D).
\\\\
\tau(P_1, \mathit{false}, D, 2i)\\
\tau(P_2, \mathit{false}, D, 2i + 1)
\end{align*}
with 
\begin{itemize}[noitemsep, nolistsep]
    \item 
    $\overline{x} = \overline{var(P_1) \cup v_2(var(P_2))}$
    \item
    $\{x_1, \dots, x_n\} = var(P_1) \cap var(P_2)$
    \item
    $v_2: var(P_1) \cap var(P_2) \rightarrow V \setminus var(P_1)$
\end{itemize}

\end{definition}

Property path patterns are given in the form 
$S \hspace{4pt} P_1 \hspace{4pt} O$, where $P_1$ is a property path expression. 
Due to the complex semantics of property paths, we have introduced a separate translation function $\tau_{PP}$ for property path expressions, which we will take a closer look at in 
Section~\ref{app:translate-paths}.

\begin{definition}[Property Path Pattern]
\label{def:PropertyPath}
Let $P_i$ be the i-th subpattern of $P$ and furthermore let $P_i = S \hspace{4pt} P_1 \hspace{4pt} O$ be a property path pattern, then $\tau(P_i, \mathit{true}, D, i)$ is defined as:
\begin{align*}
ans_i(\overline{var}(P_i), D) :- \hspace{4pt}& ans_{2i}(S, O, D).\\
\tau_{PP}(P_1, \mathit{true}, S, O, D, 2i)
\end{align*}

And $\tau(P_i, \mathit{false}, D, i)$ is defined as:
\begin{align*}
ans_i(Id, \overline{var}(P_i), D) :- \hspace{4pt}& ans_{2i}(Id_1, S, O, D).\\
\tau_{PP}(P_1, \mathit{false}, S, O, D, 2i)
\end{align*}
with $\tau_{PP}$ being the translation function for property path expressions
defined next.
\end{definition}

\subsection{Translation of Property Path Expressions}
\label{app:translate-paths}
A \emph{property path expression} (or, simply, a \emph{path expression\/})
is defined recursively as follows:
\begin{itemize}
\item if $a \in I$, then $a$ is a link path expression;
\item If $PP_1$ and $PP_2$ are property path expressions then 
\begin{itemize}
\item $\inv PP_1$ is an inverse path expression;
\item $PP_1 \mid PP_2$ is an alternative path expression;
\item $PP_1 / PP_2$ is a sequence path expression;
\item $PP_1?$ is a zero-or-one path expression;
\item $PP_1+$ is a one-or-more path expression;
\item $PP_1*$ is a zero-or-more path expression;
\end{itemize} 
\item if $\mathcal{P}$ is a set of link path expressions $p_i$ and 
inverse link path expressions $\inv p_j$, then 
$!\mathcal{P}$ is a negated path expression.
\end{itemize}

A \emph{property path pattern} is a tuple $t$ of the form $(u,p,v)$, where $u,v \in (I \cup V)$ and $p$ is a property path expression. 
Analogously to our translation function 
$\tau(P, dst, D,  \textit{NodeIndex})$ for SPARQL patterns, we 
define a translation function 
$\tau_{PP}(PP, dst, S, O, D,  \textit{NodeIndex})$ for property path expressions, where 
$S,O$, are the subject and object of the top-level property path expression that have to be
kept track of during the entire evaluation.

In the following we will only state the translations for property path expressions 
under bag semantics (i.e. $dst = false$), since for set semantics the IDs are simply left out 
or set to a constant value (e.g. $Id = []$).

\begin{definition}[Link Property Path]
\label{app:def:link}
Let $PP_i$ be the $i$-th subexpression of a property path expression $PP$ and furthermore let $PP_i = p_1$ be a link property path expression. 
Then $\tau_{PP}(PP_i, \mathit{false}, S, O, D, i)$ is defined as:
\begin{align*}
ans_i(Id, X, Y, D) :- \hspace{4pt}& triple(X, p1, Y, D).
\end{align*}
\end{definition}

\begin{definition}[Inverse Property Path]
\label{app:def:inverse}
Let $PP_i$ be the $i$-th subexpression of a property path expression $PP$ and furthermore let $PP_i = \inv PP_1 $ be an inverse property path expression. 
Then $\tau_{PP}(PP_i, \mathit{false}$, $S, O, D, i)$ is defined as:
\begin{align*}
ans_i(Id, X, Y, D) :- \hspace{4pt}& ans_{2i}(Id_1, Y, X, D).\\
\tau_{PP}(PP_1, \mathit{false}, S, O, D, 2i)
\end{align*}
\end{definition}

\begin{definition}[Alternative Property Path]
\label{app:def:alternative}
Let $PP_i$ be the $i$-th subexpression of a property path expression $PP$ and furthermore let $PP_i = PP_1 | PP_2 $ be a alternative property path expression.
Then $\tau_{PP}(PP_i, \mathit{false}, S, O, D, i)$ is defined as:
\begin{align*}
ans_i(Id, X, Y, D) :- \hspace{4pt}& ans_{2i}(Id_1, X, Y, D).\\
ans_i(Id, X, Y, D) :- \hspace{4pt}& ans_{2i+1}(Id_1, X, Y, D).\\
\tau_{PP}(PP_1, \mathit{false}, S, O, D, 2i)\\
\tau_{PP}(PP_2, \mathit{false}, S, O, D, 2i+1)
\end{align*}
\end{definition}

\begin{definition}[Sequence Property Path]
\label{app:def:sequence}
Let $PP_i$ be the $i$-th subexpression of a property path expression $PP$ and furthermore let $PP_i = PP_1/PP_2$ be a sequence property path expression.
Then $\tau_{PP}(PP_i$, $\mathit{false}, S, O, D, i)$ is defined as:
\begin{align*}
ans_i(Id, X, Z, D) :- \hspace{4pt}& ans_{2i}(Id_1, X, Y, D),\\  
& ans_{2i+1}(Id_2, Y, Z, D).\\
\tau_{PP}(PP_1, \mathit{false}, S, O, D, 2i)\\
\tau_{PP}(PP_2, \mathit{false}, S, O, D, 2i+1)
\end{align*}
\end{definition}

\begin{definition}[One-Or-More Property Path]
\label{app:def:oneOrMore}
Let $PP_i$ be the $i$-th subexpression of a property path expression $PP$ and furthermore 
let $PP_i = PP_1\hspace{-1pt}+ $ be a one-or-more property path expression.
Then $\tau_{PP}(PP_i, \mathit{false}, S, O, D, i)$ is defined as:
\begin{align*}
ans_i(Id, X, Y, D) :- \hspace{4pt}& ans_{2i}(Id_1, X, Y, D), Id = [].\\
ans_i(Id, X, Z, D) :- \hspace{4pt}& ans_{2i}(Id_1, X, Y, D), \\
& ans_{i}(Id_2, Y, Z, D), Id = [].\\
\tau_{PP}(PP_1, \mathit{false}, S, O, D, 2i)
\end{align*}
\end{definition}

\begin{definition}[SubjectOrObject]
\label{app:def:subject}
The subjectOrObject predicate defines intuitively the set of all possible subjects and objects 
occurring in a graph, i.e.:
\begin{align*}
subjectOrObject(X) :- \hspace{4pt}& triple(X, P, Y, D).\\
subjectOrObject(Y) :- \hspace{4pt}& triple(X, P, Y, D).\\
\end{align*}
\end{definition}
\begin{definition}[Zero-Or-One Property Path]
\label{app:def:zeroOrOne}
Let $PP_i$ be the $i$-th subexpression of a property path expression $PP$ and furthermore let $PP_i = PP_1? $ be a 
zero-or-one property property path expression. 
Then $\tau_{PP}(PP_i, \mathit{false}, S, O, D, i)$ consists of the following rules:
\begin{align*}
ans_i(Id, X, X, D) :- \hspace{4pt}& subjectOrObject(X), Id = [].\\
ans_i(Id, X, Y, D) :- \hspace{4pt}& ans_{2i}(Id_1, X, Y, D), Id = [].\\
\tau_{PP}(PP_1, \mathit{false}, S, O, D, 2i)
\end{align*}
Moreover, if either one of $S$ and $O$ is a variable and the other is a non-variable $t$ 
or both $S$ and $O$ are the same non-variable $t$, then the following rule is added: 
\begin{align*}
ans_i(Id, X, X, D) :- \hspace{4pt}& not \hspace{4pt} Term(X), X = t, Id = [].
\end{align*}
\end{definition}
It should be noted that, according to the SPARQL semantics of property paths\footnote{\href{https://www.w3.org/TR/SPARQL11-query/\#defn_PropertyPathExpr}{https://www.w3.org/TR/SPARQL11-query/\#defn\_PropertyPathExpr}}, 
zero-or-one, zero-or-more, and one-or-more property paths always have set semantics.
This is why the Datalog$^{\pm}$ rules for these three path expressions 
contain a body literal $Id = []$. By forcing the tuple ID to the same value whenever one of these rules
fires, multiply derived tuples are indistinguishable for our system and will, therefore, 
never give rise to duplicates.

\begin{definition}[Zero-Or-More Property Path]
\label{app:def:zeroOrMore}
Let $PP_i$ be the $i$-th subexpression of a property path expression $PP$ and furthermore let $PP_i =  PP_1* $ be a 
zero-or-more property property path expression. 
Then $\tau_{PP}(PP_i, \mathit{false}, S, O, D, i)$ consists of the following rules:
\begin{align*}
ans_i(Id, X, X, D) :- \hspace{4pt}& subjectOrObject(X), Id = [].\\
ans_i(Id, X, Y, D) :- \hspace{4pt}& ans_{2i}(Id_1, X, Y, D), Id = [].\\
ans_i(Id, X, Z, D) :- \hspace{4pt}& ans_{2i}(Id_1, X, Y, D), \\
& ans_{i}(Id_2, Y, Z, D), Id = [].\\
\tau_{PP}(PP_1, \mathit{false}, S, O, D, 2i)
\end{align*}
Moreover, if either one of $S$ and $O$ is a variable and the other is a non-variable $t$ 
or both $S$ and $O$ are the same non-variable $t$, then the following rule is added: 
\begin{align*}
ans_i(Id, X, X, D) :- \hspace{4pt}& not \hspace{4pt} Term(X), X = t, Id = [].
\end{align*}
\end{definition}

Essentially, the zero-or-more property path is a combination of the zero-or-one and one-or-more property path.

\begin{definition}[Negated Property Path]
\label{app:def:negated}
Let $PP_i$ be the $i$-th subexpression of a property path expression $PP$ and furthermore let $PP_i = !(\mathcal{P})$ be a negated property path expression. 
Then $\tau_{PP}(PP_i$, $\mathit{false}, S, O, D, i)$ is defined as:
\begin{align*}
ans_i(Id, X, Y, D) :- \hspace{4pt}& triple(X, P, Y, D), P != p_{f_1}, \dots, P != p_{f_n}.\\
ans_i(Id, Y, X, D) :- \hspace{4pt}& triple(X, P, Y, D), P != p_{b_1}, \dots, P != p_{b_m}.
\end{align*}
with
\begin{itemize}
    \item $p_{f_1}, \dots, p_{f_n} \in \{ p | p \in \mathcal{P} \}$ ... i.e. the set of negated forward predicates.
    \item $p_{b_1}, \dots, p_{b_m} \in \{ p | \inv p \in \mathcal{P} \}$ ... i.e. the set of negated backward predicates.
\end{itemize}
\end{definition}

\subsection{Translation of Query Forms}
\label{app:translate-forms}
Let $P_1$ be a graph pattern and $W$ be a set of variables.
We consider two types of query forms: 
$(\SELECT$ $W$ $P_1)$ and 
$(\ASK P_1)$. Their translation is given below.

\begin{definition}[Select]
Let $P_i$ be the i-th subpattern of P and furthermore let $P_i$ be $(\SELECT$ $W$ $P_1)$, then $\tau(P_i, \mathit{true}, D, i)$ is defined as:
\begin{align*}
ans_i(\overline{var}(W), D) :- \hspace{4pt}& ans_{2i}(\overline{var}(P_1), D). \\
\tau(P_1, \mathit{true}, D, 2i)
\end{align*}

And $\tau(P_i, \mathit{false}, D, i)$ is defined as:
\begin{align*}
ans_i(Id, \overline{var}(W), D) :- \hspace{4pt}& ans_{2i}(Id_1, \overline{var}(P_1), D). \\
\tau(P_1, \mathit{false}, D, 2i)
\end{align*}
\end{definition}

\begin{definition}[Ask]
Let $P_i$ be the i-th subpattern of P and furthermore let $P_i$ be $\ASK P_1)$, then $\tau(P_i, \mathit{true}, D, i)$ is defined as:
\begin{align*}
ans_i(HasResult) :- \hspace{4pt}& ans\_ask_i(HasResult).\\
ans_i(HasResult) :- \hspace{4pt}& not \hspace{4pt} ans\_ask_{i}(True), HasResult = false.\\
ans\_ask_i(HasResult) :- \hspace{4pt}& ans_{2i}(\overline{var}(P_1), D), HasResult = true.\\
\tau(P_1, \mathit{true}, D, 2i)
\end{align*}

And $\tau(P_i, \mathit{false}, D, i)$ is defined as:
\begin{align*}
ans_i(HasResult) :- \hspace{4pt}& ans\_ask_i(HasResult).\\
ans_i(HasResult) :- \hspace{4pt}& not \hspace{4pt} ans\_ask_{i}(True), HasResult = false.\\
ans\_ask_i(HasResult) :- \hspace{4pt}& ans_{2i}(Id_1, \overline{var}(P_1), D), HasResult = true.\\
\tau(P_1, \mathit{false}, D, 2i)
\end{align*}
\end{definition}

\section{Correctness of the Translation}
\label{app:correctness}
As was mentioned in Section \ref{sect:Correctness}, we have applied a 
two-way strategy for ensuring the correctness of our translation from 
SPARQL~1.1 to Warded Datalog$^\pm$ by carrying out an extensive empirical 
evaluation and a formal analysis. 
For the empirical evaluation, we have run 
our \name system as well as Fuseki and Virtuoso on several benchmarks which provide a good coverage of SPARQL~1.1. 
The results of our empirical evaluation are 
summarized in 
Section~\ref{sec:Compliance}. They 
give strong evidence for the correctness of the translation 
applied by \name. 
To provide yet further evidence, 
we will now formally examine the Warded Datalog$^\pm$
rules produced by our translation for the various SPARQL language constructs and compare them with the formal semantics of these 
language constructs.

As was mentioned in Section~\ref{sec:integration},
\name includes a translation engine with three methods, namely
a data translation method, a 
query translation method, and a 
solution translation method. 
The data translation is 
very straightforward. In particular, the IRIs, literals and blank nodes as well as the triples in an RDF graph are presented as 
Datalog ground facts in the obvious way. 
Recall from 
Table~\ref{tab:coverage} that, as far as query forms are concerned, 
we currently only support SELECT (which is by far the most common one) and ASK. 
The former allows one to define a projection to some of the variables in the graph pattern while the latter just asks if at least 
some mapping satisfying the graph pattern exists. 
In case of SELECT, the solution modifier can be further extended by 
a DISTINCT, ORDER BY, LIMIT, or OFFSET clause. 
The two supported solution modifiers (with the possible extensions)
are obvious 
and they are taken care of by the solution translation method of \name. 
In the sequel, we therefore restrict our discussion to the query translation method. 
As in Section~\ref{sect:TranslationEngine}, we
treat the basic translation rules and the translation of property paths in separate subsections.

\subsection{Basic Translation Rules}
\label{app:correctnessBasicRules}

First we recall some basic principles 
of defining a formal semantics of SPARQL 
\cite{DBLP:conf/semweb/AnglesG08,DBLP:books/sp/virgilio09/ArenasGP09,AnswerSetTrans}. 
At the heart of evaluating a SPARQL query is the 
evaluation of the graph pattern (GP) given in the WHERE clause
of the query. This evaluation is relative to the active graph 
$D$, which is initially the default graph (obtained by merging the 
graphs given in the FROM clause of the query) and which can 
be switched to some named graph (given by an IRI in a FROM NAMED clause of the query) via the GRAPH construct.
We write $\graph(u)$ to denote the graph with name $u$ and we write 
$\names$ to denote all names of named graphs according to the FROM NAMED clauses.

The result of evaluating a graph pattern $P$ relative to some graph $D$,
denoted by $\epag{P}{D}$,
is a multiset of partial mappings $\mu \colon V \rightarrow T$
(simply referred to as ``mappings'' henceforth), 
where
$V$ is the set of variables and $T$ is the set of terms 
(i.e., the union of IRIs, blank nodes and literals). 
It is convenient to allow also the constant
``null'' as function value to indicate by $\mu(?X) = $ ``null'' that 
$\mu$ is undefined on variable $?X$. The domain of $\mu$, denoted $\dom(\mu)$, 
is defined as the set of variables on which $\mu$ is defined.
Mappings are applied to triple patterns in the obvious way, i.e., 
let $t = (s, p, o)$ be a triple pattern and let $\var(t)$ denote the 
variables in $t$. For a mapping $\mu$ with $\var(t) \subseteq \dom(\mu)$, 
we write $\mu(t)$ to denote the triple obtained by replacing 
each variable $?X \in \var(t)$ by $\mu(?X)$.

\smallskip

\noindent
{\bf Compatibility}.
An important property when combining or comparing two mappings is compatibility. Two mappings $\mu_1,\mu_2$
are {\em compatible}, denoted $\mu_1\sim\mu_2$, if
$\mu_1(?X)=\mu_2(?X)$ holds
for all $?X \in \dom(\mu_1) \cap \dom(\mu_2)$. In this case, 
the  mapping $\mu = \mu_1 \cup \mu_2$ with 
$\mu(?X) = \mu_1(?X)$ if $?X \in \dom(\mu_1)$ and
$\mu(?X) = \mu_2(?X)$ if $?X \in \dom(\mu_2)$
is well-defined.

In Section \ref{app:translate-patterns}, we have also defined the compatibility of  two individual terms
or nulls $v_1,v_2$, namely: $v_1$ and $v_2$ are compatible if they are equal (i.e., either the same term or both `null'') or if one of them
is ``null''. Clearly, two partial mappings 
$\mu_1,\mu_2$
are compatible if and only if 
$\mu_1(?X)$ and $\mu_2(?X)$ are compatible for every variable $?X \in \var(\mu_1) \cap \var(\mu_2)$. 
If this is the case, then $\mu = \mu_1 \cup \mu_2$ is obtained as follows:
for every variable $?X$, (1) if $\mu_1(?X)=\mu_2(?X)$ (where 
$\mu_1(?X)$ and $\mu_2(?X)$ are either the same term or they are both ``null''), then 
$\mu(?X) = \mu_1(?X)=\mu_2(?X)$; and (2) if one of $\mu_1(?X),\mu_2(?X)$ is a term and the other is
``null'', then $\mu(?X)$ is set equal to the term.

We observe that the auxiliary predicate 
$comp(X_1,X_2,X_3)$
defined in Section~\ref{app:translate-patterns} realises precisely the compatibility check 
between two values $\mu_1(?X)$ and $\mu_2(?X)$ (in the first two components of $comp$) and 
yields $\mu(?X)$ in the third component.

\smallskip

\noindent
{\bf Operations on multisets of mappings}.
We consider the following operations between two sets of mappings $\Omega_1,\Omega_2$:

\smallskip

$\Omega_1 \Join \Omega_2 = 
\mset{\mu_1 \cup \mu_2 \mid \mu_1 \in \Omega_1, \mu_2 \in \Omega_2 \text{ and } \mu_1 \sim \mu_2}$ 

$\Omega_1 \cup \Omega_2 = 
\mset{ 
\mu \mid \mu \in \Omega_1 \textrm{ or } \mu \in \Omega_2 
}$ 

$\Omega_1 \setminus \Omega_2 = 
\mset{
\mu_1 \in \Omega_1 \mid \textrm{ for all } \mu_2 \in \Omega_2, \mu_1 \not\sim \mu_2 
}$ 

$\Omega_1 \lojoin  \Omega_2 = (\Omega_1 \Join \Omega_2) \cup (\Omega_1 \setminus \Omega_2)$ 

\smallskip

\noindent
Note that, of the above operations, only the union $\cup$ may alter the cardinality
of elements in the resulting multiset, namely: if a mapping $\mu$ is contained in 
both $\Omega_1$ and $\Omega_2$, then its cardinality in $\Omega$ is the sum of the 
original cardinalities in $\Omega_1$ and $\Omega_2$.

\smallskip

\noindent
{\bf Semantics of basic SPARQL constructs}.
The semantics $\epag{P}{D}$ of a graph pattern $P$ is defined recursively 
on the structure of $P$. In the base case, $P$ is a triple pattern $P = (s, p, o)$
and $\epag{P}{D}$ is defined as 
$\epag{P}{D} = \{ \mu \mid \dom(\mu) = \var(P)$ and $\mu(P) \in D\}$.
For complex graph patterns $P$, the semantics definition
$\epag{P}{D}$ is shown in 
Table~\ref{table:semanticsBasic}.

\begin{table}[!]
\caption{
Semantics of basic graph patterns. 
$P_1, P_2$ are graph patterns, 
$C$ is a filter constraint,  
$u \in I$ and $?X \in V$.}
\label{table:semanticsBasic}
\centering
\begin{tabular}{|l|l|}
\hline
\rs~Graph pattern $P$~&~~~~~~~~~~~Evaluation $\epag{P}{D}$ \\ \hline \hline 
$(\GRAPH \, u \, P_1)$   & \rs~~$\epag{P_1}{\graph(u)}$ if $u \in \names$ and $\emptyset$ otherwise \\ \hline
$(\GRAPH \, ?X \, P_1)$   & \rs~~$\bigcup_{v \in \names} 
     ( \epag{P_1}{\graph(v)} \Join \{ ?X \to v \})$ \\ 
 \hline 

$(P_1 \AAND P_2)$   & \rs~~$\epag{P_1}{D} \Join \epag{P_2}{D}$  \\ \hline  
$(P_1 \FILTER C)$  & \rs~~$\{ \mu \mid \mu \in \epag{P_1}{D} \text{ and } \mu \models C \}$ \\ \hline 
$(P_1 \OPT P_2)$   & \rs~~$\epag{P_1}{D} \lojoin \epag{P_2}{D}$ \\ \hline 
$(P_1 \OPT $    & \rs~~$
\mset{\mu \mid \mu \in  \epag{P_1}{D} \bowtie \epag{P_2}{D}
\text{ and } \mu \models C }$ $\cup$ \\ 
\mbox{} \quad  $(P_2 \FILTER C))$ & 
\rs~~$\{ \hskip-2pt  \{
\mu_1 \mid \mu_1 \in \epag{P_1}{D} \text{ and for all }  \mu_2  \in \epag{P_2}{D}$: 
\\ 
& \mbox{} \hskip 14pt either $\mu_1 \not\sim \mu_2$ 
\\
& \mbox{} \hskip 14pt or  $\mu_1 \sim \mu_2$ and $\mu_1 \cup \mu_2 \not \models C
\} \hskip-2pt  \}
$ 
\\
\hline 
$(P_1 \UNION P_2)$ & \rs~~$\epag{P_1}{D} \cup \epag{P_2}{D}$ \\ \hline 
$(P_1 \MINUS P_2)$ & \rs~~
$\{ \hskip-2pt  \{
\mu_1 \in \epag{P_1}{D}  \mid \textrm{ for all } \mu_2 \in \epag{P_2}{D}$, \\
& \mbox{} \hskip 3pt $(\mu_1 \not\sim \mu_2
\text{ or } \dom(\mu_1) \cap \dom(\mu_2) = \emptyset) 
\} \hskip-2pt  \}$ 
\\ \hline 
\end{tabular}
\end{table}

\smallskip

\noindent
{\bf Translation of basic SPARQL constructs}.
We are now ready to inspect the translations from SPARQL 1.1 to Warded Datalog$^{\pm}$
given in Figure~\ref{fig:basictranslation}
and Section~\ref{app:translate-patterns}. 
As in Section~\ref{sect:basic-rules}, we concentrate on bag semantics 
as the more complex case.

\smallskip

\noindent
{\em Triple}.
First consider the base case of graph patterns, namely a triple pattern 
$P_i = (s,p,o)$, where each of $s,p,o$ can be a term or a variable. Clearly, 
the single rule produced by our translation $\tau(P_i, \mathit{false}, D, i)$ 
in Definition~\ref{app:def:triple} produces all mappings in $\var(P_i)$
that match $(s,p,o)$ to a triple in the active graph $D$.

\smallskip

\noindent
{\em Graph}. Suppose that 
 $P_i$ is of the form $P_i = (\GRAPH$ $g$ $P_1)$. 
 According to the semantics definition in 
 Table~\ref{table:semanticsBasic}
 we have to distinguish 2 cases depending on whether $g$ is an IRI or a variable. Moreover, in the former case, 
 we have the two subcases depending on whether the IRI $g$ is the name of some named graph (i.e., it occurs in 
 $names$) or not. It is easy to verify that 
the single rule produced by our translation $\tau(P_i, \mathit{false}, D, i)$ 
in Definition~\ref{app:def:graph} covers exactly these 3 cases. 

If $g$ is an IRI that occurs in 
$\names$, then the body literal $named(g)$ of the Datalog rule will evaluate to true and the result
obtained by the mappings (on the variables $\var(P_1)$) obtained by the body literal
$ans_{2i}(Id_1, \overline{var}(P_1), g)$ are precisely the mappings obtained by evaluating 
graph pattern $P_1$ over the graph with name $g$, i.e., $\graph(g)$. In particular, 
the head variables $\overline{var}(P_i)$ coincide with the body variables $\overline{var}(P_1)$.
Note that 
the variable $Id$ in the head has the effect that every firing of the rule binds
$Id$ to a different labelled null. Hence, if $ans_{2i}(Id_1, \overline{var}(P_1), g)$ yields duplicates
(i.e., identical mappings with different bindings of $Id_1$), then these duplicates are preserved by the 
corresponding firings of the rule (producing a binding of $Id$ to a different labelled null 
for each firing of the rule). 

The rule also behaves correctly in the other 2 cases: 
if $g$ is an IRI that does not occur in 
$\names$, then the body literal $named(g)$ of the Datalog rule cannot match and the rule will never fire, thus 
producing no mapping at all, which is the correct behaviour in this case. Finally, if $g$ is a variable, 
then the body literal $named(g)$ produces mappings of $g$ to all IRIs in $\names$ and, for each such binding, 
$ans_{2i}(Id_1, \overline{var}(P_1), g)$ produces precisely the mappings obtained by evaluating 
graph pattern $P_1$ over the graph whose name is the current binding of $g$. Note that, in this case, 
the head variables 
$ \overline{var}(P_i)$
consist of the variables in $P_1$ plus the variable $g$. 
Again it is the correct behaviour that the 
rule produces bindings for this increased variable set.

\smallskip

\noindent
{\em Join}.
Suppose that $P_i$ is of the form $P_i = (P_1 \AAND P_2)$. 
By expanding the definition of the $\Join$-operator 
into the semantics definition in 
Table~\ref{table:semanticsBasic}, we 
get
$\epag{P_i}{D} = 
\{ \hskip-2pt  \{ 
\mu_1 \cup \mu_2 \mid 
\mu_1 \in \epag{P_1}{D}$, 
$\mu_1 \in \epag{P_2}{D}$, and 
$\mu_1 \sim \mu_2
\} \hskip-2pt  \}$, that is, 
the multiset of those mappings which can be 
obtained as the union of any two {\em compatible} mappings $\mu_1 \in \epag{P_1}{D}$ and
$\mu_2 \in \epag{P_2}{D}$. 

The 
rule produced by our translation $\tau(P_i, \mathit{false}, D, i)$
in Definition~\ref{app:def:join}
achieves precisely this: the two body atoms \linebreak
$ans_{2i}(Id_1, v_1(\overline{var}(P_1)), D)$
and 
$ans_{2i+1}(Id_2, v_2(\overline{var}(P_2)), D)$
yield the sets of mappings 
$\epag{P_1}{D}$ and
$\epag{P_2}{D}$. Note that the variable renamings
$v_1$ and $v_2$ make sure that there is no interference
between the evaluation of 
$\epag{P_1}{D}$ (by the first body atom) and the
evaluation of $\epag{P_2}{D}$ (by the second body atom). 
The $comp$-atoms in the body of the rule make sure 
that $\mu_1$ and $\mu_2$ are compatible on all common variables. Moreover, they bind the common variables 
$\{x_1, \dots, x_n\}$ to the correct values according 
to the definition of the $comp$-predicate, i.e., 
if $v_1(x_j)$ and $v_2(x_j)$ are bound to the same value
(i.e., either the same term or they are both set to ``null''),
then we have $x_j = v_1(x_j) = v_2(x_j)$. 
Otherwise, if one of 
$v_1(x_j)$, $v_2(x_j)$ is a term and the other is ``null'', 
then $x_j$ is set equal to the term. Finally, recall that compatibility of 
two mappings $\mu_1,\mu_2$ is defined as compatiblity 
of all common variables of the two mappings. 
Hence, the $comp$-atoms in the body of the rule
produced by our translation $\tau(P_i, \mathit{false}, D, i)$ 
indeed verify that two mappings 
$\mu_1,\mu_2$ are compatible.

\smallskip

\noindent
{\em Filter}.
Suppose that $P_i$ is of the form 
$P_i = (P_1 \FILTER C)$. 
By the semantics definition 
in Table~\ref{table:semanticsBasic}, 
$\epag{P_i}{D}$ contains those mappings 
$\mu$ of $\epag{P_1}{D}$ which satisfy the filter condition $C$.  This is precisely what the single rule 
resulting from our translation 
$\tau(P_i, \mathit{false}, D, i)$
in 
Definition~\ref{app:def:filter} achieves: the body atom 
$ans_{2i}(\overline{var}(P_1), D)$ yields all those  variable bindings that correspond to the mappings in $\epag{P_1}{D}$;  and adding the 
filter condition $C$ to the body of the rule means that the rule 
only fires for variable bindings 
(strictly speaking, for the mappings corresponding to 
these variable bindings)
for which condition $C$  evaluates to true.

\smallskip

\noindent
{\em Optional}.
Suppose that $P_i$ is of the form $P_i = (P_1 \OPT P_2)$. 
By expanding the definition of the $\lojoin$-operator 
into the semantics definition in 
Table~\ref{table:semanticsBasic}, we 
get
$\epag{P_i}{D} = 
(\epag{P_1}{D} \Join \epag{P_2}{D}) \cup
(\epag{P_1}{D} \setminus \epag{P_2}{D})$.
The translation $\tau(P_i, \mathit{false}, D, i)$
in Definition~\ref{app:def:optional}
yields three rules. 
The second rule is identical to the translation 
of a Join expression. As was argued above, it 
computes precisely the variable bindings
corresponding to the mappings in 
$\epag{P_1}{D} \Join \epag{P_2}{D}$.

It remains to show that the first and third rule taken 
together 
produce the  variable bindings
corresponding to the mappings in 
$\epag{P_1}{D} \setminus \epag{P_2}{D}$. 
The first rule is almost the same as the second one with
the only difference that it projects the join-result 
to the variables in $\var(P_1)$. In other words, it 
determines the variable bindings corresponding 
to the mappings in $\epag{P_1}{D}$ which are compatible with 
some mapping in $\epag{P_2}{D}$. Therefore, the first two 
body literals of the third rule have the following effect: 
the first literal produces all 
variable bindings corresponding 
to mappings in $\epag{P_1}{D}$ while the second (i.e., the negative) body literal
selects those variable bindings which correspond to 
mappings that are {\em not compatible} with any 
mapping in $\epag{P_2}{D}$. By setting all variables in 
$\var(P_2) \setminus \var(P_1)$ to ``null'' 
(with the remaining $m$ body atoms), the third rule
indeed produces the 
variable bindings
corresponding to the mappings in 
$\epag{P_1}{D} \setminus \epag{P_2}{D}$. 

\medskip

\noindent
{\em Optional Filter}.
Suppose that $P_i$ is an optional filter expression
of the form 
$P_i =  (P_1 \OPT (P_2 \FILTER C))$.
According to the 
semantics definition in 
Table~\ref{table:semanticsBasic}, 
$\epag{P_i}{D}$ is obtained as the union of 2 multisets: 
\begin{enumerate}
    \item the mappings $\mu$ in $\epag{(P_1 \AAND P_2)}{D}$ which 
    satisfy the filter condition $C$; 
    \item the mappings $\mu_1$ in $\epag{P_1}{D}$ for which all mappings
    $\mu_2$ in $\epag{P_2}{D}$ have one
    of the following two properties: either $\mu_1$ and $\mu_2$ are not compatible or 
    they are compatible but they combination does not satisfy the filter condition $C$.
\end{enumerate}

The translation $\tau(P_i, \mathit{false}, D, i)$
in Definition~\ref{app:def:optionalFilter}
yields three rules, which are very similar to the translation 
of Optional expressions discussed before. The only difference is that now the first and second rule have 
filter condition $C$ as additional body literals. 
Compared with the rules in case of Optional expressions, 
these additional body literals 
have the following effect:

\begin{itemize}
    \item The second rule computes 
the variable bindings
corresponding to those mappings in 
$\epag{P_1}{D} \Join \epag{P_2}{D}$ which 
satisfy the filter condition $C$. That is, 
the mappings according to item 1 above. 
\item The first rule computes those
variables bindings corresponding 
to the mappings in $\epag{P_1}{D}$ which are compatible with 
some mapping in $\epag{P_2}{D}$ {\em and} which, together 
with a compatible mapping from $\epag{P_2}{D}$ 
satisfy the condition $C$.
\end{itemize}

\noindent
Therefore, the (negative) second body literal in the third rule
has the effect that we eliminate from the multiset of mappings in 
$\epag{P_1}{D}$ (obtained via the first body atom) precisely those mappings $\mu_1$
for which there exists a compatible mapping $\mu_2$ in 
$\epag{P_2}{D}$, such that their combination satisfies the filter condition $C$.
In other words, we are left with the mappings from item 2 above. 
Analogously to Optional patterns, the variables in $\var(P_2) \setminus \var(P_1)$
are not part of the domain of these mappings. Hence, with the $null$-atoms in the
body of the third rule, we  set all these variables to ``null''.

\smallskip

\noindent
{\em Union}.
Suppose that $P_i$ is of the form $P_i = (P_1 \UNION P_2)$.
According to the semantics definition in 
Table~\ref{table:semanticsBasic}, 
$\epag{P_i}{D}$ is simply obtained as the union of the two multisets
$\epag{P_1}{D}$ and $\epag{P_2}{D}$. 
In principle, the two rules of our translation $\tau(P_i, \mathit{false}, D, i)$ 
in Definition~\ref{app:def:union} compute this union of the variable bindings 
corresponding to the mappings in $\epag{P_1}{D}$  (via the body atom 
$ans_{2i}(Id_1, \overline{var}(P_1), D)$ in the first rule) and the variable bindings 
corresponding to the mappings in $\epag{P_2}{D}$  (via the body atom 
$ans_{2i+1}(Id_2, \overline{var}(P_2), D)$ in the second rule). However, care has to be taken that 
all variable bindings obtained for $ans_{i}(Id, \overline{var}(P_i), D)$ must be defined on 
{\em all} variables in $\var(P_i)$. Therefore, variable bindings obtained from 
$ans_{2i}(Id_1, \overline{var}(P_1), D)$ have to be extended to the variables in 
$\var(P_2) \setminus \var(P_1)$ by setting the latter explicitly to ``null''. Likewise, the 
variable bindings obtained from 
$ans_{2i+1}(Id_1, \overline{var}(P_2), D)$ have to be extended to the variables in 
$\var(P_1) \setminus \var(P_2)$ by setting the latter explicitly to ``null''. 
This is achieved by the $null$-atoms in the rule bodies of the two rules. 

\smallskip

\noindent
{\em Minus}.
Suppose that $P_i$ is of the form $P_i = (P_1 \MINUS P_2)$.
According to the semantics definition in 
Table~\ref{table:semanticsBasic}, 
$\epag{P_i}{D}$ consists of those mappings $\mu_1$ of 
$\epag{P_1}{D}$ which, for any mapping $\mu_2$ of $\epag{P_2}{D}$
satisfy one of the following two conditions: either 
$\mu_1$ and $\mu_2$ are not compatible or 
$\dom(\mu_1)$ and $\dom(\mu_2)$ have no variable in common. 
In other words, a mapping $\mu_1 \in \epag{P_1}{D}$ is retained in 
$\epag{P_i}{D}$ unless there exists a mapping 
$\mu_2 \in \epag{P_2}{D}$ such that 
$\mu_1$ and $\mu_2$ are compatible and there exists at least one 
variable $x$ with $\mu_1(x) = \mu_2(x) \neq $``null''.

Similarly to our translation of Join patterns, the first rule of our translation
$\tau(P_i, \mathit{false}, D, i)$ of a Minus pattern in 
Definition~\ref{app:def:minus}
computes the variable bindings of the variables in $\var(P_1)$ and 
of $\var(P_2)$ which correspond to compatible mappings. The next $n$ rules
(all with head predicate $ans_{equal-i}$) restrict the set of compatible mappings to those whose domains have at least one variable in common, i.e., the  corresponding 
variable bindings have at least one variable on which they coincide and
they are both not ``null''. Note that the signature of $ans_{equal-i}$ is 
restricted to the variables in $\var(P_1)$. That is, $ans_{equal-i}$ contains
all variable bindings on $\var(P_1)$, which correspond to ``forbidden'' 
mappings. The last rule in our translation
$\tau(P_i, \mathit{false}, D, i)$ computes the variable bindings corresponding
to the mappings in $\epag{P_i}{D}$ by computing the variable bindings corresponding to the mappings in $\epag{P_1}{D}$ (via the first body literal) and eliminating the ``forbidden'' ones (via the negative second body literal).

\subsection{Translation Rules for Property Paths}
\label{app:correctnessPropertyPathRules}

Analogously to the previous section, we now also juxtapose the semantics of property path expressions with our translation. 

\smallskip

\noindent
{\bf Semantics of property paths}.
For a property path $PP$, we write 
$\epag{PP}{D,s,o}$
to denote the semantics of a property path $PP$
over a graph $D$ with $s,o$ denoting the 
subject and object of the top-level property
path expression. The semantics of a property path 
$PP$ is a pair $(x,y)$ of terms such that there is 
a path $PP$ from $x$ to $y$. 
Here we mainly follow the semantics definition in 
\cite{DBLP:conf/semweb/KostylevR0V15}.
The semantics 
of property paths is defined recursively, with 
link property paths (i.e., simply an IRI $p$) as the base case. 

The definition 
$\epag{PP}{D,s,o}$ for arbitrary property paths
$PP$ is given in 
Table~\ref{table:semanticsProperty}.
There we write $Distinct$ for converting a 
multiset into a set by deleting duplicates.
Moreover, we write $reach(x, PP, D,s,o)$ for the set of terms 
reachable from some start point $x$ 
by applying the path $PP$ one or more times,
where $s,o$ are again subject and object of 
the top-level property path expression.

Recall from the previous section, that the semantics 
$\epag{P}{D}$
of a graph pattern $P$ over a graph $D$ is defined as 
a multiset of (partial) mappings. If a graph pattern 
$P$ is of the form 
$P = (s\, PP\, o)$, where $PP$ is a property path, then 
$\epag{(s\, PP\, o)}{D}$ is the multiset
of mappings obtained as follows: 
\begin{tabbing}
xxxx\=
$\epag{s\, PP\, o }{D} = 
\{ \hskip-2pt  \{ 
\mu \mid$\= \kill
\> $\epag{s\, PP\, o }{D} = 
\{ \hskip-2pt  \{ 
\mu \mid dom(\mu) = \var(\{s,o\}) \mbox{ and } 
$ \\
\>\> \mbox{} $(\mu(s), \mu(o)) \in \epag{PP}{D,s,o} \} \hskip-2pt  \} $,
\end{tabbing}
where we write 
$\mu(x)$ with $x \in \{s,o\}$ for both variables and non-variables $x$ with the 
understanding that $\mu(x) = x$ if $x \not \in V$.

\begin{table}[!]
\caption{
Semantics of SPARQL property paths. 
$PP_1, PP_2$ are property paths, 
$\mathcal{P}$ a set of link property paths and inverse
link property paths, 
$p \in I$,
$s,o$ are subject and object of the top-level property path 
expression.}
\centering
\begin{tabular}{|l|l|}
\hline
\rs~Property Path $PP$~&~~~~~~~~~~~Evaluation 
$\epag{PP}{D,s,o}$ \\ \hline \hline 
$p$   & \rs~~$\{ \hskip-2pt  \{ 
(x,y) \mid (x,p,y) \in D
\} \hskip-2pt  \}
$ 
\\ \hline
$\inv PP_1 $   & \rs~~$\{ \hskip-2pt  \{ 
(x,y) \mid (y,x) \in \epag{PP}{D,s,o}
\} \hskip-2pt  \}$
 \\ 
 \hline 
$PP_1 | PP_2$   & \rs~~$\epag{PP_1}{D,s,o} \cup 
\epag{PP_2}{D,s,o}$  \\ \hline 
$PP_1/PP_2$  & \rs~~$\{ \hskip-2pt  \{ 
(x,z) \mid \exists y\colon (x,y) \in \epag{PP_1}{D,s,o}$ \\
 & \rs~~$ \mbox{}\wedge (y,z) \in \epag{PP_2}{D,s,o}
\} \hskip-2pt  \}$\\ \hline 
$PP_1\hspace{-1pt}+$   &\rs~~$\{ 
(x,y) \mid y \in reach(x, PP_1, D,s,o)
\} $ \\ \hline 
$PP_1? $    & \rs~~$
Distinct (\epag{PP_1}{D,s,o}) 
$ 
\\
& 
\rs~~$\mbox{} \cup 
\{(x,x) \mid (x,y,z) \in D\}$
\\
& 
\rs~~$\mbox{} \cup 
\{(z,z) \mid (x,y,z) \in D\}$
\\
& 
\rs~~$\mbox{} \cup 
\{(s,s) \mid s \not\in V \wedge o \in V \}$ \\
&
\rs~~$\mbox{} \cup 
\{(o,o) \mid o \not \in V \wedge s \in V\}
$ 
\\
&
\rs~~$\mbox{} \cup 
\{(s,s) \mid s \not \in V \wedge s = o\} )
$ 
\\
\hline 
$PP_1\hspace{-1pt}*$   & 
$Distinct (PP_1? \cup PP_1\hspace{-1pt}+)
$
 \\ 
\hline 
$! \mathcal{P}$ with & 
\rs~~$\{ \hskip-2pt  \{ 
(x,y) \mid \exists a \colon (x,a,y) \in D$, s.t. 
\\
$p_{f_1}, \dots, p_{f_n} \in \{ p | p \in \mathcal{P} \}
\wedge $ & 
\hskip8pt$ a \not\in \{p_{f_1}, \dots, p_{f_n} \} 
\} \hskip-2pt  \} \cup \mbox{}$
\\
$p_{b_1}, \dots, p_{b_m} \in \{ p | \inv p \in \mathcal{P} \}$
& 
\rs~~$\{ \hskip-2pt  \{ 
(x,y) \mid \exists a \colon (y,a,x) \in D$, s.t. \\
&
\hskip8pt$ a \not\in \{p_{b_1}, \dots, p_{b_m} \} 
\} \hskip-2pt  \} $

\\ \hline 
\end{tabular}
\label{table:semanticsProperty}
\end{table}

\smallskip

\noindent
\noindent
{\bf Graph pattern with a property path}.
Before we inspect the translations of property paths to Warded Datalog$^{\pm}$, 
we inspect our translation of graph patterns 
{\em using} a property path. That is, 
consider a graph pattern $P$ of the form $P = s\, PP_1\, o$, where $PP_1$ is a property path. We have recalled above the definition of 
$\epag{s\, PP_1\, o }{D}$ as a multiset of pairs of terms. 
Now suppose that 
our translation $\tau_{PP}(PP_1, \mathit{false}, S, O, D, 2i)$
of property path $PP_1$ is correct. 
(A proof sketch of this fact comes next.) 
Then the single additional rule of our translation
in Definition~\ref{def:PropertyPath},
with head atom $ans_{2i}(Id_1, S, O, D)$,  
indeed produces all mappings $\mu$ on $var(P)$ 
such that $(\mu(S), \mu(O))$ is in 
$\epag{PP_1}{D,\mu(S),\mu(O)}$. Note that 
the multiplicities of the mappings thus obtained are taken care of by
different bindings of the variable
$Id_1$.

\smallskip

\noindent
{\bf Translation of property paths}.
We are now ready to inspect the translations of property paths to Warded Datalog$^{\pm}$
given in Figure~\ref{fig:propertyPathTranslation}
and Section~\ref{app:translate-paths}. 
Again, we concentrate on bag semantics 
as the more complex case.
We proceed inductively on the structure of the property paths. In all cases, suppose that we want to evaluate property paths over a graph $D$ and let
$s,o$ denote the top-level subject and object, given in a graph pattern 
of the form 
$(s \, PP\, o)$. We start with link property paths as the base case and then cover all types of compound property path expressions.

\medskip

\noindent
{\em Link property path}.
Suppose that the property path $PP_i$ consists of a single IRI $p$, i.e., 
$PP_i = p$. Then $\epag{PP}{D,s,o}$ consists of all pairs $(x,y)$, such that 
$(x,p,y)$ is a triple in $D$. On the other hand, the single rule produced by 
our translation 
$\tau_{PP}(PP_i, false, S, O, D,  i)$ 
in Definition~\ref{app:def:link}
yields exactly these pairs of terms.

\smallskip

\noindent
{\em Inverse path}.
Consider a property path $PP_i$ of the form $PP_i = \inv PP_1$  
for some property path $PP_1$.  
Then $\epag{PP_i}{D,s,o}$ consists of all pairs $(y,x)$ such that $(x,y)$ is contained in 
$\epag{PP_1}{D,s,o}$. That is, 
$\epag{PP_i}{D,s,o}$  just swaps first and second 
component of each pair in 
$\epag{PP_1}{D,s,o}$. This is exactly what the 
single rule in our translation 
$\tau_{PP}(PP_i, \mathit{true}, S, O, D, i)$
in Definition~\ref{app:def:inverse} does.

\smallskip

\noindent
{\em Alternative path}.
Consider a property path $PP_i$ of the form $PP_i = PP_1 \mid PP_2$ 
for some property paths $PP_1$ and $PP_2$.  
Then, according to the semantics definition in 
Table~\ref{table:semanticsProperty},
$\epag{PP_i}{D,s,o}$ consists of the union of 
$\epag{PP_1}{D,s,o}$  and $\epag{PP_2}{D,s,o}$. 
The two rules in our translation 
$\tau_{PP}(PP_i, \mathit{true}, S, O, D, i)$
in Definition~\ref{app:def:alternative}
realize exactly this union. 

\smallskip

\noindent
{\em Sequence path}.
Consider a property path $PP_i$ of the form 
$PP_i = PP_1 / PP_2$ for some property paths $PP_1$ and $PP_2$.  
Then, according to the semantics definition in 
Table~\ref{table:semanticsProperty},
$\epag{PP_i}{D,s,o}$ consists of pairs 
$(x,z)$ (i.e., start and end points of paths described by $PP_i$) 
such that there exist pairs $(x,y) \in \epag{PP_1}{D,s,o}$ 
and $(y,z) \in \epag{PP_2}{D,s,o}$ (i.e., start and end points of 
paths described by $PP_1$ and  $PP_2$, respectively, 
such that the end point of a 
path according to $PP_1$ and the start point of a path according 
to $PP_2$ coincide). 
The single rule
in our translation 
$\tau_{PP}(PP_i, \mathit{true}, S, O, D, i)$
in Definition~\ref{app:def:sequence}
realizes precisely these combinations of pairs 
$(x,y) \in \epag{PP_1}{D,s,o}$ 
and $(y,z) \in \epag{PP_2}{D,s,o}$.

\smallskip

\noindent
{\em One-or-more path}.
Consider a property path $PP_i$ of the form $PP_i = PP_1+$ for some property path $PP_1$.  The semantics of the one-or-more path expression is 
essentially that of reachability via hops defined by the property 
path expression $PP_1$. That is, we get all pairs $(x,y)$ 
that are in the 
``infinite'' union 
$\epag{PP_1}{D,s,o} \cup \epag{PP_1/PP_1}{D,s,o} \cup 
\epag{PP_1/PP_1/PP_1}{D,s,o} \cup 
\epag{PP_1/PP_1/PP_1/PP_1}{D,s,o} \cup 
\dots$, with one important difference though: 
according to the SPARQL semantics of property paths\footnote{\href{https://www.w3.org/TR/SPARQL11-query/\#defn_PropertyPathExpr}{https://www.w3.org/TR/SPARQL11-query/\#defn\_PropertyPathExpr}}, 
\textit{one-or-more} property paths 
(and likewise \textit{zero-or-one} and \textit{zero-or-more}
property paths) always have set semantics. 
The
two rules
in our translation 
$\tau_{PP}(PP_i, \mathit{true}, S, O, D, i)$
in Definition~\ref{app:def:oneOrMore}
realize exactly this kind of reachability relationships. 
Moreover, neither in the 
semantics definition nor in the translation, we need to keep track 
of duplicates. Therefore, 
in the semantics definition, we define
$\epag{PP_i}{D,s,o}$ as a set (rather than a multiset). 
And in our translation, duplicates are avoided by the 
$Id = []$ body atom in both rules. As was already mentioned in 
Section~\ref{sect:translation-propertypaths}, 
this body atom has the effect that no copies of the same pair $(x,y)$ (but with different binding of $Id$)
can 
ever be produced.

\smallskip

\noindent
{\em Zero-or-one path}.
Consider a property path $PP_i$ of the form $PP_i = PP_1?$ for some property path $PP_1$.  
Then, intuitively, 
$\epag{PP_i}{D,s,o}$ consists of pairs of nodes 
that are the start and end point of a ``one-path'' (i.e., 
traversing  $PP_1$ once) plus ``zero-paths''
(i.e., identical start and end point). 
In the semantics definition 
in Table~\ref{table:semanticsProperty},
the pairs corresponding to ``one-paths'' are taken care of by the expression $\epag{PP_1}{D,s,o}$. Analogously, in our translation 
$\tau_{PP}(PP_i, \mathit{true}, S, O, D, i)$
in Definition~\ref{app:def:zeroOrOne}
these pairs are produced by the second rule. 

All remaining expressions in the semantics definition correspond to 
various ways of getting ``zero-paths'', namely: either for every term in the graph (captured by the second and third expression of the semantics 
definition) or if at least one of $s$ or $o$ is a term (captured by 
the remaining expressions of the semantics definition). In case both 
$s$ and $o$ are terms, they must be the same in order 
to constitute a ``zero-path''. 
In our translation 
$\tau_{PP}(PP_i, \mathit{true}, S, O, D, i)$
in Definition~\ref{app:def:zeroOrOne},
the first rule produces the pairs $(x,x)$ for terms $x$ occurring 
in the active graph. The last 
rule of the translation produces the remaining pairs $(t,t)$ if 
$t$ is a term that occurs as top-level subject or object of the entire 
property path expression. As with one-or-more path expressions, 
it is important 
to keep in mind that  also zero-or-one paths always 
have set semantics according to the SPARQL semantics of property paths. 
The elimination of duplicates is ensured by the $Distinct$ operator
in the semantics defintion and by the $Id = []$ body atom in the 
rules of our translation.

\smallskip

\noindent
{\em Zero-or-more path}.
Consider a property path $PP_i$ of the form $PP_i = PP_1*$ for some property path $PP_1$.  
In our semantics 
definition in 
Table~\ref{table:semanticsProperty},
we have defined 
$\epag{PP_i}{D,s,o}$ simply as the set-variant (i.e., deleting 
any duplicates) of the union of the zero-or-one path 
and the one-or-more path. Of course, in case of bag semantics 
this would be 
problematic since we thus count ``one-paths'' twice. However, since 
\textit{zero-or-more} property paths always have set semantics, 
the $Distinct$ operator applied to the union eliminates any 
duplicates anyway. 
The rules 
in our translation 
$\tau_{PP}(PP_i, \mathit{true}, S, O, D, i)$
in Definition~\ref{app:def:zeroOrMore}
are indeed obtained as the union of the rules 
that one gets for the translations
of the zero-or-one path $PP_1?$ and of the 
one-or-more path $PP_1+$. The 
$Id = []$ body atom in each of the rules makes sure that 
we never produce any copies of any pair $(x,y)$.

\smallskip

\noindent
{\em Negated  path}.
Consider a property path $PP$ of the form $PP = !\mathcal{P}$, where 
$\mathcal{P}$ is a set of ``forward'' link property path expressions 
$\{p_{f_1}, \dots, p_{f_n}\}$ and 
``backward'' link property path expressions 
$\{ \inv p_{b_1}$, $\dots, \inv p_{b_m}\}$. 
Then, according to our semantics 
definition in 
Table~\ref{table:semanticsProperty},
$\epag{PP_i}{D,s,o}$ contains those pairs $(x,y)$ for which 
either there exists a triple $(x,a,y)$ in the active graph such that 
$a$ is different from all $p_{f_j}$ 
or 
there exists a triple $(y,a,x)$ in the active graph such that 
$a$ is different from all $p_{b_j}$. 
Our translation 
$\tau_{PP}(PP_i, \mathit{true}, S, O, D, i)$
in Definition~\ref{app:def:negated}
generates two rules. Clearly, the first type of 
pairs in $\epag{PP_i}{D,s,o}$ is produced by the first rule of 
our translation and the second type
of pairs in $\epag{PP_i}{D,s,o}$ is produced by the second rule.

\section{Some Implementation Details}
\label{app:ImplementationDetails}

\paragraph{Some basic principles}
The SPARQL to Warded Datalog$^\pm$ translator was implemented in Java using the library org.apache.jena to parse SPARQL query strings and handle operations, solution modifiers, basic graph patterns, etc.\ 
appropriately. The ARQ algebra query parser\footnote{\href{https://jena.apache.org/documentation/query/}{https://jena.apache.org/documentation/query/}} of Apache Jena parses SPARQL query strings in a top-down fashion. First query forms are parsed, next solution modifiers and in the end operations starting from the outer-most operation going inward. 

In contrast, our developed SPARQL to Warded Datalog$^\pm$ parser 
analyses queries bottom-up. Thus, the translation starts at basic graph patterns and continues upwards. This setup is necessary, as the variables inside an expression need to be kept track of, when parsing it, as rules usually modify results of sub-operations. Therefore, it needs to be known which variables occur in the respective subresult predicates.

\paragraph{Datatypes, languages, and compatibility}
Our translation engine partially supports datatypes and language tags by adding two additional arguments to each variable, containing the respective information. This has implications on most SPARQL operations such as \emph{UNION}, \emph{JOIN}, and \emph{FILTER}. For example, in the case of \emph{JOIN} operations, we have extended the existing translation of \cite{Angles2016TheMS} by developing two additional comparison predicates ($compD$ and $compL$). In \cite{Angles2016TheMS} the predicate $comp$ is used for computing the compatibility between two variables. The new predicates $compD$ and $compL$ are used to compute the respective compatibility for their datatypes and language tags, which is done in the same way as for variables thus far.

Moreover, Vadalog (and Datalog in general) joins variables of the same rule by name, however the semantics of joins in SPARQL is different to the one of Datalog. It is for that reason (1)  that we have to prefix/rename variables in such a way that Vadalog's internal join strategy is prevented and (2) that we introduce the join predicate $comp$ described in Section \ref{app:translate-patterns} to realise SPARQL join semantics. 

\paragraph{Skolem functions (for bag semantics)}
The Skolem function generator lies at the heart of how we preserve duplicate results. As in the work of \cite{BagSemantic}, we introduce IDs to preserve Datalog bag semantics 
in Warded Datalog$^\pm$ set semantics. Therefore, each generated result tuple is distinguished from its duplicates by a tuple ID (referred to as TID in \cite{BagSemantic}). However, instead of simply generating nulls, our ID generation process is abstracted away by the $\mathit{getSkolF}$ function of the $\mathit{skolFG}$ object. 
We thus generate IDs as follows. Since the grounding of each positive atom in the rule body is responsible for the generation of a tuple in Datalog$^\pm$, we extract a sorted list of all variables occurring in positive atoms of the rule body $\mathit{bodyVars}$. Finally, the tuple ID is generated by assigning it to a list starting with the string ``$\mathit{f_{<ruleID>}}$'', followed by the list of positive body variables $\mathit{bodyVars}$ and a string $\mathit{label}$. The strings ``$\mathit{f_{<ruleID>}}$'' and $\mathit{label}$ were added, as we use ``$\mathit{f_{<ruleID>}}$'' to identify the translated rules of the processed operator at the current translation step, while $\mathit{label}$ provides additional information when needed. 

This setup preserves generated duplicates of SPARQL bag semantics in Warded Datalog$^\pm$ set semantics by utilizing their provenance information to make them distinguishable. Furthermore, it provides information for debugging/explanation purposes of the reasoning process, as each tuple 
carries the information which rule and grounding has led to its generation. 
As an added bonus this layer of abstraction may be used to adapt different duplicate generation semantics/strategies as might be necessary for different applications by simply exchanging the $\mathit{skolFG}$ by any self-implemented solution.

\section{Further Details of the Experimental Evaluation}
\label{appExp}

In Section \ref{sec:BenchmarkAnalysis}, we have reported on 
our analysis of various SPARQL benchmarks. One of the outcomes of this analysis 
was the selection of three concrete benchmarks, 
which we then used in Section \ref{sec:Compliance}
to 
test the standard-compliance of \name and two state-of-the-art SPARQL engines. 
In this section, we provide further details on how we have set up the benchmark analysis 
(Section \ref{app:BenchAnalysis}) and the compliance tests 
(Section \ref{app:compliance}). Following these, the most interesting part of this sections are additional benchmarks requested by reviewers, starting from Section~\ref{sec:addBench}, which provides a further overview of the following sections.

\subsection{Setup of the  Benchmark Analysis}
\label{app:BenchAnalysis}
Before we dive into the benchmark analysis, we briefly introduce its setup in this section.
Our analysis employs a similar way of counting SPARQL features as is done in  \cite{HowRepIsSPARQLBench}. We have counted each feature once per query in which it occurs, with one exception being the \emph{DISTINCT} feature. As in \cite{HowRepIsSPARQLBench}, we count the DISTINCT feature only if it is applied to the entire query. This is also in line with our interest of testing bag and set semantics in combination with different SPARQL features.

In contrast to \cite{HowRepIsSPARQLBench} we do not limit our benchmark analysis to \emph{SELECT} queries, but rather analyse all queries provided at the GitHub repository of the dice group\footnote{\href{https://hobbitdata.informatik.uni-leipzig.de/benchmarks-data/queries/}{https://hobbitdata.informatik.uni-leipzig.de/benchmarks-data/queries/}}. Therefore, we analyse also the \emph{DESCRIBE} queries of e.g. BSBM. Moreover, since the SP2Bench query file does not contain the hand-crafted \emph{ASK} queries provided on its homepage\footnote{\href{http://dbis.informatik.uni-freiburg.de/index.php?project=SP2B/queries.php}{http://dbis.informatik.uni-freiburg.de/index.php?project=SP2B/queries.php}}, we have chosen to add these to the benchmark to be able to analyse the complete query-set of SP2Bench. For these reasons, the results of overlapping SPARQL features and benchmarks from our analysis in Table \ref{tab:BenchmarkFeatureUagse} and the one of \cite{HowRepIsSPARQLBench} differ slightly.

Furthermore, note that we do not display basic features, such as \emph{Join}, \emph{Basic Graph pattern}, etc.\ in Table \ref{tab:BenchmarkFeatureUagse}, as these features are, of course,  covered by each of the considered benchmarks. Furthermore, we have chosen to omit the SPARQL features \emph{ORDER BY}, \emph{LIMIT} and \emph{OFFSET} from the table since these cannot be evaluated currently by \name due to the limitations of Vadalog and our translation engine,
as has already been mentioned in our discussion of the SPARQL feature coverage of \name 
in Table~\ref{tab:coverage}. Also \emph{ASK} was left out in the table, as ASK is not an especially challenging feature and was therefore removed for space reasons. Moreover, we have chosen to only include the \emph{REGEX} filter constraint in the feature coverage table and no other specific constraints, as the \emph{REGEX} function is argued to be of vital importance for SPARQL users in \cite{HowRepIsSPARQLBench}. For this reason, we have chosen to cover this feature by our 
translation engine in addition to the other filter constraints. Finally, we have not included the SPARQL features \emph{MINUS} and the \emph{inverted}, \emph{zero-or-one}, \emph{zero-or-more}, \emph{one-or-more} and \emph{negated property path}, as none of the selected benchmarks covers any of these SPARQL features.

\subsection{Setup of the SPARQL Compliance Tests}
\label{app:compliance}

As detailed in Section \ref{sec:BenchmarkAnalysis}, 
we have selected three benchmarks (BeSEPPI, SP2Bench and FEASIBLE (S)) for 
evaluating the standard-compliance of the chosen three systems
(\name, Jena Fuseki, and OpenLink Virtuoso). 
For our experiments, we use Apache Jena Fuseki 3.15.0 and 
Virtuoso Open Source Edition, version 7.2.5. 
The experiments are run on a Windows 10 machine with 8GB of main memory. 
In this section, we explain the setup of the standard-compliance tests that we 
performed on our \name system and the two state-of-the-art SPARQL engines
Fuseki and Virtuoso. 
Moreover, we mention some challenges with these tests and we 
provide further details on the outcome of the compliance tests.

\subsubsection{Benchmark Generation}
\label{app:BenchmarkGeneration}

In order to carry out our standard-compliance tests, we first have to 
make the queries and the data provided by the benchmarks accessible to the 
tested systems. While this turns out to be an easy task for BeSEPPI and SP2Bench, 
some care is required for the FEASIBLE (S) benchmark.

\paragraph{BeSEPPI and SP2Bench}
The BeSEPPI benchmark contains queries and a dataset for the evaluation of property path queries. Its dataset can be directly loaded into the selected 
systems and its queries can directly be executed.
The SP2Bench benchmark contains 17 hand-crafted queries and a benchmark dataset generator. For the purpose of our compliance tests, we have generated a dataset with 50k triples, which was loaded into each of the  considered systems. 

\paragraph{FEASIBLE}
The FEASIBLE benchmark contains a query generator, which generates queries for an arbitrary dataset that provides a query-log. In the case of the FEASIBLE (S) benchmark, 
we have chosen the Semantic Web Dog Food (SWDF) dataset and we 
have generated 100 queries using the SWDF query-log. 
However, some additional work was required before we could 
use the FEASIBLE (S) benchmarks for our tests.

The first complication arises from the fact that 
Vadalog uses Java sorting semantics, whereas the SPARQL standard defines its own ordering semantics. We had to remove \emph{LIMIT} and \emph{OFFSET} from each query of the FEASIBLE (S) benchmark, as queries with these features can only be reasonably evaluated (comparing the generated query results, rather than only checking if their cardinalities are equal) if the results are sorted and if each considered RDF query and storage system provides the same sorting semantics. Some queries of the generated benchmark only differed from each other in the argument of the 
\emph{LIMIT} or \emph{OFFSET} clause. Thus, after removing all \emph{LIMIT} and \emph{OFFSET} clauses, we ended up with duplicate queries. These duplicate queries were eliminated, leaving the FEASIBLE (S) benchmark with a total number of 77 unique queries.

Moreover, Vadalog does currently not support UTF-8 characters. We were therefore faced with the necessity of changing the encoding of the SWDF dataset of the FEASIBLE (S) benchmark. We have made the plausible assumption that dropping non-ASCII characters from RDF strings would not lead to vastly different results and 
we have therefore simply deleted all non-ASCII characters from the SWDF dataset. 

Furthermore, since the FEASIBLE (S) benchmark includes queries with the \emph{GRAPH} feature (which selects the graph IRI of RDF triples), we have loaded the SWDF dataset both into the default graph of each tested system and into a named graph for the FEASIBLE benchmark to be able to test the GRAPH feature.

\subsubsection{Challenges of the Evaluation Process}
\label{app:Challenges}

The evaluation of the standard compliance of the three chosen systems requires the comparison 
of query results. In case of BeSEPPI, the benchmark also provides the expected result for each query. We therefore have to compare the result produced by each of the three systems with the correct result defined by the benchmark itself. In contrast, 
FEASIBLE and SP2Bench do not provide the expected results for their queries. 
We therefore use a majority voting approach to determine the correct answer. 
That is, we compare the query results produced by the three considered systems and 
accept a result as the expected query answer if it is equal to the generated query result 
of at least two of the tested systems.

A major challenge for comparing query results (both, when comparing the result produced by one system with the expected result defined by the benchmark itself and when comparing the 
results produced by two systems) comes from \textit{blank nodes}. 
On the one hand, each system employs its own specific functionality for assigning blank node names.  Therefore, to compare blank nodes between the different result multisets, a mapping between the internal system-specific blank node names has to be found. However, finding such a mapping comes down to finding an isomorphism between two arbitrarily sized tables, containing only blank nodes, which requires exponential time in the worst case and which poses a 
serious problem for large result multisets with many blank nodes. 
We have therefore tried out a simple heuristic to find a suitable mapping between blank nodes 
by sorting the query results without considering blank node names first. We then iterate over
both results and finally, each time when a new blank node name is encountered, we save the mapping between the system-specific blank node names. Even though this is 
a very simple heuristic, it has worked quite well in many cases. 
Nevertheless, there 
are cases, where this simple procedure infers wrong blank node mappings, even though the results are semantically equivalent. Hence, due to the instability of this efficient blank node checking heuristic, we have chosen to remove the evaluation of blank nodes from our compliance tests. That is, our current evaluation test suite does for this reason not distinguish between different blank node names but, of course, it distinguishes between all other terms.

\subsubsection{Outcome of the Compliance Tests}

The outcome of our compliance tests is based on the notions 
of {\em correctness} and {\em completeness} of query results, which we 
define in the same way as done in \cite{BeSEPPI}:

\smallskip
\noindent
{\em Correctness} defines the ratio of correct tuples generated by 
the tested system  for a query. For a \emph{SELECT} query $q$ with $R_{expected}(q)$ being the expected result of $q$ and $R_{sys}(q)$ being the response of system $sys$ to the \emph{SELECT} query, \cite{BeSEPPI} defines correctness as follows:
\[
    correct(q) = 
\begin{cases}
    \frac{|R_{expected}(q) \cap R_{sys}(q)|}{|R_{sys}(q)|}, & \text{if } R_{sys}(q)\neq 0\\
    1,              & \text{otherwise}
\end{cases}
\]
It intuitively accepts a result as correct ($correct(q) = 1$) if the returned result of the considered system $R_{sys}(q)$ is a subset of the expected answer $R_{expected}(q)$. For \emph{ASK} queries we consider a result to be correct only if it exactly matches the expected answer, as done in \cite{BeSEPPI}.

\smallskip
\noindent
{\em Completeness} defines the ratio of all accepted result-tuples generated by the 
tested system  for a query. For a \emph{SELECT} query $q$ with $R_{expected}(q)$ being the expected result of $q$ and $R_{sys}(q)$ being the response of system $sys$ to the \emph{SELECT} query, \cite{BeSEPPI} defines completeness as follows:
\[
    complete(q) = 
\begin{cases}
     \frac{|R_{expected}(q) \cap R_{sys}(q)|}{|R_{expected}(q)|}, & \text{if } R_{expected}(q)\neq 0\\
    1,              & \text{otherwise}
 \end{cases}
\]
It intuitively accepts a result as complete ($complete(q) = 1$) if the expected answer $R_{expected}(q)$ is a subset of the returned result of the considered system $R_{sys}(q)$. For \emph{ASK} queries we consider a result to be complete only if it exactly matches the expected answer, as done in \cite{BeSEPPI}.

\paragraph{FEASIBLE (S)}
FEASIBLE (S) contains 77 unique queries. From these 77 queries, a total of 68 are accepted by our translation engine and were thus used for the correctness and completeness evaluation of our system. The remaining 9 queries could not be translated into Datalog$^{\pm}$ programs, since they contain features that are currently not supported by our \name  system for the following reasons: 

Three queries contain a complex expression in an \emph{ORDER BY} statement and
two queries  contain complex expressions in \emph{COUNT} aggregates. Both features 
are currently supported by \name only 
for simple expressions consisting of a single variable. 
Two queries cannot be translated currently due to the missing support of our engine for the functions \emph{ucase} and \emph{contains}.
However, these SPARQL features do not impose any conceptual hurdle and were only left out from the first version of \name since 
we considered their priority as low. 
Similarly, two queries are not accepted by our engine at the moment, as they contain the \emph{DATATYPE} feature. Again, this is mainly due to the low priority that we gave this feature. Since our translation engine already tracks datatypes and language tags of RDF terms, it should be no problem to integrate it in later versions of \name.

On the remaing 68 queries that are accepted by our system, our translation engine produces for all executed 68 queries the same result as Apache Jena Fuseki. 
In contrast, OpenLink Virtuoso 
returned an erroneous result for 14 queries 
by either 
wrongly outputting duplicates (e.g., ignoring DISTINCTs) or omitting duplicates
(e.g., by handling UNIONs incorrectly). Moreover, in 18 cases, 
Virtuoso was unable to evaluate the query and produced an error. 

\paragraph{SP2Bench}
The SP2Bench benchmark contains 17 queries in total and is specifically designed to test the scalability of of SPARQL engines. All three considered systems produce identical results for each of the 17 queries.

\paragraph{BeSEPPI}
The BeSEPPI benchmark contains 236 queries, specifically designed to evaluate the correct and complete support of property path features. We have already summarized the outcome of our compliance tests 
of the 3 considered systems on this benchmark
in Table \ref{tab:BeSEPPIAnalysisM} 
in Section \ref{sec:Compliance}.  In a nutshell, 
while \name and Fuseki follow the SPARQL standard for the evaluation of property paths, Virtuoso produces errors for 18 queries and returns incomplete results for 
another 13 queries. 

We conclude this section by a more detailed look into the 
problems Virtuoso is currently facing with property path 
expressions. 
As already observed in \cite{BeSEPPI}, 
Virtuoso produces errors for 
zero-or-one, zero-or-more and one-or-more property paths that contain two variables. Furthermore, the error messages state that the transitive start is not given. Therefore, we come to the same conclusion as \cite{BeSEPPI} that these features were most likely left out on purpose, since Virtuoso is based on relational databases and it would 
require huge joins to answer such queries. Moreover, we have noticed that the errors for inverse negated property paths (reported in \cite{BeSEPPI}) have been fixed in the current OpenLink Virtuoso release.

Virtuoso produces 10 incomplete results when evaluating 
one-or-more property path queries. As already discovered in \cite{BeSEPPI}, they all cover cases with cycles and miss the start node of the property path, indicating that the 
one-or-more property path might be implemented by evaluating the 
zero-or-more property path first and simply removing the start node from the computed result. Finally, in contrast to the results of \cite{BeSEPPI}, we have found that the current version of OpenLink Virtuoso generates wrong answers for queries that contain alternative property paths. Virtuoso generates for three alternative property path queries incomplete results, which differ from the results of Fuseki and \name by missing all duplicates, which should have been generated. 

\subsection{Additional Empirical Results}
\label{sec:addBench}

In this section we provide additional empirical results that were requested by the reviewers. We will start this section by giving an outlook of the results first. Specifically, as requested by the reviewers, we have: 
\begin{enumerate}
    \item conceptualized and implemented the translation of additional SPARQL features to cover any SPARQL feature that occurs in our benchmarks (outlined in Section \ref{app:BenchmarkCoverage}),
    \item rerun the SP2Bench benchmark on a stronger machine (for detailed description of the benchmark setting see Section \ref{app:BenchSetting}), 
    \item compared our \name system to Fuseki and Virtuoso in Section \ref{app:BenchResults}, 
    \item provided the complete benchmark results (including the loading and execution time) of Virtuoso and Fuseki in Section~\ref{app:BenchResTables}.
\end{enumerate}

\subsection{Full SPARQL Benchmark Coverage}
\label{app:BenchmarkCoverage}
As requested by the reviewer, we have implemented the remaining missing SPARQL features to fully cover any SPARQL features occurring in our selected benchmarks. Specifically, we have implemented the following features: ORDER BY with complex arguments (such as $\textit{ORDER BY}(!\textit{BOUND}(?n))$), functions on Strings such $\textit{UCASE}$, the $\textit{DATATYPE}$ function, $\textit{LIMIT}$, and $\textit{OFFSET}$. Thereby, we have extend \name to cover the $9$ previously unsupported queries of the FEASIBLE benchmark, reaching our goal of supporting any query of the selected benchmarks.

\subsection{Benchmark Setting}
\label{app:BenchSetting}

\textbf{gMark.} As suggested by the reviewer, we have evaluated \name, Fuseki, and Virtuoso on the gMark benchmark \cite{gMark}, a domain- and language-independent graph instance and query workload generator which specifically focuses on path queries, i.e., queries over property paths. Specifically, we have evaluated \name's, Fuseki's, and Virtuoso's path query performance on the \emph{test}\footnote{\url{https://github.com/gbagan/gMark/tree/master/demo/test}} and \emph{social}\footnote{\url{https://github.com/gbagan/gMark/tree/master/demo/social}} demo scenarios. Each of these two demo scenarios provides 50 SPARQL queries and a graph instance. Since the graph instances consist of triples of entity and relation ids, we had to translate the graph instance to RDF, by replacing any entity id $\alpha$ with $\textit{<http://example.org/gMark/}\alpha\textit{>}$ and any relation id $\beta$ with $\textit{<http://example.org/gMark/p}\beta\textit{>}$. Table \ref{tab:benchmarkStats} provides further details on the benchmarks that we used for evaluating a system's query execution time.

\begin{table}[]
    \centering
    \begin{tabular}{cccc}
    \toprule
         Benchmark & \#Triples & \#Predicates & \#Queries \\
         \midrule
         Social (gMark) & 226,014 & 27 & 50 \\
         Test (gMark) & 78,582 & 4 & 50 \\
         SP2Bench & 50,168 & 57 & 17 \\
         \bottomrule
    \end{tabular}
    \caption{Benchmark Statistics}
    \label{tab:benchmarkStats}
\end{table}

\textbf{Experimental Setup.} Our benchmarks were executed on a system running openSUSE Leap 15.2 with dual Intel(R) Xeon(R) Silver 4314 16 core CPUs, clocked at 3.40GHz, with 512GB RAM of which 256GB are reserved for the system under test, and 256GB for the operating system. 
For each SPARQL engine, we set the following limits: We set a time-out of 900s and, in response to the reviewer's request, we increased the available RAM per system from 8GB to 256GB. Similar to our initial benchmark setup, we started each benchmark by repeating the same warm-up queries $5$ times and by $5$ times loading and deleting the graph instance. Furthermore, we did $5$ repetitions of each query (each time deleting and reloading the dataset), following the setting of our previous benchmark experiments. In the next section, we will discuss the benchmark results.

\subsection{Benchmark Results}
\label{app:BenchResults}

\begin{figure*}[h!]
  \centering
  \includegraphics[scale=0.8]{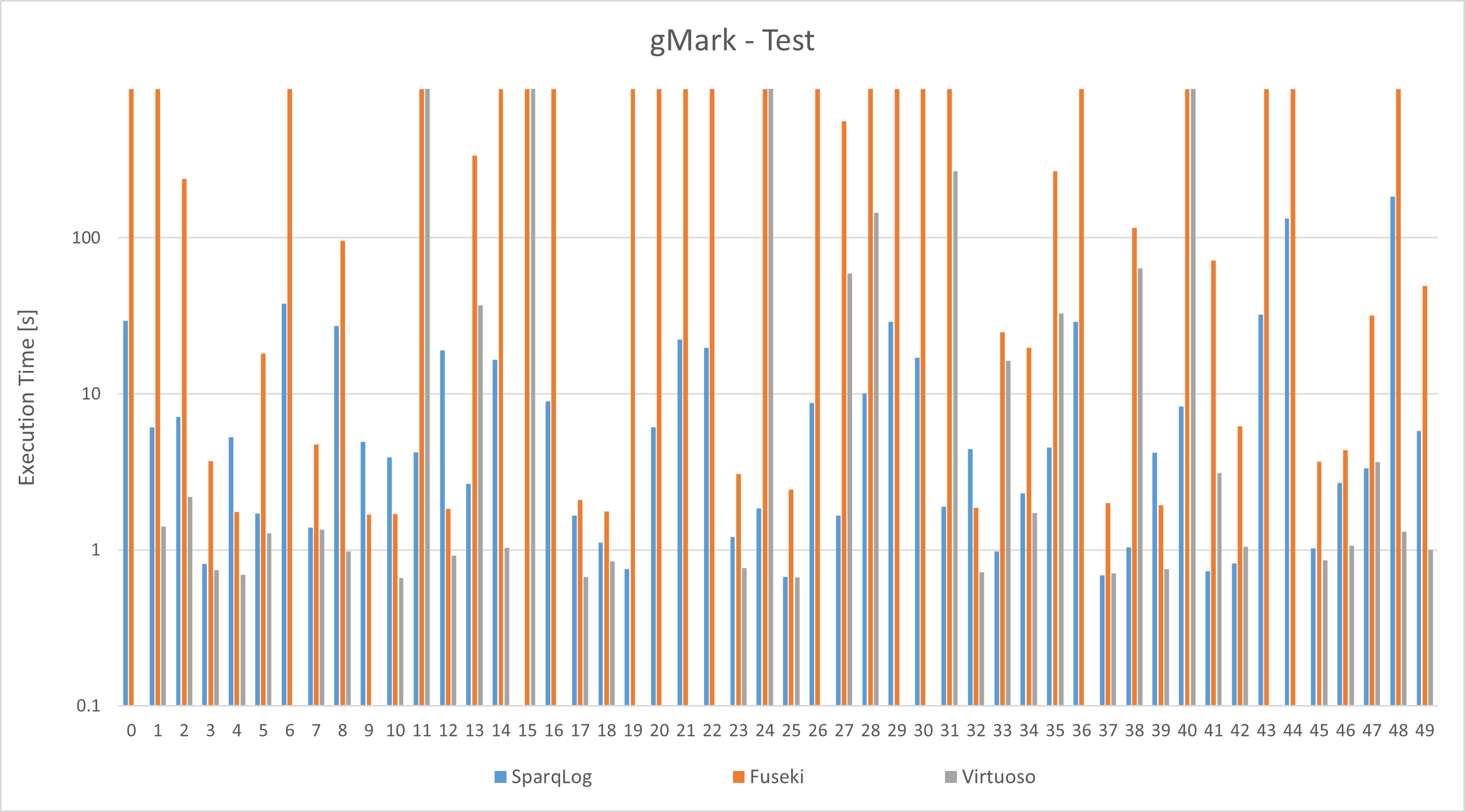}
  \caption{gMark Test Benchmark (Log Scale)}
  \label{fig:gMarkTest}
\end{figure*}

\begin{table}[]
    \centering
    \begin{tabular}{cccc}
    \toprule
         System & \name & Fuseki & Virtuoso \\
         \#Not Supported & 0 & 0 & 12\\
         \#Time- and Mem-Outs & 1 & 16 & 1 \\
         \#Incomplete Results & 0 & 0 & 16 \\
         \midrule 
         Total & 1 & 16 & 29 \\
         \bottomrule
    \end{tabular}
    \caption{Benchmark Results on Social (gMark)}
    \label{tab:gsocial}
\end{table}

\begin{table}[]
    \centering
    \begin{tabular}{cccc}
    \toprule
         System & \name & Fuseki & Virtuoso \\
         \#Not Supported & 0 & 0 & 9\\
         \#Time- and Mem-Outs & 1 & 21 & 5 \\
         \#Incomplete Results & 0 & 0 & 4 \\
         \midrule 
         Total & 1 & 21 & 18 \\
         \bottomrule
    \end{tabular}
    \caption{Benchmark Results on Test (gMark)}
    \label{tab:gtest}
\end{table}

Table \ref{tab:gsocial} and \ref{tab:gtest} reveal the results on the additional gMark benchmarks. Specifically, the tables state the number of queries of the respective benchmark that a system (1) does not support, (2) answered with a time- or mem-out (out-of-memory) exception, or (3) answered with an incomplete result. Furthermore, the tables present the total number of queries which could not be (correctly) answered by the systems. Figures~\ref{fig:gMarkSocial} and \ref{fig:gMarkTest} visualize the query execution time of the three systems per benchmarks. A bar reaching 900s represents a time-out. A missing bar represents a mem-out, a faulty result, or that a query was not supported. We have excluded query 31 from the gMark social benchmark and query 15 from the gMark test benchmark, as none of the three systems managed to answer these queries.  In the following, we compare the results of the three systems on gMark:

\textbf{Virtuoso} could not (correctly) answer $48$ of the in total $100$ queries of the gMark Social and Test benchmark. Thus, it could not correctly answer almost half of the queries provided by both gMark benchmarks, which empirically reveals its dramatic limitations in answering complex property path queries. In $20$ of these $48$ cases, Virtuoso returned an incomplete result. While in solely $3$ incomplete result cases Virtuoso missed solely one tuple in the returned result multi-set, in the remaining $17$ incomplete result cases; Virtuoso produces either the result tuple $\textit{null}$ or an empty result multi-set instead of the correct non-null/non-empty result multi-set. In the other $28$ cases Virtuoso failed either due to a time-, mem-out or due to not supporting a property path with two variables. This exemplifies severe problems with handling property path queries.

\textbf{Fuseki} reached on $37$ of the in total $100$ queries of the gMark Social and Test benchmark a time-out (i.e., took longer than $900s$ for answering the queries). Thus, it timed-out on more than a third of gMark queries, which empirically reveals its significant limitations in answering complex property path queries. 

\textbf{\name} managed to answer $98$ of gMark's (in total $100$) queries within less than $200s$ and timed-out on solely $2$ queries (see Figures~\ref{fig:gMarkSocial} and \ref{fig:gMarkTest}). This result reveals the strong ability of our system in answering queries that contain complex property paths. Furthermore, each time when both Fuseki and \name returned a result, the results were equal, even further empirically confirming the correctness of our system (i.e., that our system follows the SPARQL standard).

\textbf{SP2Bench Rerun.} We have rerun the SP2Bench benchmark using the same benchmark setting as for gMark. Thus, as requested by the reviewer we have rerun SP2Bench with significantly increased RAM and compared \name not only to Fuseki but additionally to Virtuoso. We have visualized the result in Figure \ref{fig:SP2Bench} and found that SparqLog reaches highly competitive performance with Virtuoso and significantly outperforms Fuseki on most queries.

In conclusion, these three benchmarks show that SparqLog (1) is highly competitive with Virtuoso on regular queries with respect to query execution time (see Figure \ref{fig:SP2Bench}, (2) even follows the SPARQL standard much more accurately than Virtuoso and supports more property path queries than Virtuoso (see Tables \ref{tab:gsocial} and \ref{tab:gtest}), and (3) dramatically outperforms Fuseki on query execution, while keeping its ability to follow the SPARQL standard accurately.

\subsection{Complete Benchmark Results}
\label{app:BenchResTables}

Tables~\ref{tab:gSocial}-\ref{tab:SP2Bench} display the complete results of the respective benchmarks. Specifically the tables contain the query id, the system's loading time, query execution time, and total time which is the sum of the query execution and loading time. Furthermore, the column \textit{Res Equal} indicates whether the result of Fuseki and Virtuoso is equal to the one from \name and additionally states whether each system encountered a time-out, mem-out, or not supported exception. Note that the result of Fuseki and \name is always the same when Fuseki manages to answer the query, which further exemplifies that both Fuseki and \name follow the SPARQL standard precisely.  However, Virtuoso violates the SPARQL standard often (indicated by many false entries in the \textit{Res Equal} column of the gMark benchmarks).

\clearpage

\smallskip

\begin{table*}[]
\resizebox{\textwidth}{!}{\begin{tabular}{cccccccccc}
\toprule
         & \textbf{\name}     & \multicolumn{4}{c}{\textbf{Fuseki}}                                     & \multicolumn{4}{c}{\textbf{Virtuoso}}                                      \\
         \cmidrule(lr){2-2} \cmidrule(lr){3-6} \cmidrule(lr){7-10}
\textbf{Query ID} & \textbf{Total Time}     & \textbf{Loading Time} & \textbf{Execution Time} & \textbf{Res Equal} & \textbf{Total Time} & \textbf{Loading Time} & \textbf{Execution Time} & \textbf{Res Equal} & \textbf{Total Time}    \\
0        & 5.390872 & 4.874812     & 20.37528       & TRUE              & 25.25009   & --- & --- & \textbf{FALSE}             & Faulty Result \\
1        & 3.186824 & 5.046465     & 767.9034       & TRUE              & 772.9499   & --- & --- & \textbf{FALSE}             & Faulty Result \\
2        & 19.10591 & 5.03212      & 0.026058       & TRUE              & 5.058179   & --- & --- & \textbf{FALSE}             & Faulty Result \\
3        & 11.15376 & 4.996236     & 221.0903       & TRUE              & 226.0866   & --- & --- & \textbf{FALSE}             & Faulty Result \\
4        & 21.87769 & --- & --- & \textbf{Time-Out}          & --- & --- & --- & \textbf{FALSE}             & Faulty Result \\
5        & 3.702722 & 4.737612     & 103.3389       & TRUE              & 108.0765   & 1.502375     & 21.31221       & TRUE              & 22.81458      \\
6        & 6.358109 & --- & --- & \textbf{Time-Out}          & --- & --- & --- & \textbf{Not Supported}     & --- \\
7        & 2.446714 & 4.789334     & 2.245069       & TRUE              & 7.034403   & 1.616492     & 0.310986       & TRUE              & 1.927478      \\
8        & 2.087736 & 4.822063     & 0.009277       & TRUE              & 4.83134    & 1.441408     & 0.056105       & TRUE              & 1.497513      \\
9        & 7.064603 & 4.63884      & 17.86338       & TRUE              & 22.50222   & 1.496142     & 0.002103       & TRUE              & 1.498245      \\
10       & 4.306584 & 4.889473     & 0.021995       & TRUE              & 4.911468   & --- & --- & \textbf{FALSE}             & Faulty Result \\
11       & 13.75338 & 4.852585     & 169.7659       & TRUE              & 174.6185   & --- & --- & \textbf{FALSE}             & Faulty Result \\
12       & 8.887165 & --- & --- & \textbf{Time-Out}          & --- & 1.740651     & 0.002004       & TRUE              & 1.742655      \\
13       & 63.50253 & --- & --- & \textbf{Time-Out}          & --- & --- & --- & \textbf{Not Supported}     & --- \\
14       & 2.38494  & 4.97689      & 4.229519       & TRUE              & 9.206409   & 1.578675     & 0.019961       & TRUE              & 1.598636      \\
15       & 2.439447 & 4.917312     & 1.092629       & TRUE              & 6.009942   & 1.522583     & 0.364891       & TRUE              & 1.887475      \\
16       & 2.190046 & 5.026035     & 1.49242        & TRUE              & 6.518455   & 1.566236     & 0.022168       & TRUE              & 1.588404      \\
17       & 20.94913 & --- & --- & \textbf{Time-Out}          & --- & 2.796172     & 0.233188       & TRUE              & 3.02936       \\
18       & 2.452495 & 5.109881     & 0.011308       & TRUE              & 5.121189   & 1.586266     & 0.006406       & TRUE              & 1.592672      \\
19       & 2.872853 & 5.027172     & 2.115648       & TRUE              & 7.14282    & --- & --- & \textbf{FALSE}             & Faulty Result \\
20       & 7.979547 & 4.969257     & 0.051209       & TRUE              & 5.020466   & 1.571463     & 0.025758       & TRUE              & 1.59722       \\
21       & 10.15052 & 5.041602     & 0.023377       & TRUE              & 5.064978   & 1.675484     & 0.002315       & TRUE              & 1.677799      \\
22       & 50.0754  & --- & --- & \textbf{Time-Out}          & --- & --- & --- & \textbf{Not Supported}     & --- \\
23       & 5.611648 & --- & --- & \textbf{Time-Out}          & --- & 3.276078     & 9.46E-04       & TRUE              & 3.277024      \\
24       & 4.254013 & 5.531002     & 26.98154       & TRUE              & 32.51254   & --- & --- & \textbf{Not Supported}     & --- \\
25       & 5.371955 & 4.921379     & 14.473         & TRUE              & 19.39438   & 1.642452     & 87.58693       & TRUE              & 89.22938      \\
26       & 3.551616 & 4.961935     & 0.053799       & TRUE              & 5.015734   & 1.802828     & 0.049044       & TRUE              & 1.851872      \\
27       & 58.71541 & --- & --- & \textbf{Time-Out}          & --- & --- & --- & \textbf{Not Supported}     & --- \\
28       & 2.696608 & 5.001502     & 1.842646       & TRUE              & 6.844148   & --- & --- & \textbf{FALSE}             & Faulty Result \\
29       & 6.087198 & 5.166301     & 444.8778       & TRUE              & 450.0441   & --- & --- & \textbf{FALSE}             & Faulty Result \\
30       & 19.5853  & --- & --- & \textbf{Time-Out}          & --- & --- & --- & \textbf{Not Supported}     & --- \\
31       & \textbf{Mem-Out}  & --- & --- & \textbf{Time-Out}          & --- & --- & --- & \textbf{Time-Out}          & --- \\
32       & 42.91233 & 4.727888     & 0.077733       & TRUE              & 4.80562    & --- & --- & \textbf{Not Supported}     & --- \\
33       & 157.9098 & --- & --- & \textbf{Time-Out}          & --- & --- & --- & \textbf{Not Supported}     & --- \\
34       & 49.66895 & --- & --- & \textbf{Time-Out}          & --- & --- & --- & \textbf{Not Supported}     & Faulty Result \\
35       & 19.98414 & 4.777473     & 31.81758       & TRUE              & 36.59506   & --- & --- & \textbf{FALSE}             & Faulty Result \\
36       & 5.953144 & --- & --- & \textbf{Time-Out}          & --- & --- & --- & \textbf{FALSE}             & \#REF!        \\
37       & 1.678499 & 5.169541     & 0.980466       & TRUE              & 6.150007   & 1.800935     & 0.123026       & TRUE              & 1.923961      \\
38       & 1.635076 & 4.917589     & 0.055344       & TRUE              & 4.972934   & --- & --- & \textbf{FALSE}             & Faulty Result \\
39       & 4.528049 & 4.856375     & 0.01099        & TRUE              & 4.867365   & 1.785794     & 0.012408       & TRUE              & 1.798202      \\
40       & 6.223416 & 5.106296     & 1.540247       & TRUE              & 6.646543   & 1.830541     & 0.002604       & TRUE              & 1.833145      \\
41       & 56.67597 & --- & --- & \textbf{Time-Out}          & --- & --- & --- & \textbf{Not Supported}     & --- \\
42       & 5.210878 & --- & --- & \textbf{Time-Out}          & --- & --- & --- & \textbf{Not Supported}     & --- \\
43       & 4.596554 & 4.697143     & 487.1795       & TRUE              & 491.8766   & --- & --- & \textbf{FALSE}             & Faulty Result \\
44       & 8.64179  & 4.862944     & 110.3951       & TRUE              & 115.2581   & 1.817118     & 28.31934       & TRUE              & 30.13645      \\
45       & 5.912908 & --- & --- & \textbf{Time-Out}          & --- & --- & --- & \textbf{Not Supported}     & --- \\
46       & 3.006664 & 4.718116     & 0.013131       & TRUE              & 4.731248   & 1.789845     & 0.038063       & TRUE              & 1.827909      \\
47       & 18.94113 & 4.876583     & 0.332155       & TRUE              & 5.208737   & --- & --- & \textbf{FALSE}             & Faulty Result \\
48       & 2.847446 & 4.87816      & 0.286808       & TRUE              & 5.164968   & 1.774782     & 0.042734       & TRUE              & 1.817516      \\
49       & 2.436214 & 4.765324     & 306.4633       & TRUE              & 311.2286   & --- & --- & \textbf{FALSE}             & Faulty Result \\
\bottomrule
\end{tabular}
}
\centering
\caption{gMark Social Benchmark Results}
\label{tab:gSocial}
\end{table*}

\begin{table*}[]
\resizebox{\textwidth}{!}{\begin{tabular}{cccccccccc}
\toprule
         & \textbf{\name}     & \multicolumn{4}{c}{\textbf{Fuseki}}                                     & \multicolumn{4}{c}{\textbf{Virtuoso}}                                      \\
         \cmidrule(lr){2-2} \cmidrule(lr){3-6} \cmidrule(lr){7-10}
\textbf{Query ID} & \textbf{Total Time}     & \textbf{Loading Time} & \textbf{Execution Time} & \textbf{Res Equal} & \textbf{Total Time} & \textbf{Loading Time} & \textbf{Execution Time} & \textbf{Res Equal} & \textbf{Total Time}    \\
0        & 29.3576  & --- & --- & \textbf{Time-Out}  & --- & --- & --- & \textbf{Not Supported} & --- \\
1        & 6.106013 & --- & --- & \textbf{Time-Out}  & --- & 1.087469     & 0.322955       & TRUE          & 1.410424      \\
2        & 7.146815 & 1.743377     & 237.1209       & TRUE      & 238.8643   & 0.66307      & 1.527345       & TRUE          & 2.190415      \\
3        & 0.81259  & 1.745625     & 1.969295       & TRUE      & 3.71492    & 0.674981     & 0.066597       & TRUE          & 0.741577      \\
4        & 5.301735 & 1.733303     & 0.018666       & TRUE      & 1.751969   & 0.666493     & 0.026623       & TRUE          & 0.693117      \\
5        & 1.705723 & 1.723611     & 16.47228       & TRUE      & 18.19589   & 0.671292     & 0.609226       & TRUE          & 1.280517      \\
6        & 37.87879 & --- & --- & \textbf{Time-Out}  & --- & --- & --- & \textbf{Not Supported} & --- \\
7        & 1.388953 & 1.739257     & 3.015784       & TRUE      & 4.75504    & 0.674396     & 0.676409       & TRUE          & 1.350805      \\
8        & 27.33182 & 1.693349     & 93.86912       & TRUE      & 95.56246   & 0.657102     & 0.319835       & TRUE          & 0.976937      \\
9        & 4.950027 & 1.670561     & 0.01266        & TRUE      & 1.683221   & --- & --- & \textbf{FALSE}         & Faulty Result \\
10       & 3.90854  & 1.684732     & 0.008307       & TRUE      & 1.693039   & 0.651951     & 0.010579       & TRUE          & 0.662529      \\
11       & 4.221005 & --- & --- & \textbf{Time-Out}  & --- & --- & --- & \textbf{Time-Out}      & --- \\
12       & 19.01537 & 1.82695      & 0.011788       & TRUE      & 1.838737   & 0.687586     & 0.229096       & TRUE          & 0.916682      \\
13       & 2.644757 & 1.895311     & 336.0234       & TRUE      & 337.9187   & 0.699101     & 36.34624       & TRUE          & 37.04534      \\
14       & 16.52763 & --- & --- & \textbf{Time-Out}  & --- & 1.034326     & 8.59E-04       & TRUE          & 1.035186      \\
15       & \textbf{Mem-Out}  & --- & --- & \textbf{Time-Out}  & --- & --- & --- & \textbf{Time-Out}      & --- \\
16       & 8.941485 & --- & --- & \textbf{Time-Out}  & --- & --- & --- & \textbf{FALSE}         & Faulty Result \\
17       & 1.664635 & 1.823395     & 0.270428       & TRUE      & 2.093823   & 0.622421     & 0.048782       & TRUE          & 0.671203      \\
18       & 1.118286 & 1.76035      & 0.006781       & TRUE      & 1.767131   & 0.621048     & 0.222413       & TRUE          & 0.843462      \\
19       & 0.751237 & --- & --- & \textbf{Time-Out}  & --- & --- & --- & \textbf{Not Supported} & --- \\
20       & 6.105985 & --- & --- & \textbf{Time-Out}  & --- & --- & --- & \textbf{Not Supported} & --- \\
21       & 22.34224 & --- & --- & \textbf{Time-Out}  & --- & --- & --- & \textbf{Not Supported} & --- \\
22       & 19.75693 & --- & --- & \textbf{Time-Out}  & --- & --- & --- & \textbf{Not Supported} & --- \\
23       & 1.210083 & 1.840726     & 1.230116       & TRUE      & 3.070842   & 0.643051     & 0.120556       & TRUE          & 0.763607      \\
24       & 1.852502 & --- & --- & \textbf{Time-Out}  & --- & --- & --- & \textbf{Time-Out}      & --- \\
25       & 0.672278 & 1.829289     & 0.600554       & TRUE      & 2.429843   & 0.628258     & 0.036516       & TRUE          & 0.664774      \\
26       & 8.776724 & --- & --- & \textbf{Time-Out}  & --- & --- & --- & \textbf{Not Supported} & --- \\
27       & 1.655394 & 1.686786     & 556.2064       & TRUE      & 557.8932   & 0.678793     & 58.42757       & TRUE          & 59.10637      \\
28       & 10.03315 & 1.701909     & 1022.024       & TRUE      & 1023.726   & 0.729646     & 144.8955       & TRUE          & 145.6251      \\
29       & 28.90678 & --- & --- & \textbf{Time-Out}  & --- & --- & --- & \textbf{Not Supported} & --- \\
30       & 17.07449 & --- & --- & \textbf{Time-Out}  & --- & --- & --- & \textbf{FALSE}         & Faulty Result \\
31       & 1.889749 & --- & --- & \textbf{Time-Out}  & --- & 0.767876     & 268.1806       & TRUE          & 268.9484      \\
32       & 4.435148 & 1.84552      & 0.011991       & TRUE      & 1.857511   & 0.700008     & 0.018887       & TRUE          & 0.718895      \\
33       & 0.981107 & 1.87802      & 22.94484       & TRUE      & 24.82286   & 0.672669     & 15.62866       & TRUE          & 16.30133      \\
34       & 2.307586 & 1.908785     & 17.86245       & TRUE      & 19.77124   & 0.687419     & 1.033115       & TRUE          & 1.720534      \\
35       & 4.527377 & 1.812636     & 265.5669       & TRUE      & 267.3796   & 0.687563     & 32.03738       & TRUE          & 32.72494      \\
36       & 29.09114 & --- & --- & \textbf{Time-Out}  & --- & --- & --- & \textbf{FALSE}         & Faulty Result \\
37       & 0.688185 & 1.980404     & 0.00798        & TRUE      & 1.988384   & 0.697309     & 0.01165        & TRUE          & 0.708959      \\
38       & 1.04346  & 1.871827     & 113.9019       & TRUE      & 115.7738   & 0.771054     & 63.2209        & TRUE          & 63.99196      \\
39       & 4.186533 & 1.914543     & 0.024354       & TRUE      & 1.938897   & 0.717747     & 0.038452       & TRUE          & 0.756199      \\
40       & 8.279989 & --- & --- & \textbf{Time-Out}  & --- & --- & --- & \textbf{Time-Out}      & --- \\
41       & 0.73241  & 1.898214     & 69.61925       & TRUE      & 71.51746   & 0.660548     & 2.44178        & TRUE          & 3.102328      \\
42       & 0.817126 & 1.833329     & 4.398628       & TRUE      & 6.231957   & 0.719869     & 0.328373       & TRUE          & 1.048242      \\
43       & 32.35511 & --- & --- & \textbf{Time-Out}  & --- & --- & --- & \textbf{Not Supported} & --- \\
44       & 133.7689 & --- & --- & \textbf{Time-Out}  & --- & --- & --- & \textbf{Mem-Out}       & --- \\
45       & 1.027826 & 1.753663     & 1.929043       & TRUE      & 3.682706   & 0.656611     & 0.199623       & TRUE          & 0.856234      \\
46       & 2.693819 & 1.759759     & 2.605928       & TRUE      & 4.365688   & 0.672731     & 0.392371       & TRUE          & 1.065102      \\
47       & 3.325502 & 1.833093     & 29.93909       & TRUE      & 31.77218   & 0.712115     & 2.949679       & TRUE          & 3.661795      \\
48       & 184.1826 & --- & --- & \textbf{Time-Out}  & --- & 0.892285     & 0.420084       & TRUE          & 1.312369      \\
49       & 5.807983 & 1.761885     & 47.55949       & TRUE      & 49.32137   & 0.685809     & 0.314163       & TRUE          & 0.999972     \\
\bottomrule
\end{tabular}
}
\centering
\caption{gMark Test Benchmark Results}
\label{tab:gTest}
\end{table*}

\begin{table*}[]
\resizebox{\textwidth}{!}{\begin{tabular}{cccccccccc}
\toprule
         & \textbf{\name}     & \multicolumn{4}{c}{\textbf{Fuseki}}                                     & \multicolumn{4}{c}{\textbf{Virtuoso}}                                      \\
         \cmidrule(lr){2-2} \cmidrule(lr){3-6} \cmidrule(lr){7-10}
\textbf{Query ID} & \textbf{Total Time}     & \textbf{Loading Time} & \textbf{Execution Time} & \textbf{Res Equal} & \textbf{Total Time} & \textbf{Loading Time} & \textbf{Execution Time} & \textbf{Res Equal} & \textbf{Total Time}    \\
0        & 0.255653 & 1.370513     & 0.01779        & TRUE      & 1.388303   & 0.924484     & 0.024339       & TRUE      & 0.948822   \\
1        & 0.916753 & 1.376974     & 0.531686       & TRUE      & 1.90866    & 0.813104     & 0.141013       & TRUE      & 0.954116   \\
2        & 1.103546 & 1.397222     & 0.11726        & TRUE      & 1.514483   & 0.752074     & 0.050747       & TRUE      & 0.802822   \\
3        & 0.957007 & 1.366825     & 0.028847       & TRUE      & 1.395672   & 0.805103     & 0.00634        & TRUE      & 0.811443   \\
4        & 0.788907 & 1.358605     & 0.021999       & TRUE      & 1.380604   & 0.773406     & 0.003473       & TRUE      & 0.776879   \\
5        & 2.446126 & 1.404242     & 207.2158       & TRUE      & 208.62     & 0.755929     & 1.769996       & TRUE      & 2.525925   \\
6        & 5.827524 & 1.419442     & 43.60139       & TRUE      & 45.02083   & 0.719526     & 0.035193       & TRUE      & 0.754719   \\
7        & 0.31812  & 1.455698     & 0.442014       & TRUE      & 1.897712   & 0.656873     & 0.04001        & TRUE      & 0.696883   \\
8        & 3.091841 & 1.451437     & 1.853865       & TRUE      & 3.305302   & 0.681998     & 0.132236       & TRUE      & 0.814234   \\
9        & 1.480341 & 1.497634     & 25.45474       & TRUE      & 26.95237   & 0.723982     & 0.011736       & TRUE      & 0.735719   \\
10       & 2.385485 & 1.494223     & 0.039226       & TRUE      & 1.533449   & 0.753507     & 0.017065       & TRUE      & 0.770571   \\
11       & 0.440275 & 1.494761     & 0.158715       & TRUE      & 1.653476   & 0.773565     & 0.008843       & TRUE      & 0.782408   \\
12       & 0.289779 & 1.48597      & 0.019753       & TRUE      & 1.505723   & 0.758892     & 0.007191       & TRUE      & 0.766083   \\
13       & 0.34749  & 1.576266     & 0.121115       & TRUE      & 1.697381   & 0.758498     & 0.060892       & TRUE      & 0.81939    \\
14       & 4.737758 & 1.57317      & 0.035073       & TRUE      & 1.608243   & 0.748497     & 0.002614       & TRUE      & 0.751111   \\
15       & 1.835838 & 1.552763     & 0.01046        & TRUE      & 1.563222   & 0.737808     & 0.012529       & TRUE      & 0.750336   \\
16       & 0.212753 & 1.547268     & 0.009173       & TRUE      & 1.556441   & 0.734842     & 7.08E-04       & TRUE      & 0.73555   \\
\bottomrule
\end{tabular}
}
\centering
\caption{SP2Bench Benchmark Results}
\label{tab:SP2Bench}
\end{table*}

\clearpage

\subsection{Ontological Reasoning Results}
\label{app:ontologicalReasoning}

One of the main advantages of our \name system is that it provides 
a uniform and consistent framework for reasoning and querying Knowledge Graphs. 
We therefore wanted to measure the performance of query answering in the 
presence of an ontology. Since Fuseki and Virtuoso do not provide such support, we now compare \name with Stardog, which is a commonly accepted state-of-the-art system for reasoning and querying within the Semantic Web. 
Furthermore, we have also created a benchmark based on SP2Bench's dataset that provides property path queries and ontological concepts such as subPropertyOf and subClassOf. We have provided this benchmark in the supplementary material.

\begin{figure}[h!]
  \centering
  \includegraphics[scale=0.12]{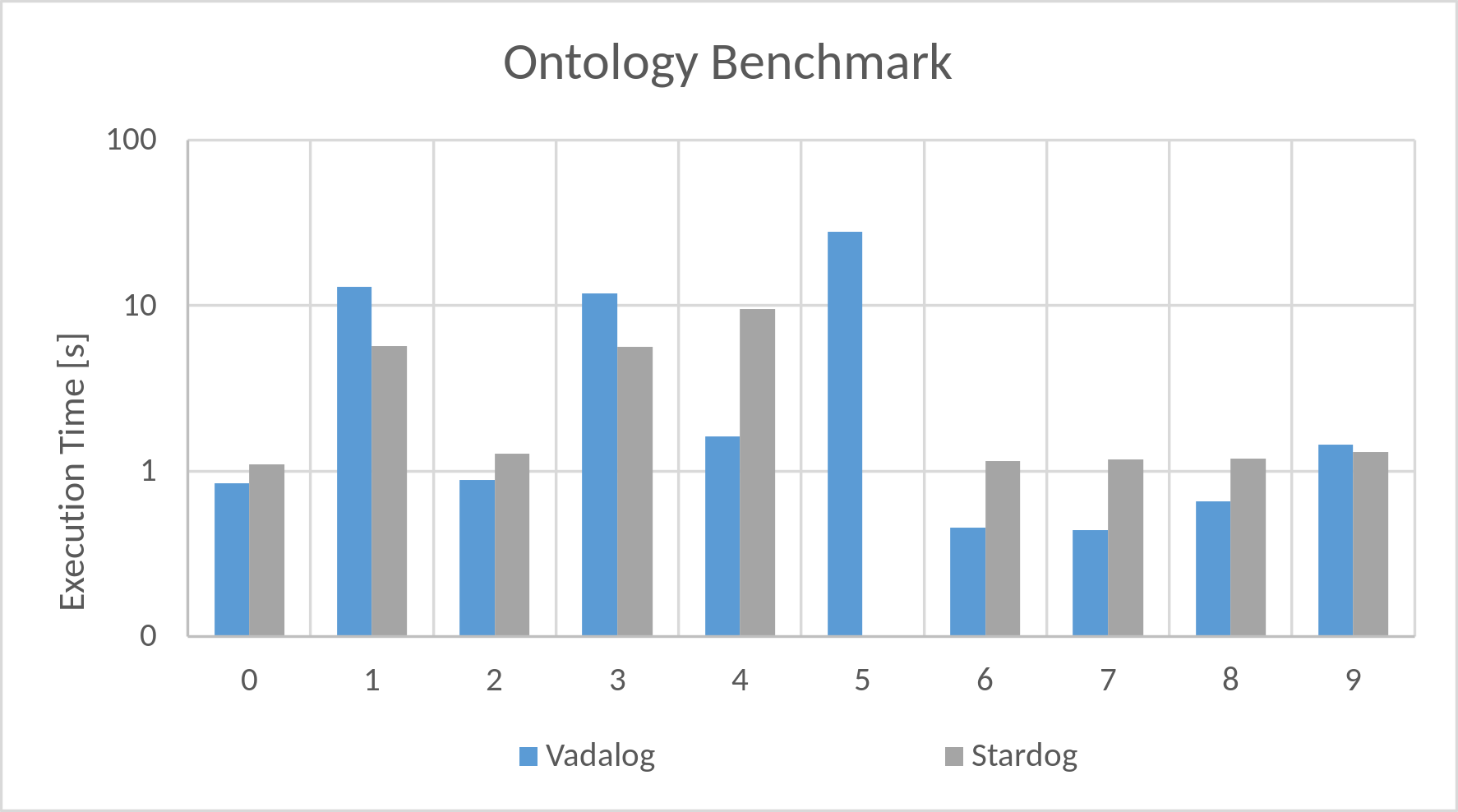} 
  \caption{Ontology Benchmark}
  \label{fig:OntologyBenchM}
\vspace{-5pt}  
\end{figure}

Figure \ref{fig:OntologyBenchM} shows the outcome of these experiments. 
In summary, we note that \name is faster than Stardog on most queries. Particularly interesting are queries 4 and 5, which contain recursive property path queries with two variables. Our engine needs on query 4 only about a fifth of the execution time of Stardog and it can even answer query 5, on which Stardog times outs (using a timeout of $900s$). On the other queries, Stardog and \name perform similarly.

\section{More on Datalog$^{\pm}$}

Recall from Section \ref{sect:sparql} that query answering under an ontology defined by 
Datalog$^{\pm}$ program comes down to solving an entailment problem. More precisely, let $Q(\vec{z})$ be a CQ with free variables $\vec{z}$ over database $D$ and let 
an ontology be expressed by Datalog$^{\pm}$  program $\Pi$. 
Then the answers to $Q(\vec{z})$ over $D$ under ontology $\Pi$ 
are defined as $\{\vec{a} \mid \Pi \cup D \models Q(\vec{a})\}$, where
$\vec{a}$ is a tuple of the same arity as $\vec{z}$ with values from the domain of $D$.

\paragraph{Canonical model and the chase}
Note that $\Pi \cup D$ can have many models. A canonical model is obtained via the {\em chase\/}, which is defined as follows: 
We say that a rule $\rho \in \Pi$ with head 
$p(\vec{z})$ is {\em applicable\/} to an instance $I$ if there exists a homomorphism $h$ from the body of $\rho$ to $I$. 
We may then carry out a {\em chase step\/}, which consists in 
adding atom $h'(p(\vec{z}))$ to instance $I$, 
where $h'$ coincides with $h$ on all variables occurring in
the body of $\rho$ and $h'$ maps each existential variable in $p(\vec{z})$
to a fresh {\em labelled null\/} not occurring in $I$.
A {\em chase sequence} for database $D$ and program $\Pi$ is a sequence of instances $I_0, I_1, \dots$ obtained by applying a sequence of chase steps, starting with $I_0 = D$. The union of instances obtained by all possible 
chase sequences is referred to as $\mathit{Chase}(D,\Pi)$. 
The labelled nulls in $\mathit{Chase}(D,\Pi)$ play the same role as
blank nodes in an RDF graph, i.e., resources for which the concrete value 
is not known. The importance of $\mathit{Chase}(D,\Pi)$ comes from the
equivalence $\Pi \cup D \models Q(\vec{a})
\Leftrightarrow \mathit{Chase}(D,\Pi) \models  Q(\vec{a})$
\cite{DBLP:journals/tcs/FaginKMP05}. 
Note however that, in general,  
$\mathit{Chase}(D,\Pi)$ is infinite. Hence, the previous
equivalence does not yield an algorithm to evaluate a CQ 
$Q(\vec{z})$ w.r.t.\ database $D$ and program $\Pi$. 
In fact, without restriction, this is an undecidable 
problem~\cite{DBLP:journals/jcss/JohnsonK84}. 
Several subclasses of  Datalog$^{\pm}$ have thus been 
presented 
\cite{DBLP:journals/tcs/FaginKMP05,DBLP:journals/jair/CaliGK13,pvldb/CaliGP10,DBLP:conf/rr/CaliGP10,BagetLMS09,BagetLMS11,Arenas2014ExpressiveLF}
that ensure decidability of CQ answering 
(see \cite{DBLP:journals/jair/CaliGK13} for an overview).

\paragraph{Bag semantics of Datalog}
In  \cite{DBLP:conf/vldb/MumickPR90},
a bag semantics of Datalog was introduced 
based on {\em derivation trees}. 
Given a database $D$ and Datalog program $\Pi$, 
a {\em derivation tree} (DT) 
is a tree $T$ with node and edge labels, such that either (1) $T$ consists of a single node
labelled by an atom from $D$ or (2) $\Pi$ contains a rule 
$\rho \colon \ H \leftarrow A_1,A_2,\ldots,A_k$ with $k>0$, 
and there exist DTs $T_1,\ldots, T_k$ whose root nodes are 
labelled with atoms $C_1, \ldots, C_k$ such that 
$A_1, \ldots,A_k$ are simultaneously matched to $C_1, \ldots, C_k$ by applying 
some substitution $\theta$, 
and $T$ is obtained as follows: 
$T$ has a new root node $r$ with label $H\theta$ and the $k$ root nodes of the DTs
$T_1,\ldots, T_k$ are appended as child nodes of $r$ in this order. 
All edges from $r$ to its child nodes are labelled
with $\rho$. Then the bag semantics of program $\Pi$ over database $D$ consists of all ground atoms
derivable from $D$ by $\Pi$, and the multiplicity 
$m \in \mathbb{N} \cup \{\infty\}$
of each such atom $A$ is the number of possible DTs
with root label $A$. Datalog with bag semantics 
is readily extended 
by stratified negation~\cite{DBLP:conf/adc/MumickS93}: 
the second condition of the definition of DTs now 
has to take
negative body atoms in a rule 
$\rho \colon H \leftarrow A_1,A_2,\ldots,A_k, \neg B_1, \dots \neg B_\ell$ 
with $k>0$ and $\ell \geq 0$ with head atom $H$ from some stratum $i$ 
into account in that we request that none of the atoms $B_1\theta, \dots B_\ell\theta$ can be derived
from $D$ via the rules in $\Pi$ from strata less than $i$.

\paragraph{Bag semantics via set semantics of Warded Datalog$^\pm$}
In \cite{BagSemantic} it was shown how Datalog with {\em bag} semantics 
can be 
transformed into Warded Datalog$^\pm$ with {\em set} semantics. The idea
is to replace 
every predicate $P(\ldots)$ by a
new version $P(. \ ;\ldots)$ with an extra, first argument 
to accommodate 
a labelled null which is interpreted as tuple ID (TID). Each rule 
in $\Pi$ of the form 

\begin{center}
$\rho \colon H(\bar{x}) \ \leftarrow \ A_1(\bar{x}_1),A_2(\bar{x}_2),\ldots,A_k(\bar{x}_k)$, with $k>0, \bar{x} \subseteq \cup_i \bar{x}_i$     
\end{center}

\noindent
is then transformed into the Datalog$^\pm$ rule  

\begin{center}
$\rho' \colon \ \exists z \ H(z;\bar{x}) \leftarrow A_1(z_1;\bar{x}_1),A_2(z_2;\bar{x}_2),\ldots,A_k(z_k;\bar{x}_k)$,     
\end{center}

\noindent
with fresh, distinct variables $z, z_1, \ldots,z_k$.
Some care (introducing auxiliary predicates) 
is required for rules with negated body atoms so as not to 
produce unsafe negation. A Datalog rule \\
$\rho \colon  H(\bar{x}) \ \leftarrow \ A_1(\bar{x}_1), \ldots, A_k(\bar{x}_k),
\neg B_1(\bar{x}_{k+1}), \ldots, \neg B_\ell(\bar{x}_{k+\ell})$ 
with, $\bar{x}_{k+1}, \ldots, \bar{x}_{k+\ell}  \subseteq \bigcup_{i=1}^k \bar{x}_i$ is replaced by 
$\ell + 1$ rules in the corresponding 
Datalog$^\pm$ program $\Pi'$: 

\smallskip\noindent
$\rho'_0 \colon  \exists z H(z;\bar{x}) \leftarrow A_1(z_1;\bar{x}_1), \ldots, A_k(z_k;\bar{x}_k)$, \\
\phantom{$\rho'_0 \colon  \exists z H(z;\bar{x}) \leftarrow $}
$\neg \mathit{Aux}_1^\rho(\bar{x}_{k+1}), \ldots, \neg \mathit{Aux}_\ell^\rho(\bar{x}_{k+\ell})$,\\
$\rho'_{i} \colon \mathit{Aux}_i^\rho(\bar{x}_{k +i}) \leftarrow B_i(z_i;\bar{x}_{k+i}), \ \ i = 1, \ldots, \ell$. 

\smallskip\noindent
The resulting Datalog$^\pm$ program $\Pi'$ is trivially warded since the rules 
thus produced contain no dangerous variables at all. Moreover, it is proved in 
\cite{BagSemantic} that an atom $P(\vec{a})$ is in the DT-defined bag semantics
of Datalog program $\Pi$ over database $D$ with multiplicity $m \in \mathbb{N} \cup \{\infty\}$, 
iff $\mathit{Chase}(D,\Pi')$ contains atoms of the form 
$P(t;\vec{a})$ for $m$ distinct labelled nulls $t$ (i.e., the tuple IDs).

\fi

\end{document}